\documentclass[12pt]{article}
\pdfoutput=1
\usepackage{jheppub}

\usepackage[mathscr]{euscript}
\usepackage[inline]{enumitem}
\usepackage{amsmath}
\usepackage{amsfonts}
\usepackage{amssymb}
\usepackage{latexsym}

\numberwithin{equation}{section}

\usepackage{bbold}

\usepackage{mathtools}

\usepackage{epsf}
\usepackage{color}
\usepackage{graphicx}
\usepackage{tikz}
\usepackage{dsfont}
\usepackage{subfig}

\usepackage[export]{adjustbox}
\usepackage{hyperref}

\usetikzlibrary{arrows,shapes,positioning}
\usetikzlibrary{decorations.markings}
\usepackage[rightcaption]{sidecap}
\tikzstyle arrowstyle=[scale=1]
\tikzstyle directed=[postaction={decorate,decoration={markings,
    mark=at position .65 with {\arrow[arrowstyle]{stealth}}}}]
\tikzstyle reverse directed=[postaction={decorate,decoration={markings,
    mark=at position .65 with {\arrowreversed[arrowstyle]{stealth};}}}]
\usetikzlibrary{positioning}

\graphicspath{{Figures/}}

\newcounter{row}

\newenvironment{imgrows}[1][\textwidth]%
  {\begin{minipage}{#1}%
   \setcounter{row}{0}%
   \stepcounter{figure}%
  }%
  {\addtocounter{figure}{-1}%
   \end{minipage}%
  }
\newcommand\imgrow
  {\par\noindent
   \refstepcounter{row}%
   \makebox[1.5em][r]{(\alph{row})}\
  }

\AtBeginDocument{}

\renewcommand{\thefootnote}{\fnsymbol{footnote}}
\renewcommand{\thanks}[1]{\footnote{#1}}
\newcommand{\starttext}{
\setcounter{footnote}{0}
\renewcommand{\thefootnote}{\arabic{footnote}}}

\newcommand{\bea}{\begin{eqnarray}}
\newcommand{\eea}{\end{eqnarray}}
\newcommand{\be}{\begin{equation}}
\newcommand{\ee}{\end{equation}}
\newcommand{\bal}{\begin{align}}
\newcommand{\eal}{\end{align}}
\newcommand{\<}{\langle}
\renewcommand{\>}{\rangle}

\def\j{\psi}
\def\jb{\bar{\psi}}
\def\mub{\bar{\mu}}

\def\scrJ{{\mathscr{J}}}

\def\cA{{\cal A}}

\def\cC{{\cal C}}
\def\cD{{\cal D}}
\def\cE{{\cal E}}

\def\cJ{{\cal J}}

\def\cQ{{\cal Q}}

\def\cS{{\cal S}}

\def\cZ{{\cal Z}}

\def\Im{{\rm Im \,}}
\def\tr{{\rm tr}}
\def\Tr{{\rm Tr}}

\def\a{\alpha}
\def\b{\beta}

\def\m{\mu}

\def\){\right)}
\def\({\left( }
\def\]{\right] }
\def\[{\left[ }

\def\no{\nonumber}

\usepackage{comment}

\def\JUM{\tilde J} 
\def\cJCSYK{\tilde \cJ} 

\makeatletter\def\l@subsubsection#1#2{}%
\makeatother

\DeclareMathOperator\arctanh{arctanh}

\title{Complex Sachdev-Ye-Kitaev model in the double scaling limit}
\author{Micha Berkooz\texorpdfstring{${}^a$}{},}
\author{Vladimir Narovlansky\texorpdfstring{${}^{a,b}$}{},}
\author{Himanshu Raj\texorpdfstring{${}^a$}{}}
\emailAdd{micha.berkooz@weizmann.ac.il}
\emailAdd{narovlansky@princeton.edu}
\emailAdd{himanshu.raj@weizmann.ac.il}
\affiliation{${}^a$Department of Particle Physics and Astrophysics, \\ Weizmann Institute of Science, Rehovot 7610001, Israel
}
\affiliation{${}^b$Princeton Center for Theoretical Science, \\  Princeton University, Princeton, NJ 08544, USA}

\abstract{
We solve for the exact energy spectrum, 2-point and 4-point functions of the complex SYK model, in the double scaling limit at all energy scales.
This model has a $U(1)$ global symmetry.
The analysis shows how to incorporate a chemical potential in the chord diagram picture, and we present results for the various observables also at a given fixed charge sector.
In addition to matching to the spectral asymmetry, we consider an analogous asymmetry measure of the 2-point function obeying a non-trivial dependence on the operator's dimension.
We also provide the chord diagram structure for an SYK-like model that has a $U(M)$ global symmetry at any disorder realization.
We then show how to exactly compute the effect of inserting very heavy operators, with formally infinite conformal dimension. The latter separate the gravitational spacetime into several parts connected by an interface, whose properties are exactly computable at all scales. In particular, light enough states can still go between the spaces. This behavior has a simple description in the chord diagram picture.

}

\begin{document}

\starttext
\setcounter{footnote}{0}

\maketitle

\setcounter{equation}{0}
\setcounter{footnote}{0}


\section{Introduction} \label{sec:intro}

The Sachdev-Ye-Kitaev (SYK) model \cite{Sachdev:1992fk,Sachdev:2010um,Kitaev:2015lct,Polchinski:2016xgd,Maldacena:2016hyu} is a quantum mechanical model ($0+1$ dimensions) of $N$ fermions with random all-to-all interactions, which can nevertheless be studied analytically in the large $N$ limit. The SYK model exhibits various interesting properties. In particular, it is a highly chaotic model, with a maximal quantum Lyapunov exponent at low temperatures \cite{Maldacena:2015waa}.
Letting $\chi _i$ be $N$ Majorana fermions, $i=1,\cdots ,N$, such that $\{\chi _i,\chi _j\}=2\delta _{ij} $, the Hamiltonian of the SYK model is given by
\begin{equation}\label{realSYKHam}
H = i^{p/2} \sum _{1 \le i_1<\cdots <i_p \le N} J_{i_1 \cdots i_p} \chi_{i_1} \cdots \chi_{i_p}.
\end{equation}
$p$ is a parameter in the model which sets the length of the all-to-all interactions in the Hamiltonian, and the $J_{i_1 \cdots i_p} $ are the random couplings.

The model can be analyzed using Feynman diagrams. At large $N$, the dominant diagrams are the so-called melonic diagrams. The sum of all those diagrams is described by a set of Schwinger-Dyson (SD) equations \cite{Sachdev:1992fk,Sachdev:2010um,Kitaev:2015lct,Polchinski:2016xgd,Maldacena:2016hyu}, and at low energies the model has a conformal regime in which the Schwinger-Dyson equations can be solved.
At low energies, the equations are invariant under a symmetry of time reparametrizations, which is broken spontaneously (and also explicitly in the full theory).
Beyond the strictly conformal regime, the modes which describe this reparametrization symmetry are lifted by the Schwarzian action (see also \cite{Bagrets:2016cdf, Bagrets:2017pwq,Mertens:2017mtv,Lam:2018pvp,Maldacena:2016upp,Jensen:2016pah,Stanford:2017thb,Kitaev:2017awl}). Various correlation functions in the Schwarzian theory have been found in \cite{Mertens:2017mtv,Bagrets:2016cdf,Bagrets:2017pwq,Maldacena:2016upp}.

From the holographic point of view, the Schwarzian action is equivalent to Jackiw-Teitelboim gravity in the bulk \cite{Almheiri:2014cka,Jensen:2016pah,Maldacena:2016upp,Iliesiu:2019xuh}.
Recently, it was shown that the partition function of Jackiw-Teitelboim gravity (on surfaces of any genus) is described by a particular double-scaled random matrix model \cite{Saad:2019lba}, and relations of random matrix theory and chaos to the SYK model were discussed in \cite{Cotler:2016fpe,Gharibyan:2018jrp}.

SYK-like models in higher dimensions have been studied in
\cite{parcollet1999non,Berkooz:2016cvq,Gu:2016oyy,Berkooz:2017efq,Murugan:2017eto,Lian:2019axs} (in such theories there are in general interesting disorder effects related to renormalization \cite{Narovlansky:2018muj,Aharony:2018mjm}).
Higher point correlation functions were obtained in
\cite{Gross:2017hcz,Gross:2017aos}, and results beyond the leading order in $N$ were obtained in
\cite{Garcia-Garcia:2016mno,Garcia-Garcia:2017pzl,a:2018kvh}.

The techniques above are applicable when the value of $p$ is taken to be a constant independent of $N$, such as $p=4$, or in the large $p$ limit (as long as $N\rightarrow \infty$ first). However, an interesting scaling was recently studied, known as double-scaled SYK \cite{erdHos2014phase,Cotler:2016fpe,Berkooz:2018qkz,Berkooz:2018jqr}, where $p$ is taken to scale as $\sqrt{N}$. In this limit the model can be solved exactly at all energy scales using combinatorial tools, and it is rigidly controlled by a quantum group symmetry which replaces the conformal symmetry at all energy scales.

In solving double-scaled SYK, the combinatorial description of the partition function, as well as of correlation functions, is given in terms of chord diagrams (which are reviewed in section \ref{sec:chord_diagrams}). The summation of those chord diagrams can be performed analytically. In particular, the 4-point function, which encodes the quantum Lyapunov exponent, was found exactly \cite{Berkooz:2018jqr}. At low energies and in the limit $p^2/N \to 0$, it indeed agrees with the result \cite{Mertens:2017mtv} found from the Schwarzian theory.

The double scaled limit is in a sense also much more universal as many different microscopic models reduce to similar chord diagrams prescription; for example, one can replace the Majorana fermions by Pauli matrices (which commute between different sites) and obtain the same set of chord diagrams, or consider the SUSY model \cite{Berkooz:2020xne} which results in modified rules, still within the chord diagrams framework. It also suppresses spin glass phases in the theory \cite{Baldwin:2019dki}.
Moreover, at low energies and when $p^2/N \to 0$, the model is still described by the Schwarzian theory, which means that it is still equivalent to Jackiw-Teitelboim gravity living in two-dimensional $AdS$ space.

In this paper, we study the version of the SYK model, where the fermions are complex rather than real such that there is a global symmetry, in the double-scaling limit. This is a variant of the original Sachdev-Ye model \cite{Sachdev:1992fk}, which can be written in terms of complex fermions, with four-fermion interactions.

\subsection{Outline and summary of new results}

We start in section \ref{sec:review} where we review the definition of the complex SYK model and mention some known results about it. In particular, the model was studied in the literature in the large $p$ limit ($N$ taken to infinity first and then $p$ is taken to be large), and we will make contact with these results.
Indeed, this large $p$ limit is obtained as a limit of double-scaled SYK as $p^2/N \to 0$.
As the combinatoric approach to double-scaled SYK reduces the calculation of various observables to a description in terms of a summation over chord diagrams, we derive the value assigned to each chord diagram in generic double-scaled SYK theories with complex fermions in section \ref{sec:chord_diagrams}. The result is simply given in terms of rules assigning to every chord and pairs of chords a particular value. Then we use these rules to evaluate observables in the theory at all energies. In section
\ref{sec:partition_function} we calculate the canonical and the grand canonical partition functions of the theory, where in the latter a chemical potential for the global $U(1)$ symmetry is turned on. In addition, we find in section \ref{sec:partition_function_fixed_charge} the exact partition function in every sector of a particular fixed charge, providing a refined information related to the charge. We verify that in the limit $p^2/N \to 0$ these results reduce to the large $p$ results from the literature. In section \ref{sec:UMSYK} we consider another SYK model with a global symmetry, namely a $U(M)$ symmetry, and use similar tools to get the partition function with chemical potential for the various Cartan generators.

We then go on to study correlation functions. In section \ref{sec:2ptFunctionSec} we find the full two-point function (in a fixed chemical potential and in a fixed charge sector). We go to the limit exhibiting a conformal behavior, as well as get small corrections to it. As a check, we match to the spectral asymmetry factor in the theory.
We also consider an analogous measure of asymmetry defined by the two-point function with these corrections, exhibiting a slightly different behavior; in particular, it is not simply determined by the charge of the operator, but depends also on its dimension.
In section \ref{sec:chaos} we calculate the four-point function. In particular, this gives the Lyapunov exponent in the limit of small $p^2/N$ and we match this with the literature.
In section \ref{sec:massive} we discuss the effect of very heavy operators. These are operators that in the low energy limit do not go to operators of a finite conformal dimension, but rather formally have an infinite dimension. They have a significant effect of separating spacetime into two spaces, as in \cite{Goel:2018ubv}, and the chord diagram picture provides a simple way to see this. We also discuss how very light states can still go between these separated spaces.
We finish with several appendices containing further details to which we refer from the main text.

\section{Review of the complex Sachdev-Ye-Kitaev model} \label{sec:review}

In this section we briefly review the complex SYK model. In subsection \ref{modelDef} we discuss the definition of the model and in subsection \ref{summary1} we review some key results about it that we will make contact with, following  \cite{Davison:2016ngz}.

\subsection{Definition of the model}\label{modelDef}


The complex Sachdev-Ye-Kitaev model \cite{Sachdev:2015efa} is a quantum mechanical model of $N$ complex fermions $\j^i$ and $\jb_i$ (where $i=1,2,...,N$) with random all-to-all interactions.\footnote{In the usual large $N$ scaling, this random model is described by Schwinger-Dyson equations, and has a tensor \cite{Klebanov:2016xxf,Klebanov:2018fzb} and a matrix \cite{Azeyanagi:2017drg,Ferrari:2019ogc} quantum mechanical counterpart models.} The fermions satisfy
\begin{equation}
\left\{ \psi^{i},\bar{\psi_{j}}\right\} =2\delta_{j}^i,\qquad\left\{ \psi^{i},\psi^{j}\right\} =\left\{ \bar{\psi}_{i},\bar{\psi}_{j}\right\} =0,\qquad i,j=1,\cdots,N\ ,\label{complex_Clifford}
\end{equation}
and the model is specified by the Hamiltonian
\begin{equation}\label{Hamiltonian}
H=\sum_{\substack{1\le i_{1}<\cdots<i_{p}\le N\\
1\le j_{1}<\cdots<j_{p}\le N
}
}J_{j_{1}\cdots j_{p}}^{i_{1}\cdots i_{p}}~\bar{\psi}_{i_{p}}\cdots\bar{\psi}_{i_{1}}\psi^{j_{1}}\cdots\psi^{j_{p}}= \sum_{I,I'} J_I^{I'}{\bar\psi}_{I'}\psi^I.
\end{equation}
In the last term above, $I$ and $I'$ denote an index set $I=\left\{ i_{1},i_{2},\cdots,i_{p}\right\} $
consisting of $p$ distinct indices $i_{1}<i_{2}<\cdots<i_{p}$. Fermions with
capital indices stand for the following product of the components
\begin{equation}\label{psiDef}
\begin{split} & \psi^{I}=\psi^{i_{1}}\psi^{i_{2}}\cdots\psi^{i_{p}}~,\\
 & \bar{\psi}_{I}=\bar{\psi}_{i_{p}}\cdots\bar{\psi}_{i_{2}}\bar{\psi}_{i_{1}}~.
\end{split}
\end{equation}
Note that we have reversed the ordering in $\bar{\psi}_I$. This notation will turn out to be convenient later. The couplings $J$ are Gaussian complex random variables satisfying $(J_{j_{1}\cdots j_{p}}^{i_{1}\cdots i_{p}})^{*}=J_{i_{1}\cdots i_{p}}^{j_{1}\cdots j_{p}}$, ensuring the Hermiticity of the Hamiltonian. Their variance is
\begin{equation}
\langle J_{j_{1}\cdots j_{p}}^{i_{1}\cdots i_{p}}J_{i_{1}\cdots i_{p}}^{j_{1}\cdots j_{p}}\rangle_{J}=J^{2}{N \choose p}^{-2} \qquad \text{(no sum)},
\end{equation}
where $J$ is a normalization constant for the disorder.\footnote{Our notations are related to those of \cite{Davison:2016ngz} as follows: $p=q/2$ , $\psi_i=\sqrt{2}f_i$ , $\(J_{j_{1}\cdots j_{p}}^{i_{1}\cdots i_{p}}\)_{\text{here}}=\frac{1}{2^p} \(J_{i_1,i_2,\cdots,i_q}\)_{\text{there}}$ , and for $N \gg p$,
$J_{\text{here}}= \frac{\sqrt{2N}}{2^p}J_{\text{there}}$.
Note that by $J^2_{\text{there}}$ we mean the one in Eq.\ (C14) of \cite{Davison:2016ngz} (that is the one used in the quoted results) which appears to us to differ from the one in Eq.\ (1.2) there (the former being half the latter).
}

The model possesses a $U(1)$ global symmetry that acts on the fermions as follows
\begin{equation}
\psi^i\to \psi^i e^{-i\phi}~,\qquad \bar{\psi}_i\to \bar{\psi}_i e^{i\phi}~.
\end{equation}
The associated conserved charge is the fermion number defined as
\be \label{U1Opdef}
Q=\frac{1}{4} \sum _{i=1}^N \left(\bar \psi_i \psi^i-\psi^i\bar \psi_i\right).
\ee
We will also use the specific charge defined by
\be \label{eq:cQ_def}
\cQ={1\over N} Q,
\ee
which takes values in the range $-1/2<\cQ<1/2$. This will be useful when comparing to existing results in the literature.

This model admits a non-trivial double scaling limit in which
\be
\lambda=\frac{p^2}{N}~~\text{fixed},\ \ \ N\rightarrow\infty\ .
\ee
In this paper we explore the complex SYK model in this limit. But first we present a short summary of known results in the usual, fixed $p$, large $N$ complex SYK model.

\subsection{Summary of known results}\label{summary1}

\subsubsection{Thermodynamics}


In the limit where $N$ is taken to infinity first, at fixed $p$, followed by a zero temperature limit $T\to 0$, the canonical free energy, $F$,\footnote{In \cite{Davison:2016ngz} the free energy is per site (divided by $N$) while here we write the full free energy.} of this model has the following low temperature expansion \cite{Davison:2016ngz} \footnote{In the results quoted here, there is in fact a problem taking the zero temperature limit \cite{TarnopolskyNotes}. However, we obtain results for a chemical potential (or charge) scaling with $N$, which is actually consistent with the careful analysis of \cite{TarnopolskyNotes}. This eliminates the problem just mentioned.}
\be \label{eq:free_energy_low_T}
F(\cQ,T) =E_0(\cQ) -T\cS(\cQ)+\cdots .
\ee
In the above expression, $E_0(\cQ)$ is a non-universal ground state energy and $\cS(\cQ)$ is the universal zero-temperature entropy (universal in the sense that it is independent of the `UV' details of the theory; for example adding higher order fermion interaction terms to the Hamiltonian does not change the result). For generic $p$ the analytic form of the ground state energy is not known. However, it can be computed analytically in a large $p$ expansion and has been found to be
\be\label{grndEn}
E_0(\cQ)\sim -\frac{J \sqrt{N}}{p} (1-4\cQ^2)^{\frac{p+1}{2}}+O(1/p^2)~.
\ee

The universal zero-temperature entropy $\cS(\cQ)$ is a symmetric function of the $U(1)$ charge $\cQ$ and has been computed analytically for any $p>2$. In a large $p$ expansion, the expression for $\cS(\cQ)$ takes the following form
\be\label{zeroTentropy}
\cS(\cQ)/N=  \frac12 \log \(\frac{4}{1-4\cQ^2}\) +\cQ \log\(\frac{1-2\cQ}{1+2\cQ} \)-\frac{\pi^2}{8}(1-4\cQ^2)\frac{1}{p^2}+O(1/p^3)~.
\ee

Because of the non-universality of $E_0(\cQ)$, the thermodynamic grand potential, $\Omega=-T\log Z(\mu,T)$,
has both universal and non-universal pieces. The universal part of $\Omega$ has been computed in \cite{Davison:2016ngz} for generic $p$ from the $G,\Sigma$ action. In the large $p$ limit, the analytic expression for $\Omega$ has been found to be\footnote{We used the convention for the chemical potential used here, which is related to \cite{Davison:2016ngz} by $\mu _{\text{there}} =-2T\mu _{\text{here}} $, and the grand potential here is the total grand potential (rather than per site).}
\be\label{SgPot}
\Omega(\mu,T)=-TN \log (2\cosh \mu)-\frac{\pi v TN}{2\(\cosh\mu\)^2}\[\tan\(\frac{\pi  v}{2}\)-\frac{\pi  v}{4}\]\frac{1}{p^2}+O(1/p^3)~.
\ee
In the above expression $v$ is the solution of the equation
\be\label{curlJDef}
\frac{\pi v}{\cos(\pi v/2)}=\frac{\cJCSYK}{T}~,~~~~\text{where} ~~~~ \cJCSYK=J \frac{2\cosh\mu}{(\cosh\m)^p} \sqrt{\frac{p^2}{N}}~.
\ee

\subsubsection{Two-point function}

A quantity that plays an important role in the complex Sachdev-Ye-Kitaev model is the so called `spectral asymmetry' factor whose thermodynamic definition is given by the charge derivative of the entropy\footnote{There is a universal relation in the usual large $N$ complex SYK model between the spectral asymmetry and the charge \cite{Davison:2016ngz,Gu:2019jub}.}
\begin{align}\label{specAsym}
\begin{split}
\cE =& \frac{1}{2\pi} \frac{d\cS}{d\cQ}~=\\
=&\frac{1}{2\pi}\log \(\frac{1-2\cQ}{1+2\cQ}\) +\frac{\pi}{2p^2}\cQ +O(1/p^3)~.
\end{split}
\end{align}
This factor reflects an asymmetry in the spectral function $A(\omega)$ which is defined as
\be\label{spfun}
A(\omega)= -\frac{1}{\pi} \Im G(\omega +i\epsilon)~,
\ee
where $G(\omega)$ is the Green's function of a single fermion, which in terms of the Euclidean time $\tau$ is defined as follows (note there is no summation over $i$ in the following equation)
\begin{equation}
\begin{split}
G(\tau)&= - \<T_\tau \psi^i(\tau)\bar{\psi}_i(0)\>~\\[5pt]
&= -\frac{1}{Z(\mu,\beta)}\Tr \[ e^{-\beta K} T_\tau \(e^{\tau K} \psi^i e^{-\tau K} \bar{\psi}_i\) \]~.
\end{split}
\end{equation}
In the above expression $T_\tau$ specifies the $\tau$ ordering, $Z(\mu,\beta)$ is the grand canonical partition function and $K$ is the sum of the Hamiltonian and the fermion number operator
\be\label{KDef}
K=H+\frac{\mu}{2\beta} \sum _{i=1}^N \left(\bar \psi_i \psi^i-\psi^i\bar \psi_i\right)~.
\ee

As shown in \cite{Davison:2016ngz}, assuming conformal invariance in the IR, the Green's function $G(\omega )$ in the frequency domain (also in the presence of a chemical potential) at zero temperature takes a scaling form
\be\label{RetG0}
G(z)= C\frac{e^{-i(\pi\Delta+\theta)}}{z^{1-2\Delta}}~,~~~\Im(z)>0
\ee
which can then be plugged in the Schwinger-Dyson equations, fixing the dimension $\Delta =1/(2p)$ and $C$. In the above, $z$ is the complexified frequency. In the $\tau$ domain, one obtains the following ratio
\be \label{eq:ration_of_2pf}
\frac{G(\tau>0)}{G(\tau<0)}=-\frac{\sin(\pi\Delta+\theta)}{\sin(\pi\Delta-\theta)}\equiv - e^{2\pi\cE}~.
\ee
When the net-charge is zero $(\cQ=0)$, $\cE$ vanishes, and otherwise $\cE$ acquires a non-zero value which in the large $p$ limit is given by \eqref{specAsym}.

In this paper, we calculate these various quantities in the double-scaled complex SYK model. We will find that at the leading order in large $p$, our results match with those in  \cite{Davison:2016ngz}.

In the large-$N$ double scaled SYK model, the natural operators to consider are those that are made out of a string of fundamental fermions whose length is of the order of $\sqrt{N}$. For such operators that carry a $U(1)$ charge $\sigma$, the ratio of the Green's functions \eqref{eq:ration_of_2pf} gets modified so that we have $\sigma $ times the spectral asymmetry. In fact, in our analysis we will be able to calculate subleading in $N$ corrections to the ratio of the coefficients (of the time dependence) in the 2-point functions, finding a different behavior where there is also a dependence on the dimension of the operator. However, it should be stressed that the relation to the spectral asymmetry only holds in the conformal regime, which is what we indeed get at leading order in $N$.
The subleading corrections go beyond the scaling regime, and enter in a slightly different measure of asymmetry that we consider.


\section{Computation using chord diagrams} \label{sec:chord_diagrams}
Using chord diagrams, we now present a method for solving the complex Sachdev-Ye-Kitaev model in the double scaling limit
\begin{equation} \label{eq:RealSYKDoubleScale}
N\rightarrow\infty,\ \ \ \lambda ={p^2\over N}~~~\text{held fixed .}
\end{equation}
The method boils down to the calculation of the traces of generic products of complex fermions, and then taking into account the Gaussian structure of the couplings. We begin in subsection \ref{RevRealSYK} by briefly reviewing the chord diagram method for solving the SYK model with Majorana fermions. The generalization to the complex SYK will then be clearer. In subsection \ref{sec:3.2} we work out the formulas for traces of fermions relevant for the complex SYK, and in subsection \ref{sec:large_N_rules} we work out their simplified large $N$ limit. The main result of this section is given in \eqref{general_trace_formula} or \eqref{general_trace_formula2}. The large $N$ rules are summarized in fig.\ \ref{fig:allowed_overlaps}. In the next sections we use these rules to compute the partition functions and correlation functions.

\subsection{Chord diagrams and the real SYK model}\label{RevRealSYK}

\begin{figure}[h]
\centering
\includegraphics[width=0.35\textwidth]{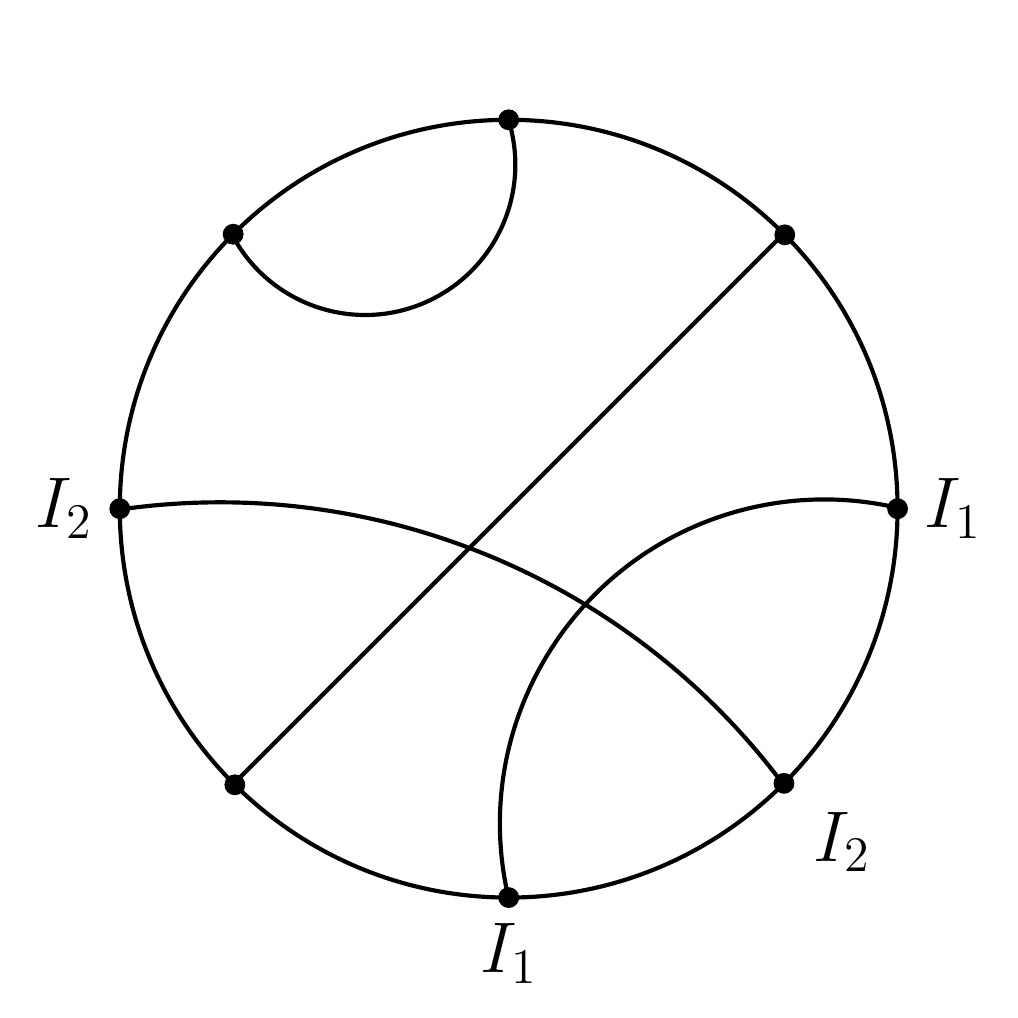}
\caption{An example of a typical chord diagram.}
\label{fig:chord_diagram}
\end{figure}

The real SYK model is a quantum mechanical model of interacting Majorana fermions whose Hamiltonian is given by \eqref{realSYKHam}. If we write the Hamiltonian in a short-hand notation as
\begin{equation}
H= i^{p/2} \sum _I J_I \chi_I~, \qquad I=\{i_1,\cdots ,i_p\}~,
\end{equation}
then the moments of the Hamiltonian are given by
\begin{equation}
m_k=\langle \tr H^k\rangle_J  = i^{kp/2} \sum_{I_1,\cdots ,I_k} \langle  \tr J_{I_1} \chi_{I_1} \cdots J_{I_k} \chi_{I_k} \rangle _J ~.
\end{equation}
The average over the Gaussian random coefficients $J_{I_j}$ is given by Wick's theorem, instructing us to sum over all pairings of the $k$ index sets $I_j$. This is represented combinatorially by chord diagrams: we draw a circle, on which we mark $k$ nodes, corresponding to the Hamiltonian insertions. The nodes are connected in pairs by chords, representing the Wick contractions. An example of a particular chord diagram is shown in fig.\ \ref{fig:chord_diagram}.

For every chord diagram, we are left with a trace over the fermions $\tr (\chi _{I_1} \chi _{I_2} \cdots \chi _{I_1} \cdots )$, where each index set $I_j$ appears twice.
This trace is then evaluated simply by commuting the different $\chi$'s, so that eventually contracted $\chi$'s are next to each other, in which case we can use $\chi_i^2=1$ and the trace becomes trivial. In the chord diagrams, this amounts to disentangling the intersections, so that each intersection, as a consequence of the fermion anti-commutation relations, gives  $(-1)^{|I_j \cap I_{j'}|}$ where $I_j,I_{j'}$ are the index sets corresponding to the intersecting chords, and $|I_j \cap I_{j'}|$ is the size of their (set) intersection. The size of the intersection $|I_j \cap I_{j'}|$ is a random variable (since we sum over the $I_j$'s), which in the large $N$ double scaled limit \eqref{eq:RealSYKDoubleScale} follows a Poisson distribution with mean $p^2/N$ \cite{erdHos2014phase}. Weighting by the Poisson probability distribution, one finds that each chord intersection contributes
\begin{equation}
\sum _{|I_j \cap I_{j'}|=k} (-1)^k \frac{(p^2/N)^k}{k!}e^{-p^2/N} = \exp\left( -2p^2/N\right) ~.
\end{equation}
Triple intersections, i.e., configurations in which there are 3 chords with $I_1\cap I_2 \cap I_3 \not= \emptyset $ are negligible in the large $N$ limit.
Therefore, the combinatorial problem that one gets is to sum over all chord diagrams, with a
 weight that depends on the number of pairwise crossings of chords, i.e.,
\begin{equation}
m_k=\sum _{\pi\in\ \text{Chord Diagrams}} \exp\left( -2p^2 \cdot \text{cr}(\pi)/N\right)
\end{equation}
where $\pi$ denotes a chord diagram, and $\text{cr}(\pi)$ is the number of pairwise crossings in the diagram.  These combinatorial sums can be evaluated using a transfer matrix technique, which captures all energy scales in the model, and gives rise to a $q$-deformation of the Schwarzian action \cite{Berkooz:2018qkz,Berkooz:2018jqr}. In a similar manner, one can extend this construction to correlation functions.


\subsection{Chord diagrams and the complex SYK model}\label{sec:3.2}

The generalization to the charged SYK model is as follows. First of all, we note that after carrying out the average over the (Gaussian distributed) couplings in the random Hamiltonian (as will be done in section \ref{sec:partition_function}), we get a sum over traces of the form
\begin{equation}\label{FermCorrA}
\tr({\bar\psi}_{{I'}_1} \psi^{I_1} {\bar\psi}_{{I'}_2} \psi^{I_2}\cdots {\bar\psi}_{{I}_1} \psi^{{I'}_1}\cdots )~.
\end{equation}
In this trace, the indices of a pair of consecutive $\bar \psi \psi$ are contracted with those of another $\bar \psi\psi$ pair (as in $\cdots{\bar\psi}_{I'} \psi^{I}\cdots{\bar\psi}_{I} \psi^{I'}\cdots$).
However, in the remainder of this section we will allow for a more general arrangement as shown in \eqref{general_trace} since it will be useful for calculating correlation functions of generic operators which we eventually do in section \ref{sec:2ptFunctionSec}.

As a starting point we will represent the complex fermions using $N$-dimensional gamma matrices. We will use the following conventions for the Pauli matrices
\begin{equation}\label{2drep}
\sigma_{+}=\sqrt{2}\left(\begin{array}{cc}
0 & 1\\
0 & 0
\end{array}\right),\qquad\sigma_{-}=\sqrt{2}\left(\begin{array}{cc}
0 & 0\\
1 & 0
\end{array}\right),\qquad\sigma_{3}=\left(\begin{array}{cc}
1 & 0\\
0 & -1
\end{array}\right),
\end{equation}
and write the complex fermions $\psi_i$ as the following tensor product
\begin{equation}
\begin{split}
 \psi^{1} &=\sigma_{+}\otimes\sigma_{3}\otimes\cdots\otimes\sigma_{3}\\
  \psi^{2}&=\mathbb{1}_2\otimes\sigma_{+}\otimes\sigma_{3}\otimes\cdots\otimes\sigma_{3}\\
  \psi^{3}&=\mathbb{1}_2\otimes\mathbb{1}_2\otimes\sigma_{+}\otimes\sigma_{3}\otimes\cdots\otimes\sigma_{3}\\
 & \vdots\\
  \psi^{N}&=\mathbb{1}_2\otimes\cdots\otimes\mathbb{1}_2\otimes\sigma_{+}
\end{split}
\label{complex_Clifford_rep}
\end{equation}
For $\bar{\psi}_i$ we simply replace $\sigma_{+}\to\sigma_{-}$ in the above formulas. Using this representation we can now work out the trace of a product of complex fermions.

\begin{figure}[h]
\centering
\includegraphics[width=0.6\textwidth]{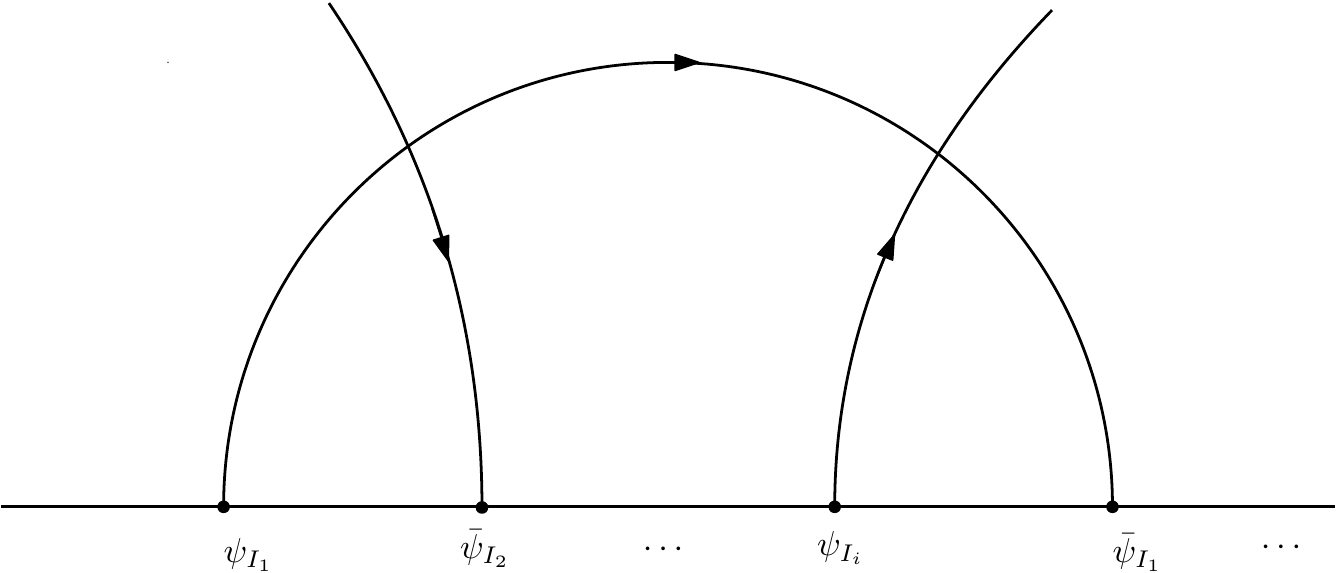}
\caption{Chord diagram representation of the generic trace \eqref{general_trace}. The chords are oriented to go from a $\psi$ insertion to a $\bar{\psi}$ insertion having the same index set $I$.}
\label{fig:general_trace}
\end{figure}

Fermions with
capital indices, $\psi^I$ and ${\bar\psi}_J$, stand for the appropriate product of components as indicated in \eqref{psiDef}. Let $p_j$ denote the length of the index set $I_{j}$. Here we have taken the length to be generic, because this generalization is needed when we compute correlation functions.
As mentioned, the quantity that we would like to evaluate in this section is a slight generalization of \eqref{FermCorrA}
\begin{equation}
\tr\left(\psi^{I_{1}}\bar{\psi}_{I_{2}}\cdots\psi^{I_i}\bar{\psi}_{I_{1}}\cdots\right)\label{general_trace}
\end{equation}
where now at each point any $\psi$ or $\bar{\psi}$ is allowed (not necessarily
alternating), with a total of $2k$ insertions, such that each of the $k$ indices
$I_{j}$ appears in one $\psi$ and one $\bar{\psi}$. The trace is
defined to be normalized\footnote{In the representation (\ref{complex_Clifford_rep}), the trace
in each tensor product factor is normalized in this way.} as $\tr~\mathbb{1}=1$. This object is represented by a chord diagram: as reviewed above, this is a circle or
a line (which is equivalent, by cutting open the circle at a point), on which $2k$ nodes are marked, such that pairs of nodes
are connected by chords. Since we have here two kinds of insertions
($\psi$ and $\bar{\psi}$), the chords are \emph{oriented}, so that each
chord has a direction; let us choose a convention where the arrow
goes from a $\psi$ insertion to a $\bar{\psi}$ insertion (see fig.\ \ref{fig:general_trace}).

The basic idea to evaluate (\ref{general_trace}) is the following.
We will consider each component $i$ of the tensor product structure
in the representation (\ref{complex_Clifford_rep}). The contribution
of $\psi^{i}$ to this component is simply $\sigma_{+}$, each $\psi^{j<i}$
contributes $\sigma_{3}$, while each $\psi^{j>i}$ gives no contribution;
for $\bar{\psi}$ it is the same, with $\sigma_{+}\to\sigma_{-}$.\footnote{Here the convention for the definition of $\psi^{I}$ and $\bar{\psi}_{I}$
in \eqref{psiDef} is useful, since the evaluation becomes easier using the identities $\sigma_{3}^{a}\sigma_{+}=\sigma_{+}$ and $\sigma_{-}\sigma_{3}^{a}=\sigma_{-}$ (for any integer power $a$).}
All the $\sigma_{3}$ commute among themselves and give only a sign,
so the first step is to determine
the sign and by this eliminate all the $\sigma_{3}$'s. Then the remaining
strings of $\sigma_{\pm}$ is straightforward to evaluate, giving
either a vanishing result, or a power of two.

\begin{figure}[h]
\subfloat[]{

\centering
\includegraphics[width=0.5\textwidth]{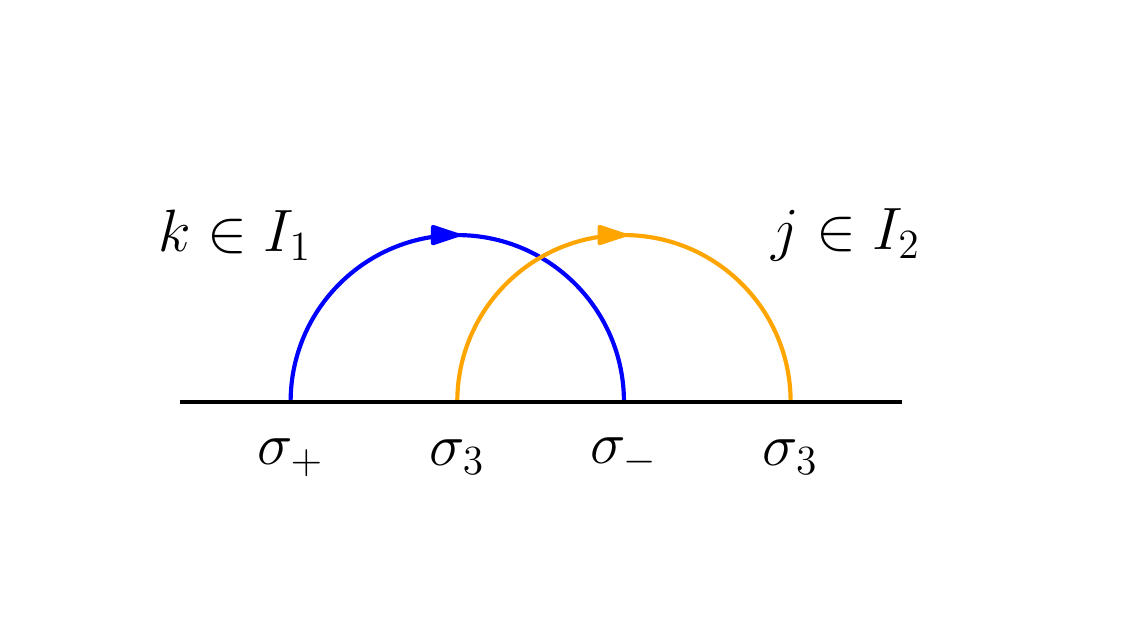}\label{fig:signs_simple}}\subfloat[]{

\centering
\includegraphics[width=0.5\textwidth]{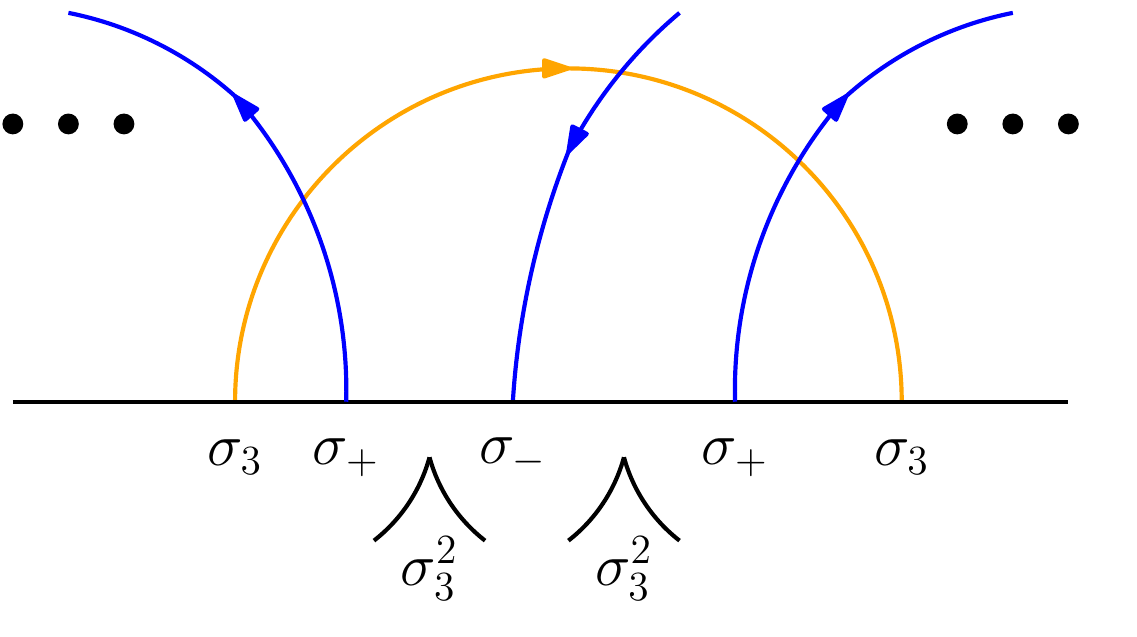}\label{fig:signs_more_general}

}\caption{Determining the sign of a diagram. Only in this figure the chords
are shown in different colors. The chords to which $j$ belongs are
shown in orange, while the ones to which $k$ belongs are in blue. In fig.\ \ref{fig:signs_more_general} we draw specific orientations for concreteness, but the argument in the text is independent of the orientations of the chords.}
\end{figure}

More explicitly, let us consider two indices $1\le j<k\le N$. For
simplicity let us discuss first the case where each of them appears
in a distinct single chord only --- this is actually the case for most of the indices appearing in chords. The case where these two chords intersect
is shown in figure \ref{fig:signs_simple}. Consider the $k$'th component of
the tensor product representation. The chord containing the $k$'th index
gives one $\sigma_{+}$ and one $\sigma_{-}$, while the other chord
containing the $j$'th index gives two $\sigma_{3}$ as shown in the
figure (note that we can consider each $i<k$ independently because
they all give $\sigma_{3}$ factors that commute).
That is, we have something of the form $\tr \left[ \cdots \sigma _+ \cdots \sigma _3 \cdots  \sigma _-  \cdots \sigma _3 \cdots\right] $.
Using the algebra of the $\sigma $ matrices,
we can get rid of the $\sigma_{3}$ insertions,
resulting in a factor of $(-1)$. If the chords were not intersecting,
then using $\sigma_{3}^{2}=1$ we would get no sign.

More generally, consider two indices $j<k$ such that each of them
can appear in any number of chords. As before, since the $\sigma_{3}$'s
commute, we can consider each chord containing $j$ in turn. As shown
in figure \ref{fig:signs_more_general}, when we consider the $k$'th tensor
product factor, by inserting $\sigma_{3}^{2}=1$ in intermediate steps,
and using the fact that $\sigma_{3}\sigma_{\pm}\sigma_{3}=-\sigma_{\pm}$,
each intersection of a chord containing $j$ with a chord containing
$k$ gives a $(-1)$.

Now we would like to combine all the signs from all the tensor product
factors. Take a pair of chords $J,K$ that intersect. For any $j\in J$
and $k\in K$, if $j<k$ we saw that when looking at the $k$'th
tensor product factor, we get a $(-1)$, and similarly for $j>k$
we get a $(-1)$ from considering the $j$'th tensor product factor;
but if $j=k$ there are no $\sigma_{3}$'s and no $(-1)$. As a result,
by eliminating all the $\sigma_{3}$'s, we find a sign which is
\begin{equation}\label{signs}
(-1)^{\sum_{\text{Chords \ensuremath{i,j} intersect}}(p_{i}p_{j}-p_{ij})}
\end{equation}
where $p_{j}=|I_{j}|$ and $p_{ij}=|I_{i}\cap I_{j}|$.

After this step, we are left only with factors of $\sigma_{\pm}$. If we have
two consecutive $\sigma_{+}$ or $\sigma_{-}$ then this just gives
zero since $\sigma_{\pm}^{2}=0$. This just corresponds to the fact that if we have two consecutive $\psi$'s or $\bar \psi's$ with common indices, we immediately get zero. Otherwise, consider again a particular
index $i$ with the corresponding tensor product factor. If it appears
in no chord, then the trace in this tensor product factor is simply
$\tr~\mathbb{1}_2=1$ (recall that the trace is normalized to one). If it appears in a single chord, then we have $\tr\left(\sigma_{+}\sigma_{-}\right)=\tr\left(\begin{array}{cc}
2 & 0\\
0 & 0
\end{array}\right)=1$  (or $\tr(\sigma_{-}\sigma_{+})$ which is
the same here and below). If the index $i$ appears in $m$ number of chords,
then we get
$\tr\left((\sigma_{+}\sigma_{-})^{m}\right)=2^{m-1}$.

As a result, a general oriented chord diagram with fixed indices
$I_{j}$, that is expression (\ref{general_trace}), equals
\begin{equation}\label{general_trace_formula}
\begin{split}
& \tr\left(\psi^{I_{1}}\bar{\psi}_{I_{2}}\cdots\psi^{I_i}\bar{\psi}_{I_{1}}\cdots\right) = \\[10pt]
& = (-1)^{\sum_{\text{Chords \ensuremath{i,j} intersect}}(p_{i}p_{j}-p_{ij})}2^{\sum_{m=2}^{\infty}(m-1)\cdot\left(\text{\# of indices \ensuremath{1\le i\le N} appearing in \ensuremath{m} chords}\right)}~,
\end{split}
\end{equation}
unless when restricted to any particular $i$ the $\psi$ and $\bar{\psi}$
do not appear in an alternating form, in which case the value of the trace is just 0.

Let us also introduce the notation $s_{ijk\cdots }$ which denotes the number of sites that appear in $I_i$, $I_j$, $I_k$ and so on, but in no other set. Note the difference between the two symbols $p_{ij}$ and $s_{ijk\cdots}$. They are related by $p_{ij} = s_{ij} +\sum _k s_{ijk} +\cdots $. In terms of $s_{ijk\cdots}$ we can write \eqref{general_trace_formula} as
\begin{equation}\label{general_trace_formula2}
\begin{split}
& \tr\left(\psi^{I_{1}}\bar{\psi}_{I_{2}}\cdots\psi^{I_i}\bar{\psi}_{I_{1}}\cdots\right) = (-1)^{\sum_{\text{Chords \ensuremath{i,j} intersect}}(p_{i}p_{j}-p_{ij})}2^{\sum_{m=2}^{\infty}(m-1)\cdot
\sum _{i_1<i_2<\cdots< i_m} s_{i_1i_2\cdots i_m}
}~.
\end{split}
\end{equation}
This result holds for any value of $N$ and $p_i$.


\subsection{Large $N$ simplification} \label{sec:large_N_rules}

A simplification occurs in the large $N$ double scaling limit. Consider the same
trace object as above, associated to a particular oriented chord diagram,
but now summing over the indices
\begin{equation}
\left(\prod_{j=1}^{k}{N \choose p_{j}}^{-1}\right)\sum_{I_{1},\cdots,I_{k}}\tr\left(\psi^{I_{1}}\bar{\psi}_{I_{2}}\cdots\psi^{I_i}\bar{\psi}_{I_{1}}\cdots\right).\label{general_trace_large_N}
\end{equation}
The combinatorial prefactor turns counting of events in the sum into
probabilities of those events. As was shown in \cite{erdHos2014phase},
in the large $N$ limit, the intersections are independently Poisson
distributed $p_{ij}=|I_{i}\cap I_{j}|\sim \text{Pois}\left(\frac{p_{i}p_{j}}{N}\right)$,
and there are no triple (or higher) intersections with probability
that goes to 1 in the large $N$ limit. Therefore, in \eqref{general_trace_formula2},
the sum over $m$ is now restricted only to $m=2$.

\begin{figure}[h]
\begin{imgrows}
\centering
\imgrow\label{fig:allowed_overlaps_1}
\raisebox
    {\dimexpr-0.5\height+0.5ex}
    {\includegraphics[width=0.8\textwidth]{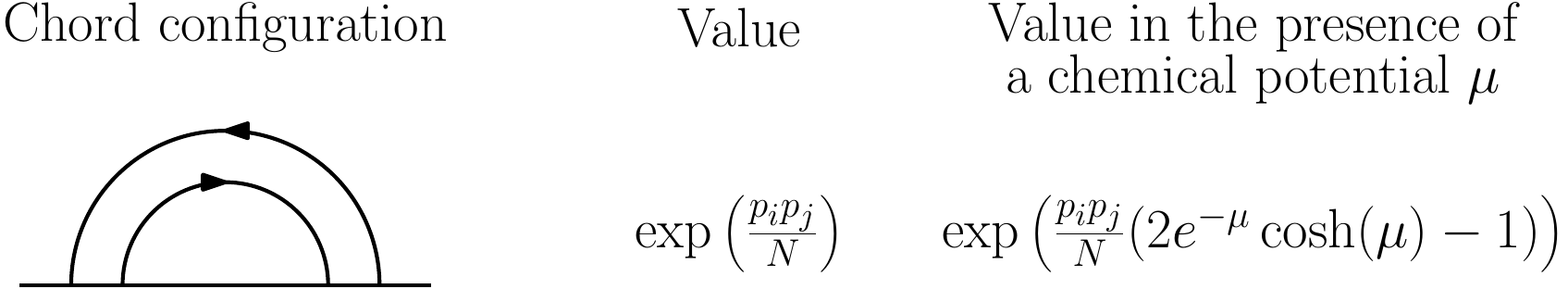}}
\vspace{2ex}

\imgrow\label{fig:allowed_overlaps_2}
\raisebox
    {\dimexpr-0.5\height+0.5ex}
    {\includegraphics[width=0.8\textwidth]{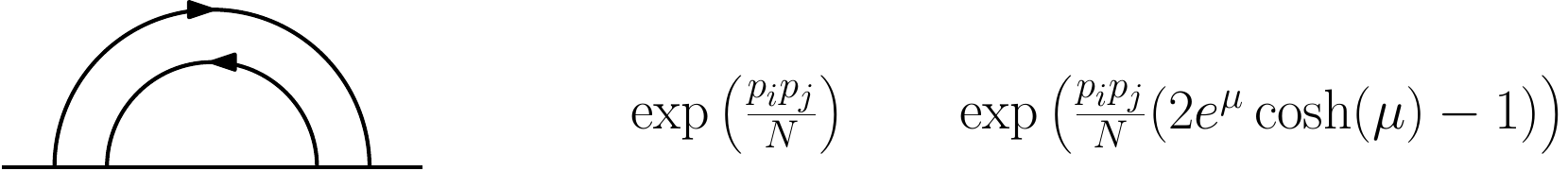}}
\vspace{2ex}

\imgrow\label{fig:allowed_overlaps_3}
\raisebox
    {\dimexpr-0.5\height+0.5ex}
    {\includegraphics[width=0.8\textwidth]{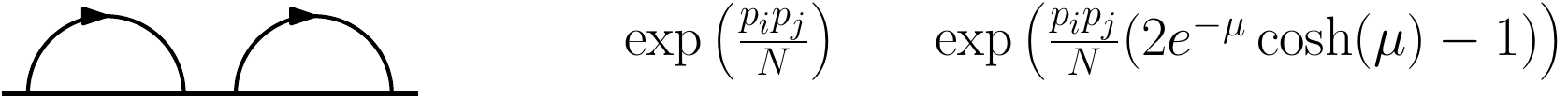}}
\vspace{2ex}

\imgrow\label{fig:allowed_overlaps_4}
\raisebox
    {\dimexpr-0.5\height+0.5ex}
    {\includegraphics[width=0.8\textwidth]{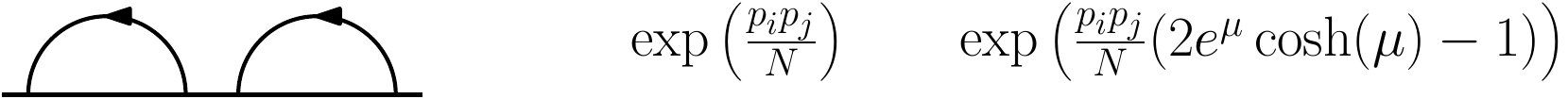}}

\imgrow\label{fig:allowed_overlaps_5}
\raisebox
    {\dimexpr-0.5\height+0.5ex}
    {\includegraphics[width=0.8\textwidth]{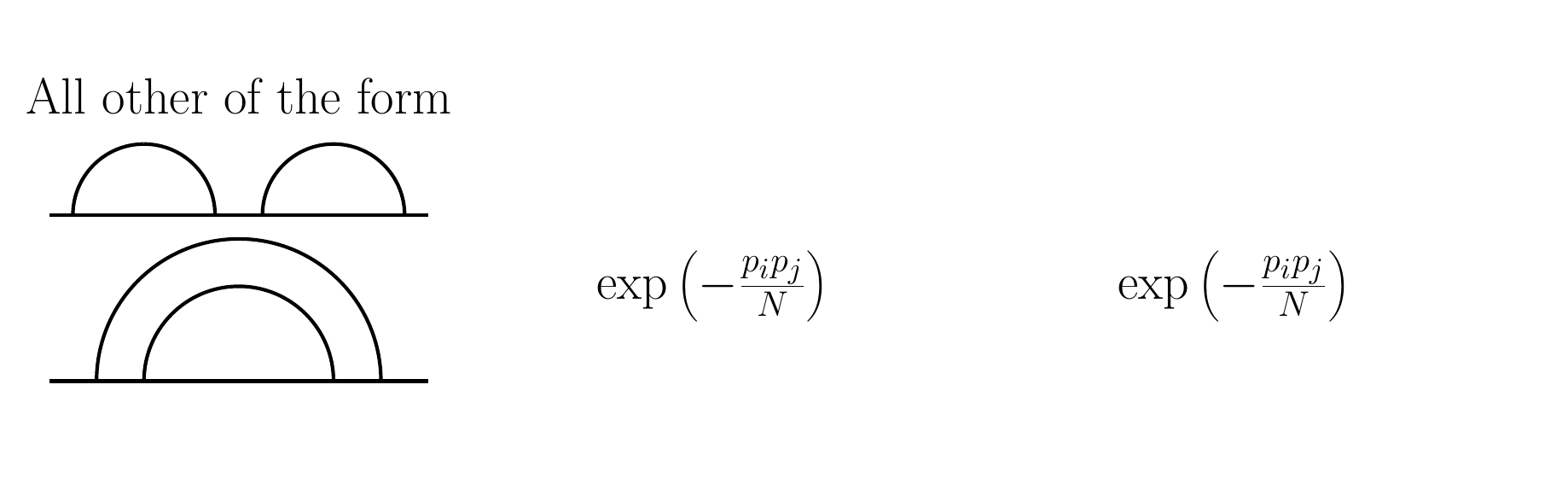}}

\imgrow\label{fig:allowed_overlaps_6}
\raisebox
    {\dimexpr-0.5\height+0.5ex}
    {\includegraphics[width=0.8\textwidth]{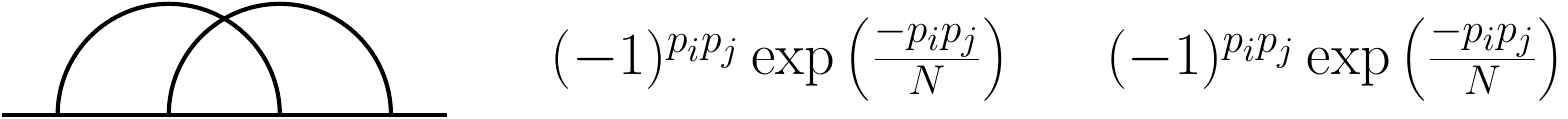}}
\vspace{2ex}

\imgrow\label{fig:allowed_overlaps_7}
\raisebox
    {\dimexpr-0.5\height+0.5ex}
    {\includegraphics[width=0.8\textwidth]{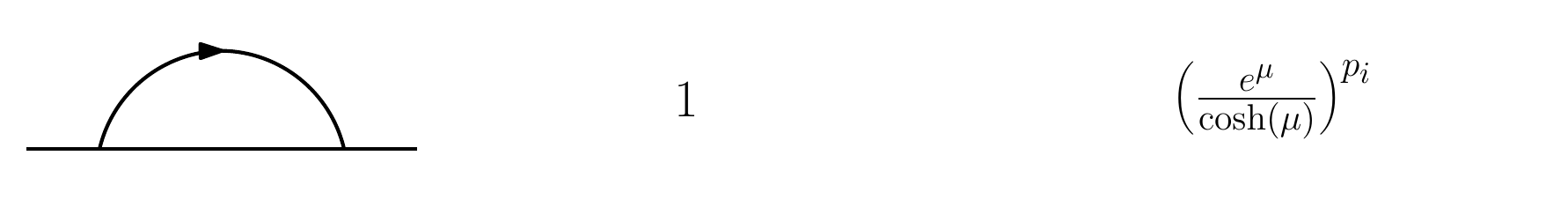}}

\imgrow\label{fig:allowed_overlaps_8}
\raisebox
    {\dimexpr-0.5\height+0.5ex}
    {\includegraphics[width=0.8\textwidth]{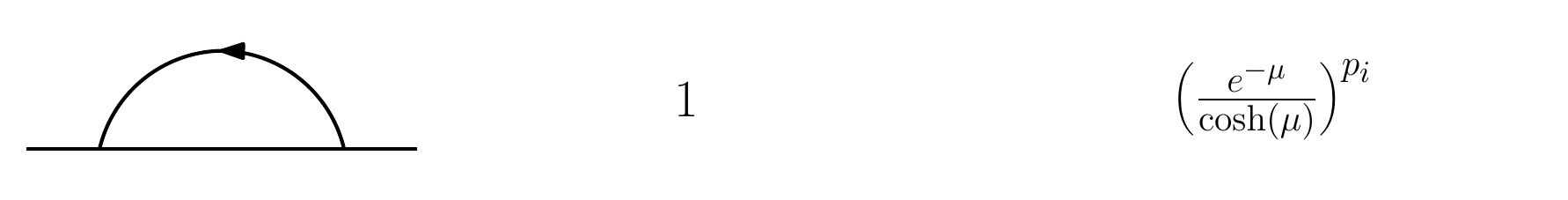}}

\imgrow\label{fig:allowed_overlaps_9}
\raisebox
    {\dimexpr-0.5\height+0.5ex}
    {\includegraphics[width=0.8\textwidth]{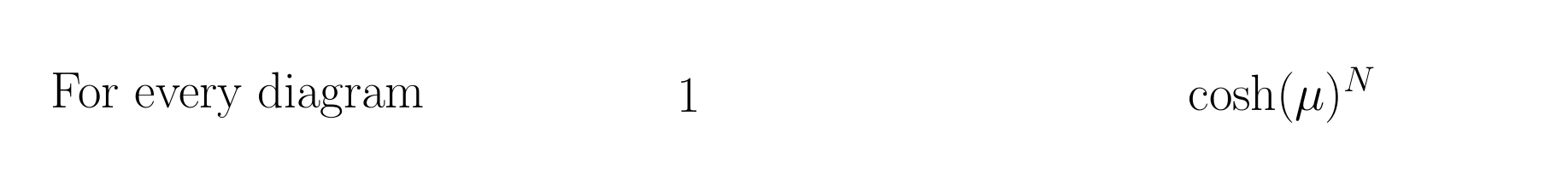}}

\end{imgrows}
\caption{Rules for evaluating \emph{oriented} chord diagrams (when an orientation is not shown, it means that it does not matter). The values when we have a chemical potential, to be used later (which are derived in section \ref{sec:partition_function}), are also shown.
}
\label{fig:allowed_overlaps}
\end{figure}

Thus, we should consider independently pairs of chords $I_{i},I_{j}$.
First, we need to make sure that we do not get a vanishing
result. For an index $i$ appearing only in a single chord, it appears
in one $\psi^{I}$ and one $\bar{\psi}_{I}$ and therefore it does
not vanish. Any index $i$ appearing in two chords (a higher number
of chords can be neglected in the large $N$ limit as was just mentioned) will necessarily
give a vanishing result if the two chords are intersecting (just because
one chord gives a $\sigma_{+}$ and a $\sigma_{-}$, and there is
one end of the other chord between these two, such that no matter
whether it is a $\sigma_{+}$ or a $\sigma_{-}$ we would have $\sigma_{\pm}^{2}=0$).
By this logic, a non-zero $p_{ij}$ is allowed only in the first four possibilities
appearing in figure \ref{fig:allowed_overlaps}. For each such pair of chords
(that are not intersecting and therefore get no $(-1)$ factors) we
need to sum over the number of elements $p_{ij}$ with the Poisson
probability distribution; their contribution from \eqref{general_trace_formula}
is
\begin{equation}
\sum_{p_{ij}=0}^{\infty}\frac{(p_{i}p_{j}/N)^{p_{ij}}}{p_{ij}!}e^{-p_{i}p_{j}/N}2^{p_{ij}}=e^{p_{i}p_{j}/N}.
\end{equation}
For all the other combinations of pairs of chords, we must have $p_{ij}=0$
and therefore they contribute simply
\begin{equation}
e^{-p_{i}p_{j}/N}
\end{equation}
from the Poisson distribution. In addition, if they intersect, they
also give a factor of $(-1)^{p_{i}p_{j}}$ from \eqref{general_trace_formula}
(for which as mentioned, $p_{ij}=0$ necessarily in the large $N$
limit).

These rules are summarized in fig.\ \ref{fig:allowed_overlaps} for convenience.
Using these rules we can now calculate physical observables in the double scaled complex SYK model. In the ensuing sections we calculate the partition function and correlation functions of generic operators.


\section{Partition function} \label{sec:partition_function}

In this section we obtain the canonical and grand canonical partition functions.

\subsection{Canonical partition function}

For simplicity, we first apply the rules that we already found in order to calculate the ensemble averaged partition function (without a chemical potential)
\be
Z(\beta) =\<\tr ~e^{-\beta H}\>_J~.
\ee
The calculation proceeds by expanding out the above quantity in moments
\be\label{traceMom0}
m_k=\<\tr ~H^k\>_J=J^{k}\sum_{CD}{N \choose p}^{-k}\sum_{I_{1}\cdots I_{k}}\tr\left(\bar{\psi}_{I_{1}}\psi^{I_{2}}\cdots\bar{\psi}_{I_{2}}\psi^{I_{1}}\cdots\right)\ ,
\ee
then evaluating each moment using the result from the previous section, and then resumming  the moments. In the last equality, we carried out the ensemble average to obtain a sum over oriented chord diagrams (abbreviated as CD).
This is the result for even $k$. For odd $k$ the moments vanish due to the ensemble average.

\begin{figure}[h]

\centering
\includegraphics[width=0.8\textwidth]{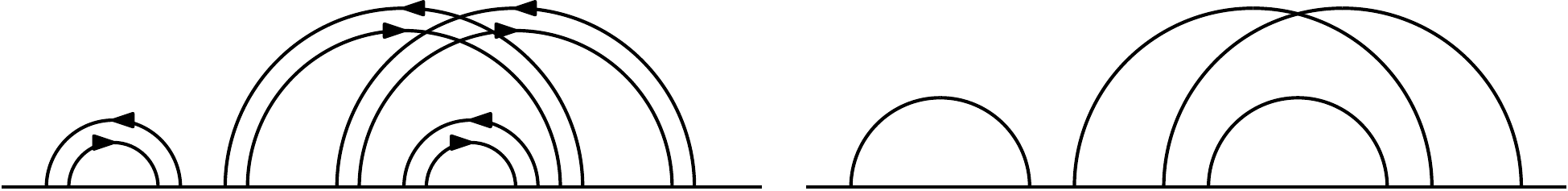}\caption{An oriented chord diagram contributing to $\langle\tr H^{8}\rangle_{J}$
and the corresponding unoriented chord diagram.}
\label{fig:oriented_to_unoriented_CD}
\end{figure}

Note the simplification here in comparison to (\ref{general_trace_large_N}),
in that the $\bar{\psi}\psi$ are contracted in pairs.
Pictorially,
the corresponding oriented chord diagrams consist of adjacent pairs
of chords of opposite orientation, so that each such pair can be replaced
by a single unoriented chord; we will refer to such an unoriented
chord as an $H$-chord (since each end of it corresponds to a single Hamiltonian
insertion). This is demonstrated in figure \ref{fig:oriented_to_unoriented_CD}.

\begin{figure}[h]

\centering
\includegraphics[width=0.7\textwidth]{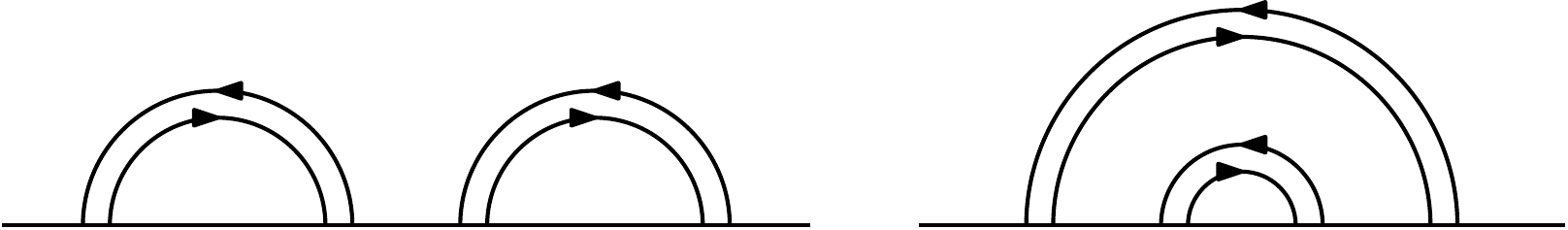}\caption{Two possibilities for non-intersecting $H$-chords.}
\label{fig:non_intersecting_H_chords}

\end{figure}

The goal is to calculate the contribution of every chord diagram in \eqref{traceMom0}, by applying
the rules of fig.\ \ref{fig:allowed_overlaps}.
To begin with, since there is an even number of intersections of oriented
chords, and all $p_{i}=p$, the sign of every diagram is positive. The remaining task is just to find the number of pairs of oriented
chords of the form of the first four pairs appearing in figure \ref{fig:allowed_overlaps} (as they are assigned the same value). Every
$H$-chord is of this form, so that we already have $k/2$ such pairs.
Every other pair of oriented chords belongs to a pair of $H$-chords. Any
intersecting pair of $H$-chords manifestly does not give anything
of the form of figure \ref{fig:allowed_overlaps_1}, \ref{fig:allowed_overlaps_2}, \ref{fig:allowed_overlaps_3} or \ref{fig:allowed_overlaps_4}, while every non-intersecting
pair of $H$-chords (see figure \ref{fig:non_intersecting_H_chords}) gives
exactly two pairs of oriented chords of the form of figure \ref{fig:allowed_overlaps_1}, \ref{fig:allowed_overlaps_2}, \ref{fig:allowed_overlaps_3} or \ref{fig:allowed_overlaps_4}.\footnote{If two $H$-chords are not intersecting, then
either they are disjoint, or one of them is embedded in the other
one. In the first case, the two pairs of chords of the same orientation belong
to figures \ref{fig:allowed_overlaps_1}-\ref{fig:allowed_overlaps_4}, while in the latter case, these are
the two pairs of chords of opposite orientation.}

Denote the number of intersections of $H$-chords by $\kappa_{H}$; then
the number of pairs of $H$-chords that do not intersect is
$\bar{\kappa}_{H}={k/2 \choose 2}-\kappa_{H}$.
The total number of pairs of oriented chords of the form of figures \ref{fig:allowed_overlaps_1}-\ref{fig:allowed_overlaps_4}
is then $\frac{k}{2}+2\bar{\kappa}_{H}$ (and the number
of pairs not of this form is just ${k \choose 2}-\left(\frac{k}{2}+2\bar{\kappa}_{H}\right)$).
Thus we arrive at
\begin{equation}\label{eq:moments0}
\begin{split}  m_k &=J^{k}\sum_{CD}\exp\left[\frac{p^{2}}{N}\left\{ \frac{k}{2}+2\bar{\kappa}_{H}-\left[{k \choose 2}-\left(\frac{k}{2}+2\bar{\kappa}_{H}\right)\right]\right\} \right]\\
 & =\left(J e^{\frac{p^{2}}{2N}}\right)^{k}\sum_{\text{unoriented }CD}\left(e^{-\frac{4p^{2}}{N}}\right)^{\kappa_{H}}  \\
 &=\left(J e^{\frac{\lambda }{2}}\right)^{k}\sum_{\text{unoriented }CD}\left(e^{-4\lambda}\right)^{ \kappa _H},
\end{split}
\end{equation}
where $\lambda = p^2/N $ is what we keep fixed as $ N \to \infty $. This is just as in the real SYK where for every intersection one gets a factor of $e^{-2p^2/N}$, whereas in the complex SYK model this factor gets modified to $e^{-4p^2/N}$.
In the next subsection we give an analytic expression for the partition function (evaluating the sum over chord diagrams) in the more general case when having a chemical potential.


\subsection{Grand canonical partition function} \label{sec:grand_partition_function}
We now move on to calculate the ensemble averaged grand canonical partition function of the complex SYK model which is given by
\be\label{GCPf1}
Z(\beta,\mu) =\<\tr ~e^{-\beta H-2\mu Q}\>_J~,
\ee
where $Q$ is defined in \eqref{U1Opdef}.
As before we will use the moment method and compute (at this stage, without any restriction on $\mu$)
\begin{equation}
m_{k}(\mu) = \<\tr~H^k e^{-2\mu  Q}\>_J ;
\end{equation}
note that we are not expanding the charge term.
We will see, however, that resummation is difficult for finite $\mu\not=0$ that is independent of $N$, since the sum over $k$ diverges.
We note that the divergence originates from very large values of $k$, where the computation is not reliable.
To remedy this we can either
\begin{enumerate*}[label=(\roman*)]
\item scale $\mu$ appropriately with $N$, or
\item go to the fixed charge basis.
\end{enumerate*}
We will do each of those below.

Computing the individual moments uses techniques similar to the ones used before. The moments now read
\begin{equation}
m_{k}(\mu) =
J^{k}\sum _{CD}  {N \choose p}^{-k}\sum_{I_{1}\cdots I_{k}}\tr\left[\bar{\psi}_{I_{1}}\psi^{I_{2}}\cdots\bar{\psi}_{I_{2}}\psi^{I_{1}}\cdots\exp\left(-\frac{\mu}{2}\sum_{i=1}^{N}(\bar{\psi}_{i}\psi^{i}-\psi^{i}\bar{\psi}_{i})\right)\right]~.
\end{equation}
In the representation (\ref{complex_Clifford_rep}), the chemical
potential term is represented by
\begin{equation}
\begin{split} & \exp\left(-\frac{\mu}{2}\sum_{i=1}^{N}(\bar{\psi}_{i}\psi^{i}-\psi^{i}\bar{\psi}_{i})\right)=\prod_{i}e^{-\frac{\mu}{2}(\bar{\psi}_{i}\psi^{i}-\psi^{i}\bar{\psi}_{i})}=\\
 & =\left(\begin{array}{cc}
e^{\mu}\\
 & e^{-\mu}
\end{array}\right)\otimes\left(\begin{array}{cc}
e^{\mu}\\
 & e^{-\mu}
\end{array}\right)\otimes\cdots.
\end{split}
\end{equation}

Similarly to section \ref{sec:large_N_rules}, let us provide the rules for computing oriented chord diagrams (that will also be valid beyond the partition function) in the large $N$ limit, where now we allow the presence of a chemical potential.

Earlier, the first step was to apply the $\sigma_{3}$'s resulting
in a sign. This clearly remains the same (we do not use the trace here, so this is unaffected by the presence of a chemical potential). Non-zero $p_{ij} $ of intersecting chords give vanishing contribution to the trace and therefore the only signs we get are $(-1)^{p_ip_j} $ for intersecting chords. In the partition function and the correlation functions that we consider here, all intersections are in pairs, so overall the sign is plus.

Then, for every site $i$, if
it does not appear in any chord, we simply have $\tr\left(\begin{array}{cc}
e^{\mu}\\
 & e^{-\mu}
\end{array}\right)=\cosh(\mu)$. If it appears in at least one chord, then if the first appearance
is in a $\psi$ (that is, the first chord in which $i$ appears goes
to the right) we get $\tr\left(\begin{array}{cc}
2 & 0\\
0 & 0
\end{array}\right)\left(\begin{array}{cc}
e^{\mu}\\
 & e^{-\mu}
\end{array}\right)=e^{\mu}$ times $2$ to the power of the number of chords it appears in minus
one; if the first appearance is in a $\bar{\psi}$ (the first chord
goes to the left), we would get instead of $e^{\mu}$, a factor of
$\tr\left(\begin{array}{cc}
0 & 0\\
0 & 2
\end{array}\right)\left(\begin{array}{cc}
e^{\mu}\\
 & e^{-\mu}
\end{array}\right)=e^{-\mu}$.

In the large $N$ limit there are no triple intersections of the index sets $I_j$, and therefore the only options we have are summarized by
\begin{itemize}
\item Each index $i$ appearing in no chord gives $\cosh(\mu )$,
\item Each index $i$ appearing in 1 chord going to the right gives $e^{\mu } $,
\item Each index $i$ appearing in 1 chord going to the left gives $e^{-\mu } $,
\item Each index $i$ appearing in 2 chords going to the right gives $2e^{\mu } $,
\item Each index $i$ appearing in 2 chords going to the left gives $2e^{-\mu } $ .
\end{itemize}

For simplicity, let us start first with the case that all the indices in the various $I_{j} $ are distinct. We can first assign a value of $\cosh(\mu )^N$ which would be the case if there were no chords. An index in a chord, if it goes to the right, should be assigned $e^{\mu } $ instead of $\cosh(\mu )$ so that a chord going to the right gets $\left( e^{\mu } /\cosh(\mu )\right) ^{p_i} $ (and $\mu  \to -\mu $ if it goes to the left). For the partition function, there are the same number
of chords of each orientation, the $e^{\pm\mu}$ factors cancel,
and altogether we simply get an extra factor of $\left(\cosh(\mu)\right)^{N-kp}$
in $m_{k}(\mu)$ relative to the case without the insertion of a chemical potential.

Now let us see how this counting changes for the general case, where the $I_j$ are not necessarily distinct.
Recall that in the large $N$ limit, the
different intersections are independent and they do not overlap, so
we can consider each $p_{ij}$ separately. The allowed non-zero $p_{ij}$
were given in figures \ref{fig:allowed_overlaps_1}, \ref{fig:allowed_overlaps_2}, \ref{fig:allowed_overlaps_3} and \ref{fig:allowed_overlaps_4}.
Let us start with figure \ref{fig:allowed_overlaps_1}.
In order to allow an intersection of the corresponding two index sets, we need to take two indices, one from each of those two chords, and turn them into one index which is common to the two chords, and one which does not appear in any chord. In the counting of the previous paragraph, the two indices were together assigned the value of one (since they are of opposite orientation). However, now they should be assigned a value of $2e^{-\mu}\cosh(\mu)$ (where the $\cosh(\mu )$ corresponds to the additional index not appearing in any chord). Therefore, the additional rule with respect to the previous paragraph, is to assign each index of this type the value of $2e^{-\mu}\cosh(\mu)$.
Similarly, each index in the second, third,
and fourth pair of figure \ref{fig:allowed_overlaps} should be assigned a value of $2e^{\mu}\cosh(\mu)$,
$2e^{-\mu}\cosh(\mu)$, $2e^{\mu}\cosh(\mu)$ respectively.

All that is left is to weight those possibilities by the Poisson distribution. For example, for the case of fig.\ \ref{fig:allowed_overlaps_1} we have
\begin{equation} \label{eq:fig4a_value_mu}
\sum_{p_{ij}=0}^{\infty}\frac{(p_ip_j/N)^{p_{ij}}}{p_{ij}!}e^{-p_ip_j/N}2^{p_{ij}}\left(e^{-\mu}\cosh(\mu)\right)^{p_{ij}}=\exp\left(\frac{p_ip_j}{N}(2e^{-\mu}\cosh(\mu)-1)\right),
\end{equation}
and similarly for the rest.
For the pairs of chords that must have $p_{ij} =0$ we simply get $\exp\left[ -p_i p_j/N\right] $. These rules for evaluating a chord diagram in the presence of a chemical potential are summarized in fig.\ \ref{fig:allowed_overlaps}, and can be used in the calculation of various observables.\footnote{As mentioned, these rules are general (valid for any arrangement of complex fermions in the trace) and in particular hold for the $\mathcal{N} =2$ supersymmetric SYK model analyzed in \cite{Berkooz:2020xne}. The large $N$ analysis there corresponds to a scaling $\mu \sim 1/\sqrt{N}$ which is mentioned below in subsection \ref{sec:partition_function_chemical_pot_half}. In this case, the chemical potential enters only in the values assigned to the entire diagram and single chords (rules in figs.\ \ref{fig:allowed_overlaps_7}-\ref{fig:allowed_overlaps_9}), with no contribution of the chemical potential coming from pairs of chords.}

We shall now specialize to the partition function, and recast this into the language of (the unoriented) $H$-chords. We have
already enumerated the possibilities for allowed $p_{ij}>0$, corresponding to figures \ref{fig:allowed_overlaps_1}, \ref{fig:allowed_overlaps_2}, \ref{fig:allowed_overlaps_3}, and \ref{fig:allowed_overlaps_4},
in terms of $H$-chords. First, for every one of the $H$-chords (and there
are $k/2$ of those), we get the pair of figure \ref{fig:allowed_overlaps_1}.
Therefore for every such pair of chords we have (setting $p_i=p_j=p$ in \eqref{eq:fig4a_value_mu})
\begin{equation}
\exp\left(\frac{p^{2}}{N}(2e^{-\mu}\cosh(\mu)-1)\right).
\end{equation}
For intersecting $H$-chords there are no $p_{ij}>0$ allowed. There
are $\bar{\kappa}_{H}$ non-intersecting pairs, of the forms shown
in figure \ref{fig:non_intersecting_H_chords}. For each of those, there are
two
pairs corresponding to two of figures \ref{fig:allowed_overlaps_1}-\ref{fig:allowed_overlaps_4} as mentioned before,
one giving again $\exp\left(\frac{p^{2}}{N}(2e^{-\mu}\cosh(\mu)-1)\right)$
while the other (in which $\mu\to-\mu$) gives $\exp\left(\frac{p^{2}}{N}(2e^{\mu}\cosh(\mu)-1)\right)$.

Putting everything together one obtains
\begin{align} \label{eq:moments_chemical_potential}
\begin{split}  m_{k}(\mu) &=J^{k}\cosh(\mu)^{N-kp}\sum_{\text{unoriented }CD}\exp\Bigg\{\frac{p^{2}}{N}\Bigg[-\left({k \choose 2}-\left(\frac{k}{2}+2\bar{\kappa}_{H}\right)\right)+\\
 & +\frac{k}{2}\left(2e^{-\mu}\cosh(\mu)-1\right)+\bar{\kappa}_{H}\left(2e^{-\mu}\cosh(\mu)-1\right)+\bar{\kappa}_{H}\left(2e^{\mu}\cosh(\mu)-1\right)\Bigg]\Bigg\}\\
 & =J^{k}\cosh(\mu)^{N-kp}\left(e^{\lambda \frac{1-\sinh(2\mu)}{2}}\right)^{k}\left(e^{\lambda \frac{\sinh^{2}(\mu)}{2}}\right)^{k^{2}}\sum_{\text{unoriented }CD}\left(e^{-4\lambda \cosh^{2}(\mu)}\right)^{\kappa_{H}}
\end{split}
\end{align}
(where recall that $\kappa _H$ is the number of intersections of $H$-chords, and ${\bar k}_H$ is the number of pairs of $H$-chords that do not intersect).
The ensemble averaged grand canonical partition function is obtained formally by resumming $m_k(\mu)$
\be\label{Zmresum}
Z(\beta,\mu)=\sum_{k=0}^\infty \frac{(-\beta)^k}{k!} m_k(\mu)~.
\ee

There are two new features of this result when compared to the case $\mu=0$. The first is that for each intersection of the $H$ chords we get the factor
\be
q(\mu)\equiv e^{-4\lambda \cosh^{2}(\mu)}
\ee
which explicitly depends upon the chemical potential. A second feature is that for every pair of $H$ chords that do not intersect we got a factor of $e^{4\lambda \sinh^{2}(\mu)}$. Due to this factor a term like $\left(e^{\lambda \frac{\sinh^{2}(\mu)}{2}}\right)^{k^{2}}$ appears in the expression of the moments. This term changes the game in two ways. The first is that the presence of a term of the form $c^{k^2} $ makes it difficult to rewrite $m_k(\mu)$ as $\int dE \rho(E) E^k$. The second is that this term blows up as $k\rightarrow\infty$, preventing us from resumming the series in a straightforward way. In the next section we consider fixed charge sectors where this problem does not occur, but for now we will proceed with a fixed chemical potential.

Otherwise, apart from these new ingredients, the sum over chord diagrams has the same form as in the double-scaled SYK model with Majorana fermions. Therefore, the analytic evaluation of this sum over the (unoriented) chord diagrams appearing in the moments \eqref{eq:moments_chemical_potential} can be done exactly in the same manner (for a review see section 2.3 of \cite{Berkooz:2018jqr}). This leads to the following result
\begin{align}
m_{k}(\mu) = & \cosh\(\mu \)^{N-kp}  \left(e^{\lambda \frac{1-\sinh(2\mu)}{2}}\right)^{k}\left(e^{\lambda \frac{\sinh^{2}(\mu)}{2}}\right)^{k^{2}}  \cdot \no\\
&\hspace{0.5in}  \int_0^\pi \frac{d\theta}{2\pi}\(q(\mu ),e^{\pm2i\theta};q(\mu )\)_\infty \(\frac{2J\cos\theta}{\sqrt{1-q(\mu )}}\)^k
\label{Zeqm2}
\end{align}
where the $q$-Pochhammer symbol is defined by
\begin{equation}
(a;q)_n=\prod _{j=0} ^{n-1} (1-aq^j)
\end{equation}
and $(a_1,a_2,\cdots ;q)_n$ stands for the product $\prod_j(a_j;q)_n$ (similarly the $\pm $ notation means the product of the corresponding terms with each sign).

In the $\lambda  \to 0$ limit, we expect this result for the partition function to agree with that in \cite{Davison:2016ngz}. This is indeed verified in appendix \ref{app:grand_potential_lambda_zero}.

\subsection{Partition function at different scalings of the chemical potential}

In general, one should specify how $\mu $ scales with $N$. We saw that for $0 \neq \mu \in \mathbb{R} $ independent of $N$, the sum over $k$ in the partition function naively diverges.
In this subsection we consider various scalings of $\mu $ of the form
\be
\mu =\frac{\bar \mu}{N^\a}~
\ee
with $\alpha >0$ and $\bar \mu$ fixed. In this case the effect of the $k^2$ term is subleading and can be ignored, and therefore there is no problem in evaluating the sum (up to subleading corrections in $N$). In the following we consider several possibilities for $\alpha $.

\subsubsection{$\alpha =1/2$} \label{sec:partition_function_chemical_pot_half}

For $\a=1/2$ we find that (in the expression below we have kept terms only up to order $O(N^0)$)
\begin{align}\label{Zeqm4}
\begin{split}
Z(\b,\mub)  &=  e^{\mub^2/2 }  \int_0^\pi \frac{d\theta}{2\pi}\(q,e^{\pm2i\theta};q\)_\infty \exp\[-\frac{2\b J \cos\theta}{\sqrt{1-q}}e^{\lambda/2}\]~\\
&= e^{\mub^2/2 } Z(\b,\mu=0)~
\end{split}
\end{align}
where $q=e^{-4\lambda } $ and $Z(\b,\mu=0)$ is the partition function of the complex SYK model with zero chemical potential.

\subsubsection{$\alpha =1/4$}
There is another interesting scaling for the chemical potential in which $\a=1/4$. In this case we find that the partition function takes the following form (again keeping terms only up to order $O(N^0)$)
\begin{align}\label{Zeqm3}
\begin{split}
Z(\b,\mub) =& \exp\(\frac12 \mub^2 \sqrt{N}-\frac{1}{12} \mub^4\)\cdot \\
& \int_0^\pi \frac{d\theta}{2\pi}\(q,e^{\pm2i\theta};q\)_\infty \exp\[-\frac{2\b J\cos\theta}{\sqrt{1-q}}\exp\(\frac{\lambda}{2}-\frac12\mub^2\sqrt{\lambda}\)\] .
\end{split}
\end{align}
The main change is that the range of energies has changed, as can be seen from the coefficient of $\beta$ in the exponent (the extremal values of the energy are given at $\cos(\theta)=\pm 1$).
Note that for $\alpha >1/4$ the change in the range of the energies is subleading in $N$.
The non-trivial $\mub$ dependent finite ($N$ independent) change in the energies happens only for $\a=1/4$.
Indeed, for $\alpha<1/4$ the energies are suppressed exponentially $\sim \exp\(-\#N^{\text{positive power}}\)$. The overall factor in front counts the effective number of states and will also appear in correlation functions. It can therefore be normalized away.


\section{Canonical partition function at fixed charges} \label{sec:partition_function_fixed_charge}
In the previous section, we coupled the chemical potential to the $U(1)$ charge operator \eqref{U1Opdef}.
The state $|0\rangle$, satisfying $\psi_i|0\rangle=0$ (note that it is not the ground state) has $U(1)$ charge $-N/2$. Then the states of charge $n$ are $\bar \psi_{i_1} \cdots \jb_{i_{n+N/2}} |0\rangle$, and the full range of charges of states in the Hilbert space goes from $-N/2$ to $N/2$ with spacings of a unit charge. For simplicity we take $N$ to be even.

In this section we extract the canonical partition function in the fixed charge ensemble by defining
\be
z=e^{2\mu}=e^{i\chi}
\ee
and projecting on the fixed charge sector
\begin{align} \label{eq:Z_beta_n}
{\cal Z}(\beta,n) = \frac{1}{2\pi i}\oint_\gamma dz~ \frac{ Z(\beta,z)}{z^{n+1}}
=\sum_{k=0}^\infty \frac{\(-\b\)^k}{k!} \[\frac{1}{2\pi}\int_{-\pi}^\pi d\chi~ e^{-in\chi}m_k(\chi)\]
\end{align}
where $\gamma$ is a closed contour encircling the origin.
To be precise, in the convention used here, the charge (corresponding to $Q$) is $(-n)$.

Note that for an imaginary chemical potential, the sum over $k$ converges. This is because in the expression of the moments, the sign in front of the $k^2$ term in the exponent is now negative. With this issue solved, we can use the moment method to compute a finite fixed charge partition function.

In the ensuing subsection we estimate the $\chi$-integral of the moments $m_k(\chi)$ by a large $N$ saddle point.
\subsection{Large $N$ evaluation} \label{sec:partition_function_charge_saddle}
We will assume that the charges $n$ scale as $N^{\alpha } $ for some $\alpha >0$.
Then, we can approximate the $\chi$-integral of the moments $m_k(\chi)$ using a saddle point
\begin{align}
\int_{-\pi}^\pi d\chi~ e^{-in\chi}m_k(\chi) =   J^k &\int_{-\pi}^\pi d\chi~ e^{-in\chi+(N-kp)\log \cos\(\chi/2\)} g(\chi)^{k^2} \cdot \no\\
& \hspace{0.3in}  \int_0^\pi \frac{d\theta}{2\pi}\(q(\chi),e^{\pm2i\theta};q(\chi)\)_\infty \(\frac{2f(\chi)\cos\theta}{\sqrt{1-q(\chi)}}\)^k
\label{chi1}
\end{align}
where we have defined the following functions
\begin{align}
\begin{split}
g(\chi)=\exp\(-\frac12 \lambda \sin\(\frac{\chi}{2}\)^2\)&~,\hspace{0.3in}f(\chi)=\exp\(\frac{\lambda}{2} e^{-i \chi}\)  \exp\(\lambda \sin\(\frac{\chi}{2}\)^2\)~,\\[5pt]
q(\chi) =\exp&\(-4\lambda \cos\(\frac{\chi}{2}\)^2\)~.
\end{split}
\end{align}
The saddle point, for large $n$ and $N$, is encoded in the term
\begin{equation} \label{eq:dominant_imaginary_chemical_potential}
C_k=\exp\left[-in\chi +(N-kp) \log \cos \frac{\chi }{2} \right]~ .
\end{equation}
We have kept the $n$ dependence since we allow for $n$ to increase with $N$ as well (the most we can have is $n\propto N$).\footnote{When $n$ is fixed, the pre-exponential terms are as important as $\exp(-in\chi)$, but then the saddle point is trivially at $\chi=0$.}  The saddle point is then given by
\begin{equation}\label{SP1}
\chi  = -2 \tan ^{-1} \left(\frac{2in}{N-kp} \right)~.
\end{equation}
Plugging this back into \eqref{eq:dominant_imaginary_chemical_potential},
keeping $k$, $\lambda$ fixed, and  $p=\sqrt{\lambda N}\rightarrow\infty$, we obtain
\begin{equation}
\begin{split}
C_k&=\exp\Bigg[-2n\tanh^{-1} \left(\frac{2n}{N} \right)- \frac{N}{2} \log \left(1-\frac{4n^2}{N^2} \right)+ \frac{\sqrt{\lambda N}  k }{2} \log \left(1-\frac{4n^2}{N^2} \right) -\\
& \qquad \qquad - \frac{2n^2 \lambda  k^2}{N^2-4 n^2}+O\left(\frac{1}{\sqrt{N}} \right) \Bigg]~.
\end{split}
\end{equation}
In addition, in \eqref{chi1} we have an explicit factor of
\begin{equation}
g(\chi )^{k^2} = \exp \left[-\frac{2n^2 \lambda  k^2}{4n^2-(N-kp)^2} \right] = \exp\left[-\frac{2n^2 \lambda  k^2}{4n^2-N^2}+O\left(\frac{1}{\sqrt{N} } \right) \right],
\end{equation}
which was evaluated at the saddle point. We notice that the bothersome $k^2$ terms in the last two expressions cancel exactly. At the saddle point, the function $f(n)$ and the parameter $q(n)$, which are now functions of the charge, are
\begin{equation}
\begin{split}
&q(n)= \exp\left( - \frac{4\lambda }{1- \frac{4n^2}{N^2} } \right)~,~~~~ f(n) = \exp\left[\frac{1}{2}\left(1-\frac{4Nn}{N^2-4n^2} \right)\lambda \right] ~.
\end{split}
\end{equation}
Putting everything together, we have the following result for the integral in \eqref{chi1}
\begin{equation} \label{eq:Z_fixed_charge}
\begin{split}
& \int_{-\pi}^\pi d\chi~ e^{-in\chi}m_k(\chi) = \sqrt{\frac{8\pi }{N} \frac{1}{1-4n^2/N^2}  } \exp\Bigg[ -2n\arctanh \left(\frac{2n}{N} \right)- \frac{N}{2} \log\left(1-\frac{4n^2}{N^2} \right) \Bigg] \cdot \\[5pt]
& \cdot \int _0^{\pi } \frac{d\theta }{2\pi } \left(q(n),e^{\pm 2i\theta } ;q(n)\right)_{\infty } \cdot \\
& \cdot \left\{\frac{2J\cos(\theta )}{\sqrt{1-q(n)} } \exp \left[ \frac{\sqrt{\lambda N} }{2} \log\left(1-\frac{4n^2}{N^2} \right)+\frac{\lambda }{2} \left(1-\frac{4Nn}{N^2-4n^2} \right) \right] \right\}^k ~.
\end{split}
\end{equation}
Since $n$ is an integer going from $-N/2$ to $N/2$, we have that $4n^2/N^2 \le 1$ which implies that the $\log$ is well defined. More importantly, the expression for the integral is of the form $$m_k=\int d\theta \rho(\theta;n) E(\theta;n)^k$$ from which we can read the range of energies and the full density of states of the model, for each charge sector $n$ separately.

A few comments are in order
\begin{enumerate}
\item   The main observation is that the model has the $q$-Gaussian density of states as in the Majorana SYK model, for each charge sector, but with a rescaled energy $E$, and --- more interestingly --- a renormalized intersection weight $q(n)$ that depends on the charge density. We expect this to carry over to the computation of any correlation function, as we will verify in sections \ref{sec:2ptFunctionSec} and \ref{sec:chaos}. In particular, when $n/N\rightarrow \pm 1/2$ then $q(n)\rightarrow 0$. This means that the model becomes more and more strongly coupled and, at least in terms of the macroscopic density of states and correlation functions, becomes more and more like the standard unitary or orthogonal RMT ensembles (i.e., anything that can be computed using a single trace in the Hilbert space at time scales that do not scale like $N$, or correspondingly, at the level of the analogue of free probability theory).

In the gravity dual this should probably be interpreted as stronger and stronger coupling in the bulk. This is to be expected since when the charge is large, the electric flux backreacts more strongly on spacetime. Here we also see that when chords intersect in the bulk the suppression factor (proportional to $q(n)$ nominally) becomes larger, until in the large charge limit $q(n)\rightarrow 0$ and particles are not allowed to cross.

\item  A second comment has to do with the range of energies in the charge $n$ sector, which is
\begin{equation}\label{eq:energy_range}
|E| \le \frac{2J}{\sqrt{1-q(n)} } \exp \left[ \frac{\sqrt{\lambda N} }{2} \log\left(1-\frac{4n^2}{N^2} \right)+\frac{\lambda }{2} \left(1-\frac{4Nn}{N^2-4n^2} \right) \right] ~
\end{equation}
which for $n \propto N$ shrinks exponentially fast with $e^{-C\sqrt{N} } $. The results here apply for any scaling of the charges $n\sim N^\a$ with $0<\a\leq 1$. The case of $\a=3/4$ is special because in this case the range of energies is finite. Defining a fixed $r$ as
\be
\frac{n}{N^{3/4}}=r~,
\ee
the range of the energies in the charge $n \sim N^{3/4}$ sector is
\be \label{Erange}
|E|\leq \frac{2J e^{\lambda/2}}{\sqrt{1-q} } \exp \left( -2r^2\sqrt{\lambda} \right) ~
\ee
where $q=e^{-4\lambda } $.

\item Finally, the prefactor in \eqref{eq:Z_fixed_charge} is easy to understand. Using the fact that $\int _0^{\pi } \frac{d\theta }{2\pi } (q,e^{\pm 2i\theta } ;q)_{\infty } =1$, the total number of states in the charge $n$ sector is (in the $N\rightarrow\infty$ limit) given by\footnote{Assuming no particular degeneracy of states at $E=0$. Also, we have restored the total number of states $2^N$ which was implicit because of our normalization $\tr~\mathbb{1}=1$.}
\begin{equation}\label{m0FC}
\begin{split}
 \frac{2^N}{2\pi } \int _{-\pi } ^{\pi } d\chi \, e^{-in\chi } m_0(\chi )=&\frac{2}{\sqrt{2\pi N}} \cdot \frac{1}{\sqrt{1-4n^2/N^2} }\\
&\hspace{-0.5in} \exp\left[N\log(2)-2n \arctanh\left(\frac{2n}{N} \right)-\frac{N}{2} \log\left(1-\frac{4n^2}{N^2} \right)\right] ~.
\end{split}
\end{equation}
We know that the exact number of states of charge $n$ (that is $\bar \psi_{i_1} \cdots \jb_{i_{n+N/2}} |0\rangle$) is given by
\begin{equation}
\binom{N}{n+N/2}~;
\end{equation}
using Stirling's formula for $N!$ and $(N/2 \pm n)!$ in the large $N$ limit, this binomial coefficient becomes the same as \eqref{m0FC}.

\end{enumerate}

From Eq.\ \eqref{eq:Z_fixed_charge} (after inserting the factor of $2^N$ that comes from restoring the normalization of the trace) and Eq.\ \eqref{eq:Z_beta_n}, we find that the partition function in the canonical ensemble reads
\begin{align}\label{Z_beta_n1}
\begin{split}
\cZ(\b,\cQ)=&\frac{2 }{\sqrt{2\pi N}} \frac{1}{\sqrt{1-4\cQ^2}  } \exp\Bigg[ N \cQ\log \left(\frac{1-2\cQ}{1+2\cQ} \right)+ \frac{N}{2} \log\left(\frac{4}{1-4\cQ^2} \right) \Bigg] \cdot \\[5pt]
&\hspace{-1in}  \int _0^{\pi } \frac{d\theta }{2\pi } \left(q(\cQ),e^{\pm 2i\theta } ;q(\cQ)\right)_{\infty } \exp\left\{ - \frac{2\b J\cos \theta}{\sqrt{1-q(\cQ)} } e^{\lambda/2}\left(1-4\cQ^2 \right)^{p/2} \exp \left(\frac{2\lambda \cQ}{1-4\cQ^2} \right)  \right\}~ ,
\end{split}
\end{align}
where (using the definition \eqref{eq:cQ_def})
\be \label{qAsFunOfQ}
\cQ=-\frac{n}{N}~,~~~~q(\cQ)=\exp \left(-\frac{4\lambda}{1-4\cQ^2} \right) ~.
\ee
The range of charges $-N/2 \leq n \leq N/2$ translates to $-1/2 \leq \cQ \leq 1/2$, which is the convention of \cite{Davison:2016ngz}.

\subsection{Fixed charge canonical ensemble in the limit $\lambda\to 0$} \label{sec424}

Our interest in this section is the limit $\lambda\to 0$ at a fixed $\cQ$ of \eqref{Z_beta_n1}.
This limit was computed in \cite{Berkooz:2018qkz} (see Eq.\ (4.13) there) where we should substitute instead of $\lambda $ there the value $\bar{\lambda}= \frac{4\lambda}{1-4\cQ^2}$.
We find that the free energy $F(T,\cQ)=-T\log \cZ(T,\cQ)$ has the following low temperature expansion in the $\lambda  \to 0$ limit
\begin{align}\label{FreeEnergy}
F(T,\cQ)= &-\frac{J}{\sqrt{\lambda}}\(1-4\cQ^2\)^{\frac p2+\frac12}\no\\[5pt]
&\hspace{-0.5in} -T\[N\(\cQ\log \left(\frac{1-2\cQ}{1+2\cQ} \right)+ \frac{1}{2} \log\left(\frac{4}{1-4\cQ^2} \right)\)-\frac{\pi ^2}{8\lambda } (1-4\cQ^2) + \cdots  \] +O(T^2).
\end{align}
We can compare this result with the findings of \cite{Davison:2016ngz}. The canonical free energy of the complex SYK model has a low temperature expansion \eqref{eq:free_energy_low_T}.
We find from \eqref{FreeEnergy} that
\begin{align}
E_0(\cQ)&=-\frac{J}{\sqrt{\lambda}}\(1-4\cQ^2\)^{\frac {p+1}{2}}~,\\
\cS(\cQ)&=N\[\cQ\log \left(\frac{1-2\cQ}{1+2\cQ} \right)+ \frac{1}{2} \log\left(\frac{4}{1-4\cQ^2}\)\]-\frac{\pi ^2}{8\lambda } (1-4\cQ^2) ~.
\end{align}
These agree with \eqref{grndEn} and \eqref{zeroTentropy}. Thus, the large $p$ limit of the model (appendix C of \cite{Davison:2016ngz}) is indeed reproduced by $\lambda  \to 0$.


\section{$U(M)$ symmetric SYK model} \label{sec:UMSYK}

In this section we consider a generalization of the $U(1)$ model discussed in the previous sections to a model with $U(M)$ global symmetry.
We will find that the basic machinery developed for solving the $U(1)$ model generally carries over to this case as well, with some differences pointed out below.
In this section we outline the calculation of the partition function of the $U(M)$ SYK model in the double scaling regime and relegate the details to appendix \ref{UMDetails}.

Consider complex fermions with two kinds of indices $\bar{\psi}_{i \a}~,\psi^{i \b}$. The index $i$ ranges from $1$ to $N$, and the index $\a$ is a $U(M)$ flavor index that ranges from $1$ to $M$. These fermions transform in the (anti) fundamental representation of $U(M)$, and satisfy the anti-commutation relation
\be\label{unga}
\{\bar{\psi}_{i \a}, \psi^{j \b}\}=2\delta_{i}^j\delta_{\a}^\b~.
\ee
Here we consider an SYK-like Hamiltonian which is invariant under a global $U(M)$ symmetry
\begin{equation}\label{UMHamilGen}
\sum J_{j_{1}\cdots j_{p}}^{i_{1}\cdots i_{p}}~\bar{\psi}_{i_{p}\a_p}\cdots\bar{\psi}_{i_{1}\a_1}\psi^{j_{1}\a_1}\cdots\psi^{j_{p}\a_p},
\end{equation}
where an unconstrained summation over all upper and lower indices is to be understood.
This model appears also in, e.g., \cite{Yoon:2017nig,Bhattacharya:2017vaz}.
For Hermiticity of the Hamiltonian we demand that $(J_{j_{1}\cdots j_{p}}^{i_{1}\cdots i_{p}})^{*}=J_{i_{1}\cdots i_{p}}^{j_{1}\cdots j_{p}}$. The Hamiltonian is, therefore, just a product of $SU(N)$ currents with random coefficients.

Depending on symmetry properties of the tensor $J$, the model has many variants. A-priori, the only symmetry property of the tensor $J_{j_{1}\cdots j_{p}}^{i_{1}\cdots i_{p}}$ is invariance under exchange of pairs $(i_m,j_m)\leftrightarrow (i_n,j_n)$, so that exchanging separately say $i_m \leftrightarrow i_n$ does not produce the same interaction term. However, for simplicity, we will consider a single coupling for every set of sites $i_1<\cdots <i_p$ and $j_1<\cdots <j_p$.
Other variants can be treated using a similar set of tools, and the results slightly differ. Concretely, the Hamiltonian that we consider is
\begin{equation}\label{UMHamil}
H=\sum_{\substack{1\le i_{1}<\cdots<i_{p}\le N\\
1\le j_{1}<\cdots<j_{p}\le N
}
}J_{j_{1}\cdots j_{p}}^{i_{1}\cdots i_{p}}~\bar{\psi}_{i_{p}\a_p}\cdots\bar{\psi}_{i_{1}\a_1}\psi^{j_{1}\a_1}\cdots\psi^{j_{p}\a_p}~,
\end{equation}
where an unconstrained summation over the upper and lower flavor index $\a$ is to be understood. Correspondingly, we assume that the variance of the random coupling $J$ is:
\begin{equation}
\langle J_{j_{1}\cdots j_{p}}^{i_{1}\cdots i_{p}}J_{i_{1}\cdots i_{p}}^{j_{1}\cdots j_{p}}\rangle_{J}=\JUM^{2}M^{-p}{N \choose p}^{-2} \ .
\end{equation}
Our goal in this section is to calculate the ensemble averaged partition function
\be\label{UMpartFun}
Z(\beta,\mu)=\<\tr \exp\(-\beta H- \sum_{\alpha =1}^{M}\mu_{\alpha } Q^{\alpha } \)\>_J
\ee
in the limit of large $N$ but finite $M$ and finite $\lambda=p^2/N$. In the above equation, $Q^{\alpha }$, defined as
\be
Q^{\alpha }=\frac12 \sum_{i=1}^N \(\jb_{i\alpha }\j^{i\alpha }-\j^{i\alpha }\jb_{i\alpha }\)~,
\ee
are generators of the Cartan subalgebra in $U(M)$.

As in the $U(1)$ model, we calculate \eqref{UMpartFun} by expanding out in moments
\be\label{UMmoments1}
m_k(\mu)=  \<\tr ~H^k e^{-\sum_{\alpha }\mu_{\alpha } Q^{\alpha }}\>_J ~.
\ee
Note that as before, we are not expanding the chemical potential terms. Averaging over the disorder pairs up the Hamiltonians, thereby giving rise to a sum over unoriented chord diagrams (such as the one in figure \ref{fig:chord_diagram}). Each unoriented chord (which we refer to as an $H$-chord) represents a contraction of the site indices across a pair of Hamiltonians. Since the structure of the site indices of the $U(M)$ Hamiltonian \eqref{UMHamil} is the same as the $U(1)$ Hamiltonian \eqref{Hamiltonian}, it follows from section \ref{sec:3.2} that each $H$-chord can be decomposed into two oriented chords. Now here comes the first point of departure with respect to the $U(1)$ model.
Consider an oriented chord and the associated chord of opposite orientation, which together constitute the unoriented $H$ chord. Each oriented chord is associated with a multi-site index; however, there are two different flavor indices on its ends, because the flavor indices are contracted separately between the two oriented chords at each end (associated to an Hamiltonian insertion). This implies that we cannot consider oriented chords separately, but rather we have to simultaneously consider the pairs of oriented chords, that is the unoriented chords, as the basic objects. Therefore, the chord diagram rules are not given in terms of a value assigned to every oriented chord and pairs of oriented chords, as was done in fig.\ \ref{fig:allowed_overlaps}, but rather in terms of unoriented chords and pairs of those (see for example fig.\ \ref{unor_all}). As before, it is enough to consider at most pairs of chords in the large $N$ limit.

The rules obtained for unoriented chords of the $U(M)$ model in the presence of a chemical potential are as follows (they are worked out in appendix \ref{UMDetails})
\begin{enumerate}
\item For every $H$-chord there is a factor of
\be\label{UMrule1}
\exp\(\frac{\lambda}{\cA(\mu)^2} \sum_{\a=1}^M \frac{e^{-2\m_\a}}{\(\cosh\mu_\a\)^4} \)~.
\ee
\item For every pair of $H$-chords that do not intersect, there is a factor of
\be\label{UMrule2}
\exp\(\frac{4\lambda}{\cA(\mu)^2} \sum_{\a=1}^M (\tanh \m_\a \cdot \text{sech} \m_\a)^2\)~.
\ee
\item For every pair of $H$-chords that intersect, there is a factor of
\be\label{UMrule3}
\exp\(-\frac{4\lambda}{\cA(\mu)^2}\sum_{\a=1}^M\frac{1}{\(\cosh\mu_\a\)^4}\)~.
\ee
\end{enumerate}
In the above formulas, $\cA(\m)$ is the following function of $\mu_\a$
\be
\cA(\m)=\sum_{\a=1}^M\frac{1}{\(\cosh\mu_\a\)^2}~.
\ee
With these rules, it is now straightforward to write down the generic moment which is found to be
\begin{align}\label{genericUMmoment}
m_k(\m)= &\JUM^k \prod_{\a=1}^M \left[ (\cosh\m_\a)^N\right] \left(\cA(\m)/M\right)^{kp/2} \exp\(\frac{k \lambda}{2\cA(\mu)^2} \sum_{\a=1}^M \frac{e^{-2\m_\a}}{\(\cosh\mu_\a\)^4} \) \cdot\no\\[10pt]
&\exp\({k/2 \choose 2}\frac{4\lambda}{\cA(\mu)^2} \sum_{\a=1}^M (\tanh \m_\a\cdot \text{sech} \m_\a)^2\)\cdot \sum_{CD} \exp\(-\frac{4\kappa_H \lambda}{\cA(\mu)}\)~.
\end{align}
In the above expression, the sum in the second line is over all unoriented chord diagrams (as in figure \ref{fig:chord_diagram}) with $k/2$ $H$-chords.
As a consistency check, setting $M=1$ reproduces the result of the $U(1)$ model \eqref{eq:moments_chemical_potential}.

For general $M$ with no chemical potentials, \eqref{genericUMmoment} becomes
\begin{align}
&m_k= \JUM^k \exp\( \frac{k\lambda}{2M}\)  \sum_{CD} \exp\(-\frac{4\kappa_H \lambda}{M}\)~.
\end{align}
This takes the same form as in Majorana SYK, where the effective $q$ in this model is
\be
q=\exp\(-\frac{4p^2}{MN}\)~ .
\ee
The factor of $MN$ in the denominator corresponds to the the total number of fermions in the system.


\section{Two-point function}\label{sec:2ptFunctionSec}

In this section we will compute two point functions in the complex SYK model, and the first issue is to discuss what are the natural operators in the theory. We will be interested in operators of the form
\begin{align}\label{Mgen}
\begin{split}
M&=\sum_{\substack{1\le i_{1}<\cdots<i_{p_M}\le N\\
1\le j_{1}<\cdots<j_{\bar{p}_M}\le N
}
}\(J^{(M)}\)^{j_{1}\cdots j_{\bar{p}_M}}_{i_{1}\cdots i_{p_M}}~\psi^{i_{1}}\cdots\psi^{i_{p_M}}\bar{\psi}_{j_{\bar{p}_M}}\cdots\bar{\psi}_{j_{1}}~\\[5pt]
&\equiv \sum_{I, I'}\(J^{(M)}\)_{I}^{I'} \j^{I} \jb_{I'}~.
\end{split}
\end{align}
The coefficients $(J^{(M)})_I^{I'} $ are again taken to be random Gaussian variables with variance
\begin{equation}
\langle (J^{(M)})_{I_1}^{I'_1} \left( (J^{(M)})_{I_2}^{I'_2}\right)^{*}\rangle_{J_M}=J_{M}^{2}{N \choose p_{M}}^{-1} {N \choose \bar p_{ M}}^{-1} \delta_{I_1 I_2}\delta _{I'_1 I'_2}~.
\end{equation}
These coefficients are uncorrelated with the random coefficients $J_I^{I'}$ in the Hamiltonian.

Given an operator $M$, we have denoted by $p_M$ the number of $\psi$'s in the operator, and by ${\bar p}_M$ the number of $\bar\psi$'s. We will refer to the sum $p_M+\bar p_M$ as the ``size'' of the operator. The difference $p_M-\bar p_M$ determines the charge. In this section, we will take the size of the operator to be double scaled as well, or more precisely
\begin{equation}
p_M,\ {\bar p}_M\propto \sqrt{N},\ \ as\ N\rightarrow\infty~.
\end{equation}
We will refer to such operators as {\it double scaled random operators}.

In section \ref{sec:chaos} we will compute 4-point functions of such operators and in section \ref{sec:massive} we will discuss ``longer'' operators. Before computing the two point function we turn to explain why this is the right class of operators.

\subsection{Why double scaled random operators?}

The rationale for requiring random couplings was discussed in \cite{Berkooz:2018qkz,Berkooz:2018jqr}. Let us briefly review the arguments there (and then somewhat rephrase them).
\begin{enumerate}
\item {\it ``Single trace'' probes should be in the same statistical class as the energy momentum tensor.}

Consider an AdS black hole, which we think about as some core set of degrees of freedom governed by a suitable random Hamiltonian. Note that an $AdS_2$ might appear as an IR of an altogether different UV theory, and the degrees of freedom with which we describe the black hole may or may not be simply related to the degrees of freedom of that UV theory. We then probe the black hole using the available bulk probes, such as single trace operators or their analogues. The Hamiltonian in quantum mechanics, or the local energy-momentum tensor in higher dimensions, is one such operator. Probing with the full Hamiltonian does not provide any more information beyond the partition function, but the local energy momentum operator does (if the theory is higher dimensional). Since the full Hamiltonian is a random operator when acting on a suitable set of degrees of freedom describing the BH, we can expect that the local energy momentum operator will also be effectively described by some random (local) operator acting on these d.o.fs.

But the local energy-momentum tensor is just one operator in a tower of single trace operators with which we can probe the system. In ${\cal N}=4$ SYM we can use its primary $\text{tr}(X^2)$ to probe the black hole, or we can just as well use any other of the $\text{tr}(X^n)$ operators. If the former is a random operator on the states of the black hole, we can expect that all single trace operators would be random operators. If we have some idea about the statistical ensemble of the Hamiltonian, we can expect that the ensemble for all the other single trace operators will be of a similar nature. This lands us on the proposal for the observables above, after we allow for a more general charge and mass.

\item {\it Universality of the observables}

It is important to note that there is actually a large number of models with different microscopic details, yet with a double scaling limit which reduces to the same set of chord partition functions (for example, the model originally discussed in \cite{erdHos2014phase}). This is consistent with expectations from gravity, where, if one is interested in the AdS$_2$ part of space-time, one glues it to an external region in order to break conformal invariance. There is a broad range of possibilities for such different UV spaces (one usually thinks about it in the other direction --- $AdS_2$ appears as the IR near horizon limit of many different backgrounds). The probes which are legitimate in the entire theory are really defined on the boundary of spacetime, i.e., the boundary of the UV region, which is not strict $AdS_2$ physics. Different random microscopic models, or different embeddings of $AdS_2$ in bigger spaces, come with an altogether different set of particles and hence observables. So on top of the Hamiltonian we are actually instructed to be able to define a large set of observables (1) which are independent of the microscopics of the $AdS_2$ model, (2) for which we freely specify quantum numbers such as mass and spin, and (3) which obey factorization of correlation functions. Our class of random operators is precisely like that.

\end{enumerate}

\subsubsection{Double scaled random operators as consistent truncations}

We can also phrase this choice of operators in the language of consistent truncations. By a consistent truncation one means truncating the set of fields of GR/String theory to a smaller set of fields, setting all the others to zero. We require that the fields that we kept close under the equations of motion. I.e., none of the fields that we have set to zero will have a tadpole for any configuration of the fields that we kept (at least for those that satisfy the equations of motion).

Suppose we are given a random Hamiltonian and as many as we want (but finite) number of random operators of any length, all with random coefficients. We will refer to these operators as {\it the basic set of operators}. Suppose we consider any additional operator, with random coefficients (with zero mean) which are drawn independently of any of the coefficients in the basic set. It is obvious then that the 1-point function of the additional operator is zero, in any trace which contains as many insertions of the Hamiltonian and basic set operators as we want. So in any correlator of operators from the basic set we will not excite any quanta of any independent random operator outside the basic set, which is just the statement that independent random operators can be consistently truncated. Going back to the picture above that different realizations come with different particles (observables), we can just refer to them as different consistent truncations of a richer theory.


\subsection{Two-point function in a fixed chemical potential}\label{charged operator}

The two point function that we will be interested in is
\begin{equation}\label{2ptFunM}
G(\tau)=
 - \frac{1}{ Z(\beta,\mu)} \langle \Tr \[ e^{-\beta K} T_\tau \(e^{\tau K} M e^{-\tau K} \bar{M}\) \]\rangle_{J,J_M}~.
\end{equation} $T_\tau$ specifies the $\tau$ ordering of the product and $K$, defined in \eqref{KDef}, is the sum of the Hamiltonian and the fermion number operator. The notation $\langle \cdots \rangle _{J_M} $ stands for the average over the random operators and $\langle \cdots \rangle_{J,J_M}$ for the average over the ensemble of both the Hamiltonian and operator couplings.

The calculation of the two-point function proceeds by calculating the moments as follows. Define the moments
\begin{equation}\label{2ptFunMom}
\begin{split}
m_{k_1,k_2}(\mu)\equiv \exp\(\frac{2\tau\mu(\bar{p}_M-p_M)}{\beta} \)  \langle \tr M H^{k_1} \bar M H^{k_2} \exp \left[- \frac{\mu }{2} \sum _{i=1} ^N [\bar \psi_i,\psi^i] \right]\rangle_{J,J_M}
~.
\end{split}
\end{equation}
Then the 2-point function for $\tau >0$ is
\begin{equation} \label{eq:2pf_expansion}
\begin{split}
& G(\tau ) \overset{\tau>0}{=} - \frac{1}{Z(\beta,\mu )} \sum _{k_1,k_2=0} ^{\infty } \frac{(-\tau )^{k_1} }{k_1!} \frac{(\tau -\beta )^{k_2} }{k_2!} m_{k_1,k_2} (\mu )
\end{split}
\end{equation}
and for $\tau <0$ we just take the result for \eqref{eq:2pf_expansion}, plug $\tau \to -\tau $, exchange $p_M \leftrightarrow \bar p_M$, and multiply by $(-1)^{p_M + \bar p_M}$.

Performing the average over the disorder, the moments become (we denote $k=k_{1}+k_{2}$ below)
\begin{equation} \label{eq:basic_2pf}
\begin{split} & \langle\tr M H^{k_{1}}\bar{M} H^{k_{2}} \exp \left[- \frac{\mu }{2} \sum _{i=1} ^N [\bar \psi_i,\psi^i] \right] \rangle_{J,J_M}=J^{k} J_{M}^{2}\sum_{CD}{N \choose p}^{-k}{N \choose p_{M}}^{-1}{N \choose \bar p_{M}}^{-1} \cdot \\
& \qquad  \cdot \sum_{I_{1}\cdots I_{k}}\sum_{IJ}\tr\left[ \psi^{I} \bar \psi_J \cdot  \bar{\psi}_{I_{1}}\psi^{I_{2}}\cdots\bar{\psi}_{I_{3}}\psi^{I_{4}}\cdots\psi^J\bar{\psi}_{I}\cdots \exp \left[- \frac{\mu }{2} \sum _{i=1} ^N [\bar \psi_i,\psi^i] \right] \right]~. \end{split}
\end{equation}

\begin{figure}[h]
\centering
\includegraphics[width=0.8\textwidth]{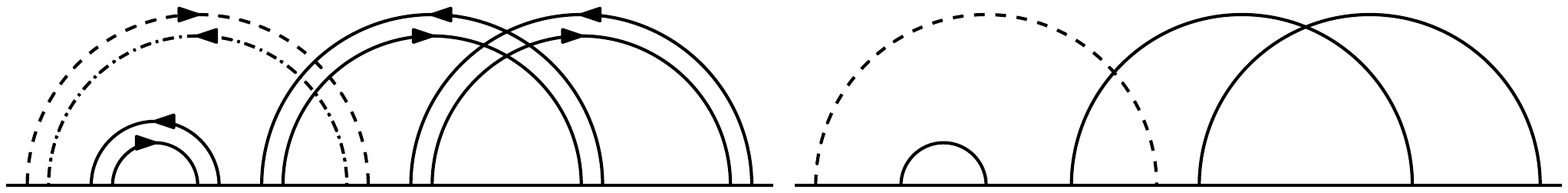}\caption{A diagram contributing to the moments of the 2-point function with
$k=6$, and the corresponding unoriented diagram.}
\label{fig:2pf_example}
\end{figure}

This time, in addition
to the $H$-chords that comprise a pair of oriented chords with opposite
orientation, there are also two oriented chords representing the
$I$ index and $J$ index contractions of the external operators; we distinguish them by drawing those as dashed chords (see figure \ref{fig:2pf_example} for an example). Note the two dashed chords correspond to index sets of different sizes in general. Nevertheless, we can just as well assign a (dashed) unoriented chord, to be referred to as an $M$-chord, corresponding to this pair of oriented operator chords.

The sum over the index sets $I,J,I_1,\cdots ,I_k$ with the binomial coefficients and the trace in \eqref{eq:basic_2pf} (in the presence of a chemical potential) was evaluated in general in section \ref{sec:grand_partition_function}, with the resulting rules for a given chord diagram shown in fig.\ \ref{fig:allowed_overlaps}.
Once again, at large $N$,
there are no minus signs since each oriented dashed chord (just as the solid chords) intersects
an even number of oriented solid chords.
For the 2-point function, if all $I,J,I_1,\cdots,I_k$ are distinct then the trace equals $\left(e^{\mu } \right)^{p_M-\bar p_M} \cosh(\mu )^{N-kp-p_M-\bar p_M} $ (corresponding to the contributions from figs.\ \ref{fig:allowed_overlaps_7}-\ref{fig:allowed_overlaps_9}). The full answer is obtained by multiplying this by the value assigned to every pair of oriented chords according to fig.\ \ref{fig:allowed_overlaps}. All that remains is to read the number of pairs of oriented chords of each sort from the number of total unoriented chords ($H$-chords and the dashed $M$-chord) and the intersections of them. We denote again by $\kappa _H$ the number of intersections of $H$-chords, and by $\kappa _{HM} $ the number of intersections of the $M$-chord with $H$-chords. This is done in appendix \ref{app:details_2pf}.
Putting all the ingredients there together, we obtain for the moments
\begin{equation} \label{eq:2pf_moments_chemical_potential}
\begin{split}
&m_{k_1,k_2}(\mu )=
J_M^2 J^k e^{\mu(1-2T\tau) (p_M-\bar p_M)} (\cosh(\mu ))^{N-kp-p_M-\bar p_M} \cdot \\[5pt]
& \cdot \exp \Bigg\{ \frac{p^2}{N} \left[\frac{k^2}{2}\sinh(\mu)^2-\frac{k}{2} \left(\sinh(2\mu )-1 \right) \right] - \frac{pp_M}{N} \left[k_1 \frac{1-e^{2\mu } }{2} +k_2 \frac{1-e^{-2\mu } }{2} \right] - \\[5pt]
& - \frac{p\bar p_M}{N} \left[k_1 \frac{1-e^{-2\mu } }{2} +k_2 \frac{1-e^{2\mu } }{2} \right]
+ \frac{p_M\bar p_M}{N} e^{2\mu } \Bigg\} \\[5pt]
& \cdot \sum _{CD} \exp\left\{ -4\kappa _H \frac{p^2}{N} \cosh(\mu )^2-2 \kappa _{HM} \frac{p(p_M+\bar p_M)}{N} \cosh(\mu )^2    \right\} \ .
\end{split}
\end{equation}

\subsection{Two-point function in a fixed charge sector} \label{sec:2pf_fixed_charge}

In this subsection we compute the 2-point function in the model, where the external states in the trace have a given \emph{fixed} charge $(-n)$. In order to do that, we change the time evolution of the operators in the previous subsection from $K$ to the usual Hamiltonian $H$, and keep $K$ only in the overall exponential inside the trace. That is, we start with $\tr \left[ e^{-\beta K} e^{\tau H} M e^{-\tau H} \bar M\right] $ and perform the same projection to a fixed charge as in the partition function. This results simply in the omission of the $\exp\left[2\tau\mu({\bar p_M}-p_M)/\beta\right]$ term in \eqref{2ptFunMom}.

As done for the partition function, in order to perform fixed charge projection we first change variables $2\mu =i\chi $, which results in\footnote{Note that we denote these moments by $m_{k_1,k_2} $ as well, but this is distinguished from the definition in \eqref{2ptFunMom} by the omission of the first exponential factor; this is implied in the notation by the argument being $\chi $.}
\begin{equation}
\begin{split}
&m_{k_1,k_2}(\chi  ) \equiv
\langle \tr M H^{k_1} \bar M H^{k_2} \exp \left[- \frac{i\chi}{4} \sum _{i=1} ^N [\bar \psi_i,\psi^i] \right]\rangle_{J,J_M}= \\
&= J_M^2 J^k e^{i\chi
(p_M-\bar p_M)/2} (\cos(\chi /2 ))^{N-kp-p_M-\bar p_M} \cdot \\[5pt]
& \cdot \exp \Bigg\{ \frac{p^2}{N} \left[-\frac{k^2}{2}\sin\left( \frac{\chi }{2} \right) ^2-\frac{k}{2} \left(i \sin(\chi  )-1 \right) \right] - \frac{pp_M}{N} \left[k_1 \frac{1-e^{i\chi  } }{2} +k_2 \frac{1-e^{-i\chi  } }{2} \right] - \\[5pt]
& - \frac{p\bar p_M}{N} \left[k_1 \frac{1-e^{-i\chi  } }{2} +k_2 \frac{1-e^{i\chi  } }{2} \right]
+ \frac{p_M\bar p_M}{N} e^{i\chi  } \Bigg\} \\[5pt]
& \cdot \sum _{CD} \exp\left\{ -4\kappa _H \frac{p^2}{N} \cos(\chi /2 )^2-2 \kappa _{HM} \frac{p(p_M+\bar p_M)}{N} \cos(\chi /2 )^2    \right\} .
\end{split}
\end{equation}
As before we calculate $\int d\chi~ e^{-in\chi }~  m_{k_1,k_2} (\chi )$ (where $n$ is an integer assuming $N$ is even) via a saddle point approximation, along the lines of section \ref{sec:partition_function_charge_saddle}. The terms in the exponent that scale with a positive power of $N$ are
\be\label{chiExp}
\exp\[-in\chi +i\frac{\chi}{2}(p_M-\bar{p}_M)
+(N-kp-p_M-\bar{p}_M)\log \cos \frac{\chi }{2}\] ~.
\ee
This has a saddle point (at least for non-finite $n$) at
\begin{equation}\label{saddlePT}
\chi = 2i \tanh^{-1}\( \frac{(p_M-\bar{p}_M)
-2n}{N-kp-p_M-\bar{p}_M} \)\equiv 
2i \tanh^{-1}\( \frac{X}{N-kp-p_M-\bar{p}_M} \)
\end{equation}
where $X$ is defined to be
\begin{align}
X&=
p_M-\bar{p}_M
-2n~.
\end{align}
Plugging the saddle point expression \eqref{saddlePT} for $\chi$ back in \eqref{chiExp} gives
\begin{equation}\label{chiExp2}
\begin{split}
&\exp\Bigg[ -\frac{1}{2} (N-kp-p_M-\bar{p}_M) \log \left(1-\frac{X^2}{(N-kp-p_M-\bar{p}_M)^2}\right)-\\
& \qquad \qquad \qquad\qquad\qquad\qquad\qquad- X \tanh ^{-1}\left(\frac{X}{N-kp-p_M-\bar{p}_M}\right) \Bigg] = \\[5pt]
=&\exp\Bigg[-\frac{1}{2} N \log \left(1-\frac{X^2}{N^2}\right)-\frac X2 \log\(\frac{1+X/N}{1-X/N}\)+\frac{1}{2} N  \log \left(1-\frac{X^2}{N^2}\right)\epsilon-\\
& \qquad \qquad -  \frac{N  X^2}{2 \left(N^2-X^2\right)}\epsilon^2+O(N\epsilon^3)\Bigg]~,
\end{split}
\end{equation}
where we have expanded in small $\epsilon$ defined as
\be
\epsilon \equiv \frac{kp+p_M+\bar{p}_M}{N}
\ee
and the correction go at most as $O\left( \frac{1}{\sqrt{N}} \right) $ (here and below).

There is also the term in $m_{k_1,k_2} (\chi )$
\begin{align}\label{chiExp3}
\exp \bigg\{  -\frac{p^2k^2}{2N} \sin(\chi/2)^2 \bigg\}=\exp\left[\frac{p^2k^2}{2N} \frac{X^2}{N^2-X^2}+O(N^{-1/2} )\right]~.
\end{align}
The 2-point function at a fixed charge sector must be finite when we sum over $k$ (as the 2-point function itself is finite). Indeed, the $k^2$ terms in \eqref{chiExp2} and \eqref{chiExp3} cancel exactly.

Completing the saddle point calculation with the second derivative term, and performing some simplifications, we find
\begin{align}
\begin{split}
&\frac{1}{2\pi}\int d\chi e^{-in\chi}m_{k_1,k_2}(\chi) =\\
&= 2\frac{J_M^2 \(Je^{p^2/(2N)}\)^k}{\sqrt{2\pi N}} \exp\bigg[-\frac{1}{2} (N+1) \log \left(1-\frac{X^2}{N^2}\right)-\frac X2 \log\(\frac{1+X/N}{1-X/N}\) \\
&+\frac{1}{2} (kp+p_M+\bar{p}_M)  \log \left(1-\frac{X^2}{N^2}\right)-\frac{ X^2/N^2}{2 \left(1-X^2/N^2\right)}\frac{1}{N}((p_M+\bar{p}_M)^2+2kp(p_M+\bar{p}_M))\\
&+\frac{kp^2}{N} \frac{X/N}{1-X^2/N^2} +\frac{p_M\bar{p}_M}{N} \(\frac{N-X}{N+X}\)-\frac{p\bar{p}_M}{N}\( -\frac{X}{N-X} k_1 +\frac{X}{N+X}k_2\)\\
&+\frac{pp_M}{N} \(-\frac{X}{N+X}k_1 + \frac{X}{N-X}k_2\)
\bigg] \cdot \sum _{CD}\exp \left[\frac{-1}{1-X^2/N^2 } \left(\frac{4p^2}{N} \kappa _H+\frac{2p(p_M+\bar{p}_M)}{N} \kappa _{HM} \right)\right].
\end{split}
\end{align}
With the following definition
\be
q(n)=\exp \left[-\frac{4p^2}{N} \frac{1}{1-4n^2/N^2 } \right]~,~~\tilde{q}(n)=\exp \left[-\frac{2p(p_M+\bar{p}_M)}{N}\frac{1}{1-4n^2/N^2 } \right]
\ee
the sum over chord diagrams is the same as in the 2-point function analysis in \cite{Berkooz:2018qkz,Berkooz:2018jqr},
giving for the moments in the fixed charge sector
\begin{align}\label{MomExpand1}
\begin{split}
& \frac{1}{2\pi}\int d\chi e^{-in\chi}m_{k_1,k_2}(\chi) =  \frac{2J_M^2 }{\sqrt{2\pi N}} e^{S_M}\int _0^{\pi } \prod _{j=1,2} \left[\frac{d\theta _j}{2\pi } \left(q(n),e^{\pm 2i\theta _j} ;q(n)\right)_{\infty } \right]  \\[5pt]
&  \frac{\left(\tilde q(n)^2;q(n)\right)_{\infty } }{\left(\tilde q(n) e^{i(\pm \theta _1 \pm \theta _2)} ;q(n)\right)_{\infty } }  \(\frac{  2J e^{p^2/(2N)} \cos\theta_1 }{\sqrt{1-q(n)}}  e^{A+B_1}\)^{k_1} \(\frac{  2J e^{p^2/(2N)} \cos\theta_2 }{\sqrt{1-q(n)}}  e^{A+B_2}\)^{k_2}~,
\end{split}
\end{align}
where the quantities $S_M,A,B_1,B_2$ are defined as follows
\begin{align}
S_M & = -\frac{1}{2} (N+1) \log \left(1-\frac{X^2}{N^2}\right)-\frac X2 \log\(\frac{1+X/N}{1-X/N}\)\no\\
+&\frac{1}{2} (p_M+\bar{p}_M)  \log \left(1-\frac{X^2}{N^2}\right)-\frac{ 4n^2}{2 \left(N^2-4n^2\right)}\frac{(p_M+\bar{p}_M)^2}{N}+\frac{p_M\bar{p}_M}{N} \(\frac{N+2n}{N-2n}\)~,\\
A&= \frac p2 \log \left(1-\frac{X^2}{N^2}\right)-\frac{p(p_M+\bar{p}_M)}{N} \frac{ 4n^2}{ N^2-4n^2}-\frac{p^2}{N} \frac{2nN}{N^2-4n^2}~,\\
B_1&= \frac{pp_M}{N} \frac{2n}{N-2n} -\frac{p\bar{p}_M}{N}\frac{2n}{N+2n} ~, \\
B_2&= -\frac{pp_M}{N} \frac{2n}{N+2n} +\frac{p\bar{p}_M}{N}\frac{2n}{N-2n}~.
\end{align}
From the expression for the moments \eqref{MomExpand1} we obtain the result  for the 2-point function of operators of size $ \sim O(N^{1/2} )$ in a charge $(-n)$ sector (keeping terms to the order in which we are working)
\begin{align}\label{twoPtFunGen}
\begin{split}
&\langle \tr e^{-\beta H} M(\tau ) \bar M(0)\rangle_{J,J_M}^{(-n)} =  \frac{2J_M^2 }{\sqrt{2\pi N}} e^{S_M}\int _0^{\pi } \prod _{j=1,2} \left[\frac{d\theta _j}{2\pi } \left(q(n),e^{\pm 2i\theta _j} ;q(n)\right)_{\infty } \right]  \\[5pt]
&  \frac{\left(\tilde q(n)^2;q(n)\right)_{\infty } }{\left(\tilde q(n) e^{i(\pm \theta _1 \pm \theta _2)} ;q(n)\right)_{\infty } } \exp \(-\frac{  2J\tau e^{p^2/(2N)} \cos\theta_1 }{\sqrt{1-q(n)}}  e^{A+B_1}-\frac{  2J(\beta-\tau) e^{p^2/(2N)} \cos\theta_2 }{\sqrt{1-q(n)}}  e^{A+B_2}\) ~.
\end{split}
\end{align}
Normalizing \eqref{twoPtFunGen} by the partition function gives the result for the two-point function for $\tau >0$ (the analog of \eqref{2ptFunM} for fixed charge).
For $\tau<0$ we need the expression for $\langle \tr e^{-\beta H} \bar{M}(0) M(\tau) \rangle_{J,J_M}^{(-n)}$ (multiplied by $(-1)$ for a fermionic operator $M$) which is obtained from \eqref{twoPtFunGen} by the replacement $p_M\leftrightarrow \bar{p}_M$ and $\tau  \to -\tau $ as mentioned above.
In the following subsection, we evaluate the integrals in \eqref{twoPtFunGen} in the conformal regime.


\subsubsection{Conformal limit}

The expression \eqref{twoPtFunGen} is the two-point function of the operator $M$ in a fixed charge sector for a generic value of $p^2/N$. In this subsection we consider the $p^2/N \to 0$ limit corresponding to the usual SYK within the conformal regime
\be
\sqrt{1-q(n)}\to \sqrt{\lambda(n)}~,~~~~\lambda(n)=\frac{4p^2}{N}\frac{1}{1-\frac{4n^2}{N^2}}~.
\ee
In the real SYK case, the coupling $ \cJ_{MS}$ used in \cite{Maldacena:2016hyu} is related to the $\cJ$ used in \cite{Berkooz:2018jqr} by $\cJ_{MS} = \sqrt{\lambda }\cJ$ (as $q \to 1$). Since here $J$ is always accompanied by an $\exp\left[p^2/(2N)\right]$ factor, we define
\be \label{eq:scrJ_def}
\mathscr J = \sqrt{1-q(n)} J e^{p^2/(2N)}~.
\ee
This definition of $\mathscr J$ depends on the charge, and is simply a convenient definition; it is not meant to provide an alternative definition for the coupling in the complex SYK model.

The $p^2/N \to 0$ limit is described conveniently using the variables $y$ related to $\theta $ through $\theta_i=\pi-\lambda(n) y_i$.
Under the parametrization
\begin{equation}
\tilde{q}=q^l, \qquad l=\frac{p_M+\bar p_M}{2p} ~,
\end{equation}
when $\lambda(n)\to 0$ we have the following limits for the $q$-Pochhammer symbols (as in \cite{Berkooz:2018jqr})
\begin{align}
\int_0^\pi \frac{d\theta}{2\pi} (q(n),e^{\pm2i\theta};q(n))_\infty &\longrightarrow \frac{\lambda(n)^3(q(n);q(n))_\infty^3}{2\pi^2}\int_0^\infty dy~ 2y \sinh(2\pi y)~,\\[10pt]
\frac{(\tilde{q}(n)^2;q(n))_\infty}{(\tilde{q}(n)e^{i(\pm\theta_1\pm\theta_2)};q(n))_\infty} &\longrightarrow \frac{\lambda(n)^{2l-3}}{(q(n);q(n))_\infty^3}\frac{\Gamma(l\pm iy_1\pm iy_2)}{\Gamma(2l)}~.
\end{align}
In this limit, the expression for the two-point function becomes
\begin{align}\label{twoPtFunGen1}
\begin{split}
& \langle \tr e^{-\beta H} M(\tau)\bar{M}(0) \rangle_{J,J_M}^{(-n)} =  \frac{2J_M^2 }{\sqrt{2\pi N}} e^{S_M} \(\frac{\lambda(n)^3(q(n);q(n))_\infty^3}{2\pi^2}\)^2 \frac{\lambda(n)^{2l-3}}{(q(n);q(n))_\infty^3}\\
& \int_0^\infty dy_1dy_2~ 4y_1 y_2  \sinh(2\pi y_1)\sinh(2\pi y_2) \frac{\Gamma(l\pm iy_1\pm iy_2)}{\Gamma(2l)}\\
&\cdot \exp \(\frac{  2\scrJ \tau  \cos(\lambda y_1) }{\lambda(n)}  e^{A+B_1}+\frac{  2\scrJ(\beta-\tau)  \cos(\lambda y_2) }{\lambda(n)}  e^{A+B_2}\)~.
\end{split}
\end{align}
We evaluate the integrals following \cite{Lam:2018pvp}; we parametrize $y_1=y_2+\omega$, and we will see that there will be a saddle point at large $y_2$ so that we will also use $\omega  \ll y_2$. Therefore we have
\begin{align}
\<\tr~ &e^{-\b H} M(\tau)\bar M(0)\>_{J,J_M}^{(-n)}=\frac{2J_M^2 }{\sqrt{2\pi N}} e^{S_M}  \frac{\lambda(n)^{2l+3}(q(n);q(n))_\infty^{3}}{4\pi^4} \times\no\\
&\int_0^\infty dy_2 d\omega~ (y_2+\omega) y_2  \exp(2\pi(y_2+\omega))\exp(2\pi y_2) \frac{\Gamma(l\pm i(y_1+y_2))}{\Gamma(2l)}\Gamma(l\pm i \omega)\times\no\\
&\cdot \exp \Bigg\{ \frac{  2\scrJ}{\lambda(n)} \left[  \tau  \cos(\lambda y_2+\lambda \omega)   e^{A+B_1}+ (\beta-\tau)  \cos(\lambda y_2) e^{A+B_2}\right] \Bigg\}~.
\end{align}
If we assume that $l$ is an integer, then we get the following simplification
\be
\Gamma(l\pm i(y_1+y_2))\approx \frac{(2y_2)^{2l-1} \pi}{\sinh\pi(y_1+y_2)}\approx 2^{2l}y_2^{2l-1} \pi\exp(-\pi(2y_2+\omega))~.
\ee
With this simplification we have
\begin{align}
&\<\tr~ e^{-\b H} M(\tau)\bar M(0)\>_{J,J_M}^{(-n)}=\frac{2J_M^2 }{\sqrt{2\pi N}} e^{S_M}  \frac{\lambda(n)^{2l+3}(q(n);q(n))_\infty^{3}}{\pi^3} \times\no\\
&\int dy_2 d\omega~ 2^{2l-2}y_2^{2l+1}   \frac{\Gamma(l\pm i \omega)}{\Gamma(2l)}\times\no\\
&\exp\Bigg\{ 2\pi y_2+\pi\omega+\frac{  2\scrJ e^A}{\lambda(n)} \left[   \tau  \cos(\lambda y_2)   e^{B_1}+ (\beta-\tau)  \cos(\lambda y_2) e^{B_2}- \omega \lambda \tau  \sin(\lambda y_2)   e^{B_1}+\cdots \right] \Bigg\} .
\end{align}
If $1\ll y_2\ll1/\lambda(n)$, then we can expand the cosine functions. There is a saddle point for the $y_2$ integral in this regime. The $y_2$ dependent piece of the exponent is
\be
2\pi y_2 - \frac{  2\scrJ e^A}{\lambda(n)}  \( \tau e^{B_1}+(\beta-\tau) e^{B_2}\) \lambda^2y_2^2/2+\cdots ~.
\ee
The saddle point is at
\begin{equation}
y_2^* = \frac{\pi }{\scrJ e^A \lambda \left[\tau e^{B_1} +(\beta -\tau )e^{B_2} \right]}~ .
\end{equation}
We look at the range
\begin{equation}
1 \ll \beta \scrJ \ll 1/\lambda, \qquad \tau\scrJ \ll 1/\lambda
\end{equation}
(corresponding to low temperatures but not comparable to $N$) so that the conditions on $y_2$ above are indeed satisfied. Doing the saddle point estimation for the $y_2$ integral together with going to real time $\tau =it$ and changing the integration variable $\omega  = \omega' \frac{\beta }{2\pi } $, we get\footnote{\label{fn:beta_stricter_range}Strictly speaking, requiring in addition that $ \beta \scrJ \gg 1/\lambda ^{1/3}$ would guarantee that we can drop the quartic and higher order terms in the expansion of $ \cos(\lambda y_2)$.}
\begin{align}
\<\tr~ &e^{-\b H} M(t)\bar M(0)\>_{J,J_M}^{(-n)}=\frac{2J_M^2 }{\sqrt{2\pi N}} e^{S_M}  \frac{\lambda(n)^{2l+3}(q(n);q(n))_\infty^{3}}{\pi^3}  \(\frac{\pi }{\scrJ e^{A}\lambda \left[it  e^{B_1} +(\beta -it )e^{B_2} \right]}\)^{2l+3/2}\no\\
&2^{2l-2} \exp\[\frac{  2\scrJ e^A}{\lambda(n)} (it ~e^{B_1} + (\beta-it) e^{B_2})+\frac{\pi^2}{\scrJ e^A\lambda \left[it ~e^{B_1} +(\beta -it )e^{B_2} \right]}\] \no\\
&\frac{\beta}{2\pi}\int_{-\infty}^\infty d\omega~    \frac{\Gamma\left( l\pm i \frac{\beta \omega }{2\pi } \right) }{\Gamma(2l)} \exp\Bigg\{ \frac{\omega \beta }{2} -it \frac{e^{B_1-B_2} \omega }{1-\frac{it}{\beta } +\frac{it}{\beta } e^{B_1-B_2} } \Bigg\}~.
\end{align}
The last integral is the one obtained in \cite{Lam:2018pvp} where the time is renormalized. In the limit of low temperature we have $t \ll \beta $, so the last line becomes
\begin{equation}
(-1)^{l} \beta \left(\frac{\beta }{2\pi } \right)^{2l-1} \frac{1}{\left(t e^{B_1-B_2} \right)^{2l} } ~,
\end{equation}
and in total we find
\begin{equation} \label{eq:2pf_fixed_charge_conformal_regime}
\begin{split}
& \langle \tr~ e^{-\beta H} M(t )\bar M(0)\rangle_{J,J_M}^{(-n)}  = \frac{2J_M^2 }{\sqrt{2\pi N}} e^{S_M}
 \frac{\lambda(n) ^{2l+3}(q(n);q(n))_{\infty } ^3 }{\pi ^3}
\left[\frac{1 }{\scrJ\lambda \beta e^{A+B_2} } \right]^{2l+3/2}
\frac{\pi ^{5/2} }{2}  \cdot \\
&\qquad \cdot \exp \Bigg\{ \frac{2\scrJ e^A}{\lambda (n)} \left[it e^{B_1} +(\beta -it)e^{B_2} \right]+ \frac{\pi ^2}{\scrJ\lambda e^A \left[it e^{B_1} +(\beta -it)e^{B_2} \right]} \Bigg\} \cdot \\
& \qquad \qquad \cdot  (-1)^{l} \beta ^{2l} \frac{1}{\left(te^{B_1-B_2} \right)^{2l} } ~.
\end{split}
\end{equation}
This result exhibits a conformal behavior, plus small corrections to the scaling form, coming from the second line in \eqref{eq:2pf_fixed_charge_conformal_regime}.

As explained in \cite{Davison:2016ngz} and reviewed in \eqref{eq:ration_of_2pf}, the spectral asymmetry factor can also be obtained through the Green's function of a single fundamental fermion in a conformal regime
\begin{equation} \label{eq:spectral_asymmetry_ratio}
e^{2\pi \cE} = \frac{\langle \tr~ e^{-\beta H} M(t )\bar M(0)\rangle_{J,J_M}^{(-n)} }{\langle \tr~ e^{-\beta H} \bar M(-t ) M(0)\rangle_{J,J_M}^{(-n)} }  \qquad \text{where $t>0$, for $M$ a single fermion}.
\end{equation}

As mentioned above, $G(t<0)$ is obtained by exchanging $p_M \leftrightarrow \bar p_M$ and $t \to -t$.
Let us allow for the moment $|n|$ to scale up to the maximal $N/2$, and remember that $p_M \sim O(N^{1/2} )$.

Before concentrating on the conformal regime, let us consider a slightly different measure of asymmetry obtained from correlation functions. Specifically we take the same ratio of correlation functions as in \eqref{eq:spectral_asymmetry_ratio}, with generic operators, but consider the $t \to 0$ limit using the IR correlation function \eqref{eq:2pf_fixed_charge_conformal_regime}.
This gives (note that $B_1|_{p_M \leftrightarrow \bar p_M}=B_2$)
\begin{equation}
\begin{split}
& \frac{\langle \tr~ e^{-\beta H} M(t )\bar M(0)\rangle_{J,J_M}^{(-n)} }{\langle \tr~ e^{-\beta H} \bar M(-t ) M(0)\rangle_{J,J_M}^{(-n)} } \to
\exp \Bigg[ S_M - S_M |_{p_M \leftrightarrow \bar p_M} + 4l(B_2-B_1)+ \\
& + \left( 2l+\frac{3}{2} \right) \left( (A+B_2)|_{p_M \leftrightarrow \bar p_M} - A- B_2\right)   +\frac{2\scrJ \beta }{\lambda (n)} \left( e^{A+B_2} -e^{(A+B_2)|_{p_M \leftrightarrow \bar p_M} } \right) + \\
& +  \frac{\pi ^2}{\scrJ \lambda (n)\beta } \left( e^{-(A+B_2)} -e^{-(A+B_2)|_{p_M \leftrightarrow \bar p_M} } \right)  \Bigg] ~.
\end{split}
\end{equation}
To evaluate this carefully, we note that
\begin{equation}
A+B_2 = \frac{p}{2} \log\left( 1-\frac{4n^2}{N^2} \right) +\frac{p}{2} \log\left( 1+\frac{4n(p_M-\bar p_M)-(p_M-\bar p_M)^2}{N^2-4n^2} \right)  - \frac{2np(p+p_M-\bar p_M)}{N^2-4n^2} ~.
\end{equation}
There are several regimes, depending on the charge $n$:

\noindent 1. For large charges, by which we mean $n \sim O(N^{\alpha } )$ with $\alpha > 3/4$, we have that $e^{A+B_2} \to 0$ strongly,; however, the $e^{-(A+B_2)} $ terms diverge strongly with $N$.

\noindent 2. Otherwise, for moderate charges, by which we mean $n \sim O(N^{\alpha } )$ with $\alpha \le 3/4$, the term $e^{A+B_2} $ is finite, but then $e^{A+B_2} -e^{(A+B_2)|_{p_M \leftrightarrow \bar p_M} } \sim o(1/N)$, and the same holds for the $e^{-(A+B_2)}$ terms.

\noindent Therefore, we are left with (up to $O\left( N^{-1/2} \right) $ corrections)
\begin{equation}
\begin{split}
& \frac{\langle \tr~ e^{-\beta H} M(t )\bar M(0)\rangle_{J,J_M}^{(-n)} }{\langle \tr~ e^{-\beta H} \bar M(-t ) M(0)\rangle_{J,J_M}^{(-n)} } \to \exp \Bigg[ (p_M-\bar p_M) \log \left( \frac{N+2n}{N-2n} \right) - \frac{4n}{N^2-4n^2} (p_M^2-\bar p_M^2)+\\
& +  \frac{\pi ^2}{\scrJ \lambda (n)\beta } \left( e^{-(A+B_2)} -e^{-(A+B_2)|_{p_M \leftrightarrow \bar p_M} } \right)  \Bigg]~.
\end{split}
\end{equation}

In order to understand this behavior we note the following.
If we concentrate on the scaling $n \sim O(N^{3/4}) $, the successive $k$'th moments are reliable, since there are no $k$ dependent terms that grow or decay with $N$. Moreover, there are pieces that go as $N^{-1/4} $ which can be trusted since we only dropped terms that scale as $N^{-1/2} $ (as $p$, $p_M$, and $\bar p_M$ are of this size) in our approximations in analyzing the large $N$ double-scaled limit.
Therefore, we now concentrate on moderate charges.

This leads us to the following asymmetry measure
\begin{equation}
\begin{split}
& \frac{\langle \tr~ e^{-\beta H} M(t )\bar M(0)\rangle_{J,J_M}^{(-n)} }{\langle \tr~ e^{-\beta H} \bar M(-t ) M(0)\rangle_{J,J_M}^{(-n)} } \to \exp \Bigg\{ (p_M-\bar p_M) \Bigg[ \log \left( \frac{N+2n}{N-2n} \right) - \frac{4n}{N^2-4n^2} (p_M+\bar p_M) \Bigg] \Bigg\}~.
\end{split}
\end{equation}
The argument of the exponent is proportional to the charge of the operator ($\bar p_M-p_M$). While in a conformal regime this is the only dependence on the operator \cite{Sachdev:2015efa,Davison:2016ngz}, here we also get a dependence on the dimension (or size) of the operator ($p_M+\bar p_M$).

Finally, we go back to the definition in \eqref{eq:spectral_asymmetry_ratio}. As mentioned, the second line in \eqref{eq:2pf_fixed_charge_conformal_regime} has a time dependence that gives a correction to the scaling behavior. While we will see below that these time dependent terms should be present in the correlation function physically, for $N \to \infty $ they go to zero (for any charge $n$) and we obtain a conformal behavior. Extrapolating to the case of a 2-point function of a single fermion $p_M=1$, $\bar p_M=0$, we have (at the order that we can trust for moderate charges)\footnote{As before, the time independent part in the second line of \eqref{eq:2pf_fixed_charge_conformal_regime} does not contribute to the ratio as $N \to \infty $.}
\begin{equation}
\cE\approx \frac{1}{2\pi } \log \frac{1+2n/N}{1-2n/N} = \frac{1}{2\pi } \log \frac{1-2\cQ}{1+2\cQ} .
\end{equation}
This is the leading contribution for large $p$ of the result \eqref{specAsym} that was found in \cite{Davison:2016ngz}.

\subsubsection{Verifying the time dependent terms}

One may be puzzled about the result \eqref{eq:2pf_fixed_charge_conformal_regime} not taking a simple conformal form, because of the appearance of the second line in this formula. (Recall that this result includes contributions subleading in $N$.) In this short subsection we perform a simple check showing that this time dependence must be there.

If we have external states of charge $(-n)$, then $ \langle \tr~ e^{-\beta H} M(t )\bar M(0)\rangle_{J,J_M}^{(-n)}$ will contain contributions of the form $\exp \left[it (E_n-E_m)\right]$ from intermediate states, where the charge corresponding to the intermediate states with energy $E_m$ is $p_M-\bar p_M-n$ (while that of $E_n$ is $(-n)$). We found that the range of energies at a given charge is given by \eqref{eq:energy_range} (being most reliable for charges that scale as $N^{3/4} $; however we even allow here the general case $n \sim O(N)$ so that it will imply the other scalings).
Since we are at low temperature, the dominant contributions from states of energies $E_n$ and $E_m$ will come from the lowest energies at the given charges. Therefore the difference between the lowest energies from the range \eqref{eq:energy_range} at these charges is given by
\begin{equation}
\begin{split}
& \frac{\lambda(n) }{2\scrJ} (E_{n,min} -E_{m,min} )=\\
&\qquad \qquad= \exp\left[\frac{p}{2} \log\left(1-\frac{4n^2}{N^2}\right) -\frac{2np^2}{N^2-4n^2} +\frac{4np(p_M-\bar{p}_M)}{N^2-4n^2} +O(N^{-1/2}) \right] \\
&\qquad \qquad  -  \exp \left[ \frac{p }{2} \log\left(1-\frac{4n^2}{N^2} \right)-\frac{2np^2}{N^2-4n^2} \right] .
\end{split}
\end{equation}

We should compare this energy difference to the coefficient of $(it)$ in the exponent in the second line of \eqref{eq:2pf_fixed_charge_conformal_regime} (being the dominant time behavior at low temperature) which is
\begin{equation}
\begin{split}
& \frac{2\scrJ}{\lambda (n)} \left(  e^{A+B_1} -e^{A+B_2} \right) =\\
& \qquad = \frac{2\scrJ}{\lambda (n)}  \exp\left[\frac{p}{2} \log \left(1-\frac{4n^2}{N^2} \right)-\frac{2np^2}{N^2-4n^2} +\frac{4np(p_M-\bar{p}_M)}{N^2-4n^2} +O(N^{-1/2} )\right] - \\
& \qquad   -\frac{2\scrJ}{\lambda (n)}  \exp\left[\frac{p}{2} \log\left(1-\frac{4n^2}{N^2} \right)-\frac{2np^2}{N^2-4n^2} +O(N^{-1/2} ) \right] =\\
& \qquad =E_{n,min} -E_{m,min} .
\end{split}
\end{equation}
We see that it matches exactly with the energy differences. This verifies that this exponential dependence on $t$ in the correlator (which goes to zero as $N^{-1/4} $ for $n \sim O(N^{3/4} )$) must indeed be there.


\section{Four-point function} \label{sec:chaos}

\subsection{Four-point function of neutral operators} \label{sec:4pf_chemical_potential}

In section \ref{sec:2ptFunctionSec} we considered the random operator $M$, and we can just as easily allow several such operators (as in \cite{Berkooz:2018jqr}). In doing this, we introduce a flavor index $A$ and consider the operators
\begin{align}
\begin{split}
M_A&= \sum_{I, I'}\(J_A^{(M)}\)_{I}^{I'} \j^{I} \jb_{I'}
\end{split}
\end{align}
having independent Gaussian random couplings
\begin{equation}
\langle (J_A^{(M)})_{I_1}^{I'_1} \left( (J_B^{(M)})_{I_2}^{I'_2}\right)^{*}\rangle_{J_M}=\delta_{AB} J_{M_A}^{2}{N \choose p_{M_A}}^{-1} {N \choose \bar p_{ M_A}}^{-1} \delta_{I_1 I_2}\delta _{I'_1 I'_2}~.
\end{equation}

In this section we calculate the 4-point function of neutral operators in the presence of a chemical potential (and then mention fixed charge sectors). Taking two flavors of operators $M_1,M_2$, there are two channels that we can consider, namely $\langle M_1 M_1 M_2 M_2\rangle $ and $\langle  M_1 M_2 M_1 M_2\rangle $ (in this schematic notation we omit the time dependence, and by expectation value mean disorder averaged and thermal expectation value). The 4-point function of an operator from one flavor will decompose into these channels. The latter channel, referred to as the crossed one, is the one encoding the quantum Lyapunov exponent and we concentrate on it.

\begin{figure}[h]
\centering
\includegraphics[width=0.9\textwidth]{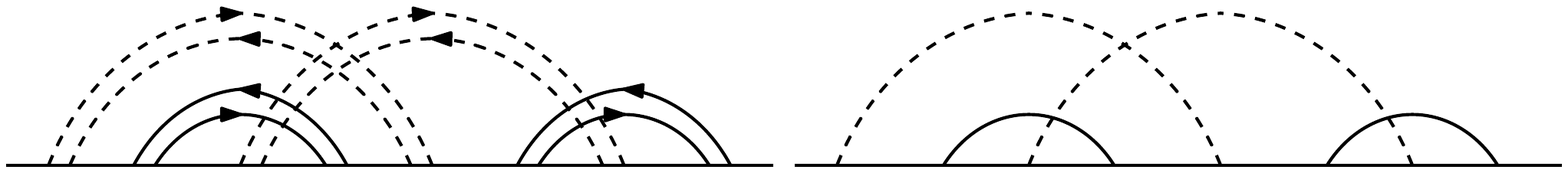}
\caption{An example of a chord diagram contributing to the crossed 4-point function. On the left is the oriented chord diagram, and on the right is the corresponding unoriented chord diagram.}
\label{fig:4pf_example}
\end{figure}

As we are restricting to neutral operators we have $p_{M1}=\bar p_{M1}$ and $p_{M2}=\bar p_{M2}$. As before, the crossed 4-point function is determined by the moments, which are given by
\begin{equation}
\begin{split}
& m_{k_1,k_2,k_3,k_4} = \langle \tr \left[M_1 H^{k_1} M_2 H^{k_2} M_1 H^{k_3} M_2 H^{k_4} e^{-2\mu Q} \right] \rangle_{J,J_M} = \\
&=J^k J_{M1} ^2 J_{M2} ^2 \sum _{CD}  \binom{N}{p} ^{-k} \binom{N}{p_{M1} } ^{-2} \binom{N}{p_{M2} } ^{-2}  \cdot \\
& \qquad \cdot \sum_{I_1,\cdots ,I_k} \sum _{I,J,I',J'} \tr\left[ (\psi^I \bar \psi_J) \bar \psi_{I_1} \psi^{I_2} \cdots (\psi^{I'} \bar \psi_{J'} ) \cdots (\psi^J \bar \psi_I) \cdots (\psi^{J'} \bar \psi_{I'} )  \cdots e^{-2\mu Q}  \right] .
\end{split}
\end{equation}
The moments reduce to a sum over chord diagrams (abbreviated CD in the equation). We mark the pair of $M_1$ operators and connect them by a dashed chord (as before, it is a single chord in the unoriented case), and similarly for the pair $M_2$. These two dashed chords cross in this channel. In the sum over chord diagrams we are instructed to sum over all possible configurations of solid chords.
See fig.\ \ref{fig:4pf_example} for an example of a chord diagram.

The evaluation of these moments again follows from the rules of fig.\ \ref{fig:allowed_overlaps}. In appendix \ref{app:details_2pf} we count the appearances of every rule of fig.\ \ref{fig:allowed_overlaps} according to the unoriented chord diagram, where similarly to before we denote by $\kappa _H$ the number of intersections of $H$-chords, $\kappa _{HM_1} $ the number of intersections between $H$ and $M_1$ chords, and similarly for $M_2$.

Combining the ingredients in appendix \ref{app:details_2pf}, we get
\begin{equation} \label{eq:4pf_moments}
\begin{split}
& m_{k_1,k_2,k_3,k_4} = J^k J_{M1} ^2 J_{M2} ^2 (\cosh \mu )^{N-kp-2p_{M1} -2p_{M2} } \cdot \\
& \qquad \cdot \exp \Bigg\{ \frac{p^2}{N} \left[ \frac{k^2}{2} \sinh(\mu )^2- \frac{k}{2} \left( \sinh(2\mu )-1\right) \right]  + \\
& \qquad + \frac{pp_{M1} }{N} 2k \sinh(\mu )^2 + \frac{pp_{M2} }{N} 2k \sinh(\mu )^2 + \frac{p_{M1} ^2}{N} e^{2\mu } +\frac{p_{M2} ^2}{N} e^{2\mu } -\frac{4p_{M1} p_{M2} }{N}  \Bigg\} \cdot \\
& \qquad \cdot \sum _{CD} \exp \Bigg\{ -4\cosh(\mu )^2 \frac{p^2}{N} \kappa _H - 4\cosh(\mu )^2 \frac{pp_{M1} }{N} \kappa _{HM1} -4 \cosh(\mu )^2 \frac{pp_{M2} }{N} \kappa _{HM2} \Bigg\} .
\end{split}
\end{equation}

In order to make sense of this (so that there is no problem from the $k^2$ term as discussed before), we can take for example $\mu \sim O(N^{-1/4}) $ and drop $O(N^{-1/2} )$ corrections, leaving us with
\begin{equation}
\begin{split}
& m_{k_1,k_2,k_3,k_4} = J^k J_{M1} ^2 J_{M2} ^2 (\cosh \mu )^{N-kp-2p_{M1} -2p_{M2} } \cdot \\
& \qquad \cdot \exp \Bigg\{ \frac{p^2}{N}  \frac{k}{2} \left[ 1- \sinh(2\mu )\right]  +
 \frac{p_{M1} ^2}{N} e^{2\mu } +\frac{p_{M2} ^2}{N} e^{2\mu } -\frac{4p_{M1} p_{M2} }{N}  \Bigg\} \cdot \\
& \qquad \cdot \sum _{CD} \exp \Bigg\{ -4\cosh(\mu )^2 \frac{p^2}{N} \kappa _H - 4\cosh(\mu )^2 \frac{pp_{M1} }{N} \kappa _{HM1} -4 \cosh(\mu )^2 \frac{pp_{M2} }{N} \kappa _{HM2} \Bigg\} .
\end{split}
\end{equation}
For many purposes we can ignore constants common to all the moments, as for example they do not affect the Lyapunov exponent.
This expression for the 4-point function is then the same as in real SYK \cite{Berkooz:2018jqr} (see Eq.\ (2.11) there) with the replacements
\begin{equation} \label{eq:relation_to_real_SYK}
\begin{split}
& J_{\text{real SYK}} \leftrightarrow J \cosh(\mu )^{-p} \exp \left[  \frac{p^2}{N} \frac{1}{2} \left(1- \sinh(2\mu )\right) \right] \\
&q_{\text{real SYK}} \leftrightarrow  \exp \left[ -4 \frac{p^2}{N} \cosh(\mu )^2 \right]  \\
& \tilde q^{(1)} _{\text{real SYK}} \leftrightarrow \exp \left[  - 4 \frac{pp_{M1} }{N} \cosh(\mu )^2 \right]  \\
& \tilde q^{(2)} _{\text{real SYK}} \leftrightarrow \exp \left[  - 4 \frac{pp_{M2} }{N} \cosh(\mu )^2 \right] .
\end{split}
\end{equation}

Similarly to subsection \ref{sec:2pf_fixed_charge}, starting from \eqref{eq:4pf_moments} we can go to a fixed charge sector. With the appropriate prefactor common to all the moments (that we do not quote), the 4-point function is the same as in the real double-scaled SYK (for the final expression for the 4-point function see \cite{Berkooz:2018jqr}), with the replacements (recall $\cQ=-n/N$)
\begin{equation} \label{eq:relation_to_real_SYK_fixed_charge}
\begin{split}
& J_{\text{real SYK}} \leftrightarrow J \exp\left[ \frac{p^2}{2N} +\frac{2p^2}{N} \frac{\cQ}{1-4\cQ^2} +\frac{p}{2} \log \left( 1-4\cQ^2\right) \right] \\
&q_{\text{real SYK}} \leftrightarrow  \exp \left[ -4 \frac{p^2}{N}\frac{1}{1-4\cQ^2}\right]  \\
& \tilde q^{(1)} _{\text{real SYK}} \leftrightarrow \exp \left[  - 4 \frac{pp_{M1} }{N} \frac{1}{1-4\cQ^2} \right]  \\
& \tilde q^{(2)} _{\text{real SYK}} \leftrightarrow \exp \left[  - 4 \frac{pp_{M2} }{N}  \frac{1}{1-4\cQ^2} \right] .
\end{split}
\end{equation}

\subsection{Lyapunov exponent}

As before, we can make contact with the usual large $N$ SYK (in which $p$ is independent of $N$) by taking the $p^2/N \to 0$ limit.
In the double scaled SYK model with real fermions, it was found that the chaos exponent has the following dependence on $q$ \cite{Berkooz:2018qkz}
\be\label{LypRealSYK}
\lambda_L=2\pi T-\frac{4\pi T^2}{J_{\text{real SYK}}\sqrt{-\log q_{\text{real SYK}}}}+\cdots~.
\ee

Plugging the relations \eqref{eq:relation_to_real_SYK} (in the regime of small $p^2/N$ the exponent in the expression for $J_{\text{real SYK} }$ can be set to 1), we find the following Lyapunov exponent
\be
\lambda_L=2\pi T-\frac{2\pi T^2}{\cosh\mu}\frac{(\cosh\mu)^p}{J}\sqrt{\frac{N}{p^2}}+\cdots .
\ee
This agrees with the result in \cite{Bhattacharya:2017vaz} (after translating eq.\ (4.9) there to the conventions here).

Alternatively, from \eqref{eq:relation_to_real_SYK_fixed_charge} and \eqref{LypRealSYK}, one can read off the Lyapunov exponent in a fixed charge $\cQ$
\be
\lambda_L=2\pi T-\frac{2\pi T^2}{J} \sqrt{\frac{N}{p^2}}\(1-4\cQ^2\)^{(1-p)/2}+\cdots ~.
\ee


\section{Heavy Operators} \label{sec:massive}

In this section we discuss how to use heavy operators to disconnect spacetimes, and how light enough particles can still go between them. More generally, given that we have good control over double-scaled SYK models at any energy scale, we can provide precise answers to questions that mix the UV and IR degrees of freedom in the theory. In such a setting one excites the theory using a high energy operator, and then examines the response of the low energy gravitational background. Generally, such operators might be singular objects in the language of the IR degrees of freedom. These include, for example, singularities, the `end of the world brane' (as in \cite{Kourkoulou:2017zaj}) and processes that glue universes such as in \cite{Goel:2018ubv}. But with control over the full theory we can carry out ``precision measurements'' of such objects.

In this section we will discuss some exact results for the simplest of such objects and study the behavior of massive operators, whose dimension (to the extent that it can even be defined in the language of the low energy theory) is parametrically large, or even infinite.  Since for neutral operators, the canonical correlation functions in complex SYK take the form of those in Majorana SYK, we will discuss this in the ordinary Majorana SYK model \cite{Berkooz:2018jqr}, and relate it to \cite{Goel:2018ubv}.

The situation is therefore that we are interested in the Hamiltonian \eqref{realSYKHam} consisting of $p$ fermions, and an additional operator
\begin{equation} \label{eq:RealSYKOp}
M=i^{p_M/2}\sum_{\substack{1\le i_{1}<\cdots<i_{p_M}\le N}
}\(J^{(M)}\)_{i_{1}\cdots i_{p_M}}~\chi_{i_{1}}\cdots\chi_{i_{p_M}}
\end{equation}
with $p_M$ fermions, which satisfy
 \begin{equation}
p=\sqrt{\lambda N/2}, \ \ \ p_M=\sqrt{\lambda_M N/2}, \ \ \ \lambda,\lambda_M\ \text{fixed},\ N\rightarrow\infty
\end{equation}
and we use the definitions
\begin{equation}
q=e^{-\lambda}, \qquad q_{M}=e^{-\sqrt{\lambda \lambda_M}}~.
\end{equation}
Note that the notations in this section, such as $p$ and $\lambda $, stand for quantities in the Majorana SYK model and should be distinguished from those in previous sections.  The similarity in the analysis shows how different physical systems are described by a similar chord diagram description.

For a simple gravitational interpretation we then take $\lambda,E\rightarrow 0$,\footnote{We can also take $\lambda$ fixed and $E\rightarrow 0$.} but we still have the freedom of what to do with $\lambda_M$ (or $q_M$).
Generally, since the scaling dimension of the Hamiltonian is 1 in the IR, then the scaling dimension of an operator of length $p_M$ will be $\sim p_M/p=\sqrt{\lambda_M/\lambda}$. In the limit of $\lambda\rightarrow 0$ and $p_M/p$ fixed, the operator becomes a conformal operator of dimension $p_M/p$. Such an operator does not have a radical effect on spacetime --- it bends the trajectory of the ``boundary particle''  but in a controlled and computable fashion. A more interesting set of questions happens when $\lambda_M$ is not taken to zero at the same time as $\lambda$, so $q_M$ does not go to one, but rather $q_M$ is taken to be fixed. In this case one ends up with an operator whose conformal dimension formally tends to infinity, or more precisely, one cannot assign to it a conformal dimension.

Nevertheless, exact computations in the presence of such objects are just as straightforward. We will see that their effect, as seen within the low energy limit, is to split the $AdS_2$ space into two spaces, touching at an interface, as argued in \cite{Goel:2018ubv}. If $\lambda _M$ is taken to be even larger, this interface can shrink to a point. We will also see that, even though that point is singular (from the point of view of the low energy theory), one can compute how correlation functions ``go through it''. Hence, we can have a large amount of control over the transmission through this space-time singular locus. In this limit, $q_M$ simply controls how many quanta of correlations go through the singularity.

\begin{figure}[h]
\centering
\includegraphics[width=0.5\textwidth]{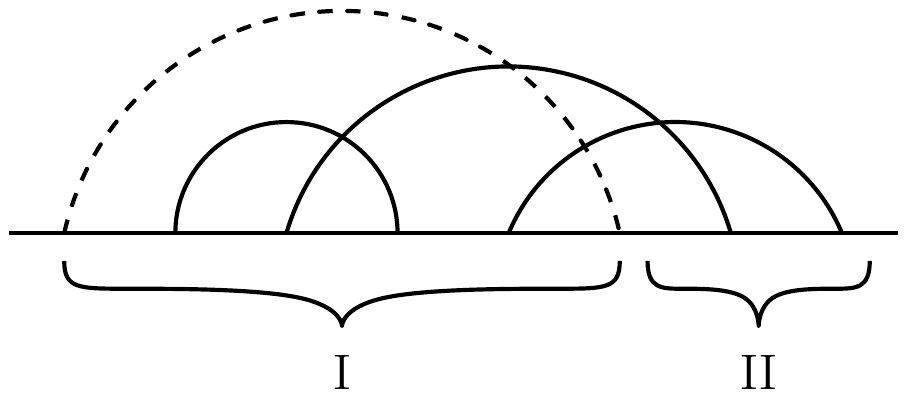}
\caption{An example demonstrating the chord diagram structure of the 2-point function in Majorana SYK.}
\label{fig:two_point_func}
\end{figure}

The limit that we will be interested in here is thus
\begin{equation}
q\rightarrow 1,\ q_M\ \text{fixed}~.
\end{equation}
It is rather intuitive to understand what happens in this limit. Recall the basic structure of chord diagrams for the 2-point function, fig.\ \ref{fig:two_point_func}. The intersections of chords of the Hamiltonian with the additional chord associated with the operator are assigned a weight $q_M^n$ where $n$ is the number of Hamiltonian chords crossing the operator chord. In the limit of $q_M\rightarrow 0$ the 2-point function splits into two independent regions consisting of only Hamiltonian chords, i.e., spacetime splits into two. Finite $q_M$ therefore controls, in a very simple way, how the two spaces are partially connected --- in the limit $q_M\rightarrow 0$ they disconnect and in the limit $q_M\rightarrow 1$ they connect (with $M$ becoming a weakly coupled particle on spacetime). In subsection \ref{subsec:Disc} we make more precise the notion of these partially connected spaces, and in subsection \ref{sec:conn_universes} we turn on probe operators in those spaces by studying the 4-point function, which describes a particle going from one space to the other.

\subsection{Disconnecting universes with massive operators}\label{subsec:Disc}

Consider first the basic formula of a 2-point function for Majorana SYK with coupling $\cJ$ defined by $\langle J_{i_1 \cdots i_p} ^2\rangle _J= \binom{N}{p} ^{-1} \cJ^2$ (with $J_{i_1 \cdots i_p} $ the coefficients in \eqref{realSYKHam}) as in \cite{Berkooz:2018qkz,Berkooz:2018jqr}\footnote{This is what one obtains for neutral operators in complex SYK from \eqref{eq:2pf_moments_chemical_potential} (up to an overall constant) with $q=\exp(-4p^2/N)$, $\mu =0$, $ q_M=\exp(-4pp_M/N)$, and with $\cJ \to J \exp(p^2/(2N))$.}
\begin{align}
\begin{split}
G(\beta, \tau) =&\frac{1}{Z(\b)} \int_0^\pi \frac{d\theta_1}{2\pi} \frac{d\theta_2}{2\pi} w(\theta_1) w(\theta_2)   \exp\[-\frac{2\cJ}{\sqrt{1-q}}\((\b-\tau)\cos\theta_1+\tau\cos\theta_2\)\]\\[5pt]
&\times \frac{({q_{M}^2};q)_\infty}{({q_{M}} e^{i(\pm\theta_1\pm\theta_2)};q)_\infty }\ ,
\end{split}
\end{align}
where the measure factor $w(\theta)$ is defined as usual as
\begin{equation}
w(\theta)=(q,e^{\pm2i\theta};q)_\infty~.
\end{equation}

We would like to evaluate this expression in the limit where $\lambda,\,E\to 0$ and $q_M\ \text{fixed}$. We first go to convenient low energy parameters $y_i$ defined by
\begin{equation}
\theta_i=\pi-\lambda y_i~,
\end{equation}
to obtain (see Eq.\ (5.12) in \cite{Berkooz:2018jqr} for the limit of the measure)
\begin{align}\label{cr1}
G(\beta, \tau)=\frac{1}{Z(\beta)} \(\frac{\lambda^3(q;q)_\infty^3}{2\pi^2}\)^2(q_M^2;q)_\infty\int_0^\infty \int_0^\infty dy_1 dy_2~ y_1 y_2 \exp[W(y_1,y_2)]
\end{align}
where $W(y_1,y_2)$ is a function of $y_1,y_2$ given by
\begin{equation}
\begin{split}
&W= \log(2\sinh(2\pi y_1))+\log(2\sinh(2\pi y_2))+\frac{2\cJ}{\sqrt{\lambda}}(\beta-\tau) \cos(\lambda y_1)+\frac{2\cJ}{\sqrt{\lambda}}\tau \cos(\lambda y_2)\\
&-\sum_{k=0}^\infty \log\[1+q_M^2q^{2k}-2q_M q^k\cos(\lambda(y_1+y_2))\]- \sum_{k=0}^\infty \log\[1+q_M^2q^{2k}-2q_M q^k\cos(\lambda(y_1-y_2))\] .
\end{split}
\end{equation}

In the limit $\lambda\to 0$, we can further approximate the $y_1$ and $y_2$ integrals via a saddle point approximation. The saddle point equations are
\begin{align}
0=&\partial_{y_1} W = 2 \pi  \coth (2 \pi  y_1) - 2\cJ \sqrt{\lambda} (\beta-\tau) \sin (\lambda  y_1)\no\\
&-\sum_{k=0}^\infty \frac{2 \lambda  q_M q^k \sin \left(\lambda  \left(y_1+y_2\right)\right)}{1+q_M^2 q^{2 k}-2 q_M q^k \cos \left(\lambda  \left(y_1+y_2\right)\right)}-\sum_{k=0}^\infty \frac{2 \lambda  q_M q^k \sin \left(\lambda  \left(y_1-y_2\right)\right)}{1+q_M^2 q^{2 k}-2 q_M q^k \cos \left(\lambda  \left(y_1-y_2\right)\right)},
\end{align}
and a similar equation with $(y_1\leftrightarrow y_2,\ \beta-\tau\rightarrow \tau)$.
We approximate each of the sums using an integral, which is possible when $\lambda  \to 0$,
giving
\begin{align}
\sum_{k=0}^\infty \frac{2 \lambda  q_M q^k \sin \left(\lambda  \left(y_1\pm y_2\right)\right)}{1+q_M^2 q^{2 k}-2 q_M q^k \cos \left(\lambda  \left(y_1\pm y_2\right)\right)} \approx 2\tan^{-1} \(\frac{q_M\sin \left(\lambda  \left(y_1\pm y_2\right)\right)}{1-q_M\cos \left(\lambda  \left(y_1\pm y_2\right)\right)}\)~.
\end{align}
In a similar manner we can approximate
\be
(q_M^2;q)_\infty \approx \exp\(-\frac{1}{\lambda}\text{Li}_2(q_M^2)\)~.
\ee
The saddle point equations become
\begin{equation}
\begin{split}
&2 \pi  \coth (2 \pi  y_1) - 2\cJ \sqrt{\lambda} (\beta-\tau) \sin (\lambda  y_1)-\\
& \qquad \qquad- 2\tan^{-1}\(\frac{2 q_M \sin \left(\lambda  y_1\right) \left(\cos \left(\lambda  y_2\right)-q_M \cos \left(\lambda  y_1\right)\right)}{q_M^2 \cos \left(2 \lambda  y_1\right)-2 q_M \cos \left(\lambda  y_1\right) \cos \left(\lambda  y_2\right)+1}\)=0~,\\[5pt]
&2 \pi  \coth (2 \pi  y_2) - 2\cJ \sqrt{\lambda}  \tau \sin (\lambda  y_2)-\\
& \qquad \qquad - 2\tan^{-1}\(\frac{2 q_M \sin \left(\lambda  y_2\right) \left(\cos \left(\lambda  y_1\right)-q_M \cos \left(\lambda  y_2\right)\right)}{q_M^2 \cos \left(2 \lambda  y_2\right)-2 q_M \cos \left(\lambda  y_1\right) \cos \left(\lambda  y_2\right)+1}\)=0~.
\end{split}
\end{equation}

Restricting further to low energies $1\ll  y_1,\,y_2 \ll 1/\lambda$
and denoting ${\cJ_{MS}}=\sqrt{\lambda}\cJ$,
the solution to the saddle point equations goes to
\be \label{eq:massive_disconn_saddle_sol}
\begin{split}
y_1^* &= \frac{\pi}{\lambda}\frac{1}{\cJ_{MS}(\beta-\tau)+\frac{2q_M}{1-q_M}}\\
y_2^* &= \frac{\pi}{\lambda}\frac{1}{\cJ_{MS}\tau+\frac{2q_M}{1-q_M}}
\end{split}
\ee
(and we require that
$1 \ll \cJ_{MS} \beta,\, \cJ_{MS}\tau \ll 1/\lambda$ for $y_{1,2}$ to be in the right range).

We should plug the above saddle point expressions into the $y_1$ and $y_2$ integrals.
To do this we need to first expand $\cos(\lambda y_{i})$ inside the $\log$ in the expression for $W(y_1,y_2)$:\footnote{See the comment in footnote \ref{fn:beta_stricter_range}, with $\scrJ$ there replaced here by $\cJ_{MS}$.}
\begin{align}
W(y_1^*,y_2^*)=&~2\pi y_1^* + 2\pi y_2^* +\frac{2\cJ_{MS}}{\lambda}\beta - \frac{\cJ_{MS}}{\lambda}(\beta-\tau) (\lambda y_1^*)^2-\frac{\cJ_{MS}}{\lambda}\tau (\lambda y_2^*)^2\no\\
&- 4 \sum_{k=0}^\infty \log \left( 1- q_M q^k \right) - \sum_{k=0}^\infty \frac{2q_M q^k}{\left(1-q_M q^k\right)^2} \(\(\lambda y_1^*\)^2+\(\lambda y_2^*\)^2\) + O(\lambda y_i)^3~.
\end{align}
In the above we approximate the $k$ sum with an integral and using \eqref{eq:massive_disconn_saddle_sol} we get
\begin{align}
W(y_1^*,y_2^*)=&~2\pi y_1^* + 2\pi y_2^* +\frac{2\cJ_{MS}}{\lambda}\beta - \frac{\cJ_{MS}}{\lambda}(\beta-\tau) (\lambda y_1^*)^2-\frac{\cJ_{MS}}{\lambda}\tau (\lambda y_2^*)^2\no\\
&+\frac{4}{\lambda} \text{Li}_2(q_M) - \frac{1}{\lambda}\frac{2q_M}{1-q_M} \(\(\lambda y_1^*\)^2+\(\lambda y_2^*\)^2\) + O(\lambda y_i^*)^3\no\\
=&~\pi y_1^* + \pi y_2^* +\frac{2\cJ_{MS}}{\lambda}\beta+\frac{4}{\lambda} \text{Li}_2(q_M)~.
\end{align}
Putting it all together, the result of the saddle point estimate of the $y_i$ integrals is
\be
\begin{split}
G(\beta,\tau) =& \frac{1}{Z(\beta)} \(\frac{\lambda^3(q;q)_\infty^3}{2\pi^2}\)^2 \exp\(\pi y_1^* + \pi y_2^* +\frac{2\cJ_{MS}}{\lambda}\beta+\frac{4}{\lambda} \text{Li}_2(q_M)-\frac{1}{\lambda} \text{Li}_2(q_M^2)\)\no\\
& \(\frac{\pi}{\lambda}\)^3\(\frac{1}{\cJ_{MS}(\beta-\tau)+\frac{2q_M}{1-q_M}} \cdot \frac{1}{\cJ_{MS}\tau+\frac{2q_M}{1-q_M}}\)^{3/2}~.
\end{split}
\ee
Similarly, the partition function is\footnote{This is obtained by taking just one integral in the computation above (the one with $\beta $) and plugging $q_M \to 0$. This also indeed agrees (up to a $\beta $ independent prefactor) with Eq.\ (4.13) in \cite{Berkooz:2018qkz}.}
\be
Z(\beta)= \frac{\lambda^3(q;q)_\infty^3}{2\pi^2} \(\frac{\pi}{\beta \cJ_{MS} \lambda}\)^{3/2}\exp\(\frac{2\beta\cJ_{MS}}{\lambda}+\frac{\pi^2}{\beta\cJ_{MS}\lambda}\)
\ee
where approximately
\be
(q;q)_\infty\approx \exp\(-\frac{\pi^2}{6\lambda}\)~.
\ee
Together we have the final result
\be
\begin{split}
&G(\beta,\tau)= \frac{1}{\sqrt{4\pi}} \(\frac{\beta \cJ_{MS}\lambda}{(\cJ_{MS}(\beta-\tau)+\frac{2q_M}{1-q_M})(\cJ_{MS}\tau+\frac{2q_M}{1-q_M})}\)^{3/2}\\
&\exp\[\frac{\pi^2}{\lambda}\frac{1}{\cJ_{MS}(\beta-\tau)+\frac{2q_M}{1-q_M}}+\frac{\pi^2}{\lambda}\frac{1}{\cJ_{MS}\tau+\frac{2q_M}{1-q_M}}-\frac{\pi^2}{\beta\cJ_{MS}\lambda}+\frac{4}{\lambda} \text{Li}_2(q_M)-\frac{1}{\lambda} \text{Li}_2(q_M^2)-\frac{\pi^2}{2\lambda}\].
\end{split}
\ee

\begin{figure}[h]
\centering
\includegraphics[width=0.5\textwidth]{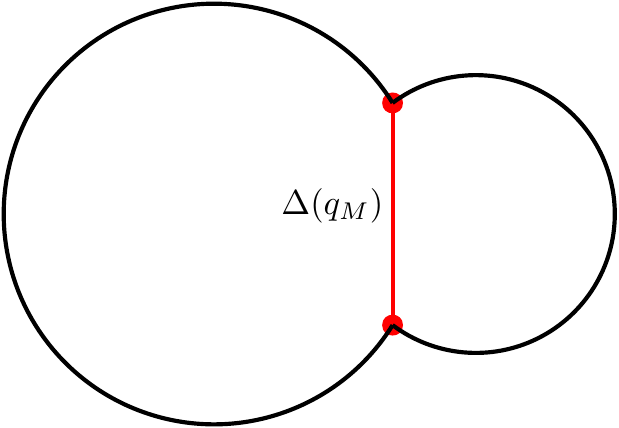}\caption{Pictorial representation of the two-point function of a massive operator. The black curves represent the thermal circle deformed by the presence of this heavy operator (indicated by red dots). $\Delta(q_M)$ represents the geodesic distance between the two insertion points.}
\label{disconnUniverses}
\end{figure}

We get that the un-normalized 2-point function $\tilde G=Z(\beta )G$ is just (up to a $\beta$ and $\tau $ independent constant) the product of thermal partition functions
\begin{equation}
\tilde G(\beta ,\tau ) =\exp\left( \frac{4}{\lambda } \text{Li}_2(q_M)-\frac{1}{\lambda } \text{Li}_2(q_M^2)-\frac{8}{\lambda } \frac{q_M}{1-q_M}\right) Z(\beta _1)Z(\beta _2)
\end{equation}
with inverse temperatures
\be\label{modLengths}
\begin{split}
\beta_1 &= (\beta-\tau) +\Delta(q_M)~,\\
\beta_2 &=\tau+\Delta(q_M)~,
\end{split}
\ee
where
\be\label{shift}
\Delta(q_M)=\frac{1}{\cJ_{MS}}\frac{2q_M}{1-q_M} .
\ee
The effect of non-vanishing finite $q_M\neq 1$ is to increase the values of $\beta_1$ and $\beta_2$ by the same amount $\Delta(q_M)$. This effect can also be seen in the standard SYK model by using the saddle point equations (2.9) and (2.10) of \cite{Goel:2018ubv}. In that paper the $k_i$ are our $y_i$ and the shift in $\tau_i$ is controlled by $1/\ell$, with $\ell$ being the scaling dimension of the operator whose two-point function is under consideration. This is pictorially represented in figure \ref{disconnUniverses}. The formulas, however, do not agree exactly in all the regimes of $\ell /C$ in their language vs.\ $q_M$ in our language.

\subsubsection{Comments on the $q_M\rightarrow 0$ limit, or touching space-times}

The limit $q_M \to 0$ corresponds to even heavier operators.
In this limit, the distance $\Delta(q_M)\rightarrow 0$. This means that the spaces only touch at a point. Otherwise, time evolution in each space in general proceeds on its own. More precisely, this is true for any probe operator of fixed conformal dimension (but light enough particles can go between the two universes, i.e. particles close to the BF bound in the $AdS$ language). For finite mass operators, in general, if we put several probe operators in the two spaces, we can consider two sets of chord diagrams --- the ``within universes" diagrams where operators (and Hamiltonians) are contracted within each part of the trace separately, and ``between universes'' diagrams where operators (and Hamiltonians) are also contracted between the two parts of the trace. In terms of figure \ref{fig:two_point_func} the first class of diagrams have chords that stay within region I or region II, and the second class have chords going from I to II. The latter are suppressed by at least $q_M$ relative to the first. So for $q_M \to 0$, the evolution of all operators happens within each spacetime.

This can also be understood in terms of a putative convergence of operators in our statistical class to the standard large $N$ limit of $\beta$-ensemble random matrices. Recall that \cite{erdHos2014phase} if one computes the moments of double scaled operators, then they converge to those of the semi-circle when $p/\sqrt{N}\rightarrow\infty$. The operator $M$ (recall that we are now in the Majorana SYK model) is given by \eqref{eq:RealSYKOp},
and it has $ N \choose{p_M}$ independent degrees of freedom. For $p_M\sim N/2$, it has the same order of magnitude of parameters as a full random $2^{N/2} \times 2^{N/2}$ Hermitian matrix, just written in a different basis of operators. In this case its correlators satisfy planarity constraints and cannot intersect any other contractions of matrices. In other words, long operators tend to act as $\beta$-ensembles random matrices, and the latter split spacetime into fragments. We see that (at the level of macroscopic observables computed in a single trace), the same is true even for much smaller operators $p_M\sim A \sqrt{N}$ when $A$ is large.

In the next subsection we will make the computation above more precise by inserting an additional light operator on each side. We will see that we can compute the leading transmission through the zero size neck quite easily. In particular we will see which operators make it between the universes easily and which get stuck at the singularity.

\subsection{Connecting universes} \label{sec:conn_universes}

In this section we consider the 4-point function of two heavy operators $M_M$ and two light operators $M_L$.
From Eq.\ (4.11) of \cite{Berkooz:2018jqr} we have the expression for the crossed four-point function as follows
\be\label{crossed-4pffun}
\begin{split}
&\<\tr e^{-\beta H} M_L(\tau_1)M_M(\tau_2)M_L(\tau_3)M_M(\tau_4)\>_{J,J_M}  = q_{ML} \int _0^{\pi }  \prod _{j=1} ^4 \left\{ \frac{d\theta _j}{2\pi } (q,e^{\pm 2i\theta _j} ;q)_{\infty }\right\} \times\\
&  \exp\left( -\frac{2\cJ}{\sqrt{1-q} }\((\tau_2-\tau_3) \cos(\theta_2) + (\tau_1-\tau_2)\cos(\theta_3)+ (\tau_3-\tau_4) \cos(\theta_1) + (\beta-\tau_1+\tau_4)\cos(\theta_4)\) \right) \times \\
&   \frac{\left( q_{L}^2,  q_{L}^2;q \right)_{\infty } }{\left( q_{L} e^{i(\pm  \theta _{1} \pm \theta _{2} )},  q_{L} e^{i(\pm  \theta _{3} \pm \theta _{4} )} ;q \right)_{\infty } } \times \\
& \frac{(q_M^2;q)_{\infty } }{(q_M e^{i(\pm \theta _2 \pm \theta _3)} ;q)_{\infty } }  \sum_{n=0}^\infty \frac{q^n_M}{(q_{L}^2;q)_n(q;q)_n} Q_n(\cos\theta_1|q_{L} e^{\mp i\theta_2};q) Q_n(\cos\theta_4| q_{L} e^{\mp i\theta_3};q)
\end{split}
\ee
where
\begin{align}
\begin{split}
&{q}_{ML}=e^{-\sqrt{\lambda_M\lambda_L}}~,~~~~{q}_{L}=e^{-\sqrt{\lambda\lambda_L}}~,~~~~{q}_{M}=e^{-\sqrt{\lambda\lambda_M}}~,\\
&2p^2=\lambda N~,~~~~2p_M^2=\lambda_M N~,~~~~2p_L^2=\lambda_L N~
\end{split}
\end{align}
and $Q_n$ are the Al Salam-Chihara polynomials defined in terms of the basic hypergeometric series ${}_r\phi _s$ through
\begin{equation}
Q_n(\cos(\theta)|a,b;q) = \frac{(ab;q)_n}{a^n} {}_3\phi _2\left[  \begin{matrix}
q^{-n},ae^{\pm i\theta } \\ ab,0
\end{matrix};q,q\right] .
\end{equation}

We now focus on the limit
\be
\lambda\to 0~,~~~~\lambda_M\to\infty~,~~~~\lambda \lambda_M ~\text{fixed\ large}
\ee
such that we go to a gravity regime (the limit on $\lambda$) and we can arrange the connection between the spaces in an expansion in $q_M$.
We then also take the limit
\be
\lambda_L\to 0~,~~~~\lambda_L\lambda_M \equiv \lambda_{LM}^2~\ \text{fixed,\ not\ small}\ ,
\ee
such that the new light field can go between the spaces without suppression.

\begin{figure}[h]
\centering
\includegraphics[width=0.8\textwidth]{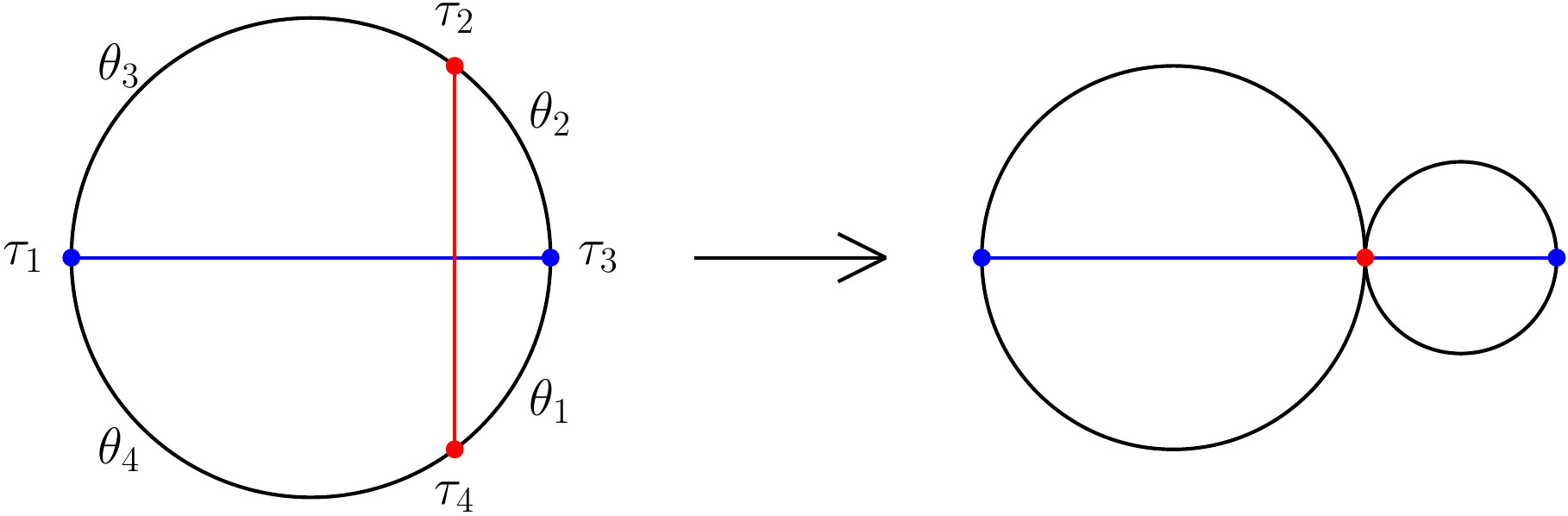}
\caption{The effect of very heavy operators, shown in red, causing the 4-point function to collapse. The light operators are shown in blue.}
\label{fig:connectingU_qM_zero}
\end{figure}

When $q_M=0$ only the $n=0$ term (which is one) survives in the last line of \eqref{crossed-4pffun} and the 4-point function factorizes into a product of two 2-point functions
\be
\begin{split}
&\<\tr e^{-\beta H} M_L(\tau_1)M_M(\tau_2)M_L(\tau_3)M_M(\tau_4)\>_{J,J_M}  = q_{ML} \times\\
& \int _0^{\pi }  \prod _{j=1,2}  \left\{ \frac{d\theta _j}{2\pi } w(\theta_j) \right\} \exp\left( \frac{-2\cJ(\tau_3-\tau_4) \cos\theta_1 -2 \cJ(\tau_2-\tau_3)\cos\theta_2}{\sqrt{1-q} } \right)   \frac{\left( q_L^2;q \right)_{\infty } }{\left( q_L e^{i(\pm  \theta _{1} \pm \theta _{2} )};q \right)_{\infty } }  \\
& \int _0^{\pi }  \prod _{j=3,4} \left\{ \frac{d\theta _j}{2\pi } w(\theta_j) \right\} \exp\left( \frac{-2\cJ(\tau_1-\tau_2) \cos\theta_3 -2 \cJ(\beta-\tau_1+\tau_4)\cos\theta_4}{\sqrt{1-q} } \right)   \frac{\left( q_L^2;q \right)_{\infty } }{\left( q_L e^{i(\pm  \theta _{3} \pm \theta _{4} )};q \right)_{\infty } }  ~.
\end{split}
\ee
This result can be interpreted as follows. The presence of the massive operator $M_M$ for which $\lambda_M\to \infty$ such that $q_M\to 0$, creates a background where the thermal circle which was of length $\beta$ gets deformed into two osculating circles of lengths
\be
\beta_1=\beta-\(\tau_2-\tau_4\)~,~~~~\beta_2=\tau_2-\tau_4~,
\ee
see fig.\ \ref{fig:connectingU_qM_zero}.
The light operator $q_L$ then probes this background and measures the boundary lengths $\beta_1$ and $\beta_2$ of these two spaces. This is seen in the expression for the crossed four-point function which factorizes into a product of two thermal two-point functions with inverse temperatures $\beta_1$ and $\beta_2$.

The main point, however, is that this correlator is finite even in the limit $q_M\rightarrow 0$. In this limit, no Hamiltonian chords go from one space to the other so the two of them gravitationally decouple. Nevertheless light states (corresponding to operators whose dimension is close to zero, which is allowed in quantum mechanics), can go between the spaces quite easily.


\newpage
\section*{Acknowledgements}
We would like to thank O.~Aharony, N.~Brukner, M.~Isachenkov, P.~Narayan, M.~Rangamani, A.~Raz, M.~Rozali, G.~Tarnopolsky, G.~J.~Turiaci, and H.~Verlinde for useful discussions. This work is supported by an ISF center of excellence grant (2289/18). MB is the incumbent of the Charles and David Wolfson Professorial chair of theoretical Physics. HR acknowledges the support from the PBC postdoctoral fellowship program as well as the Israel Science Foundation center for excellence grant (grant number 1989/14) and by the Minerva foundation with funding from the Federal German Ministry for Education and Research.

\appendix


\section{Grand potential in the $\lambda  \to 0$ limit} \label{app:grand_potential_lambda_zero}
In this section we will calculate the grand potential $\Omega=-T\log Z(T, \mu)$ in the $\lambda \to 0$ limit and compare it to the large $p$ result in \cite{Davison:2016ngz}. We will work in the scaling $\mu=\frac{\bar{\mu}}{N^{1/4}}$, and keep only terms up to $O(N^0)$. The partition function in this scaling was found in \eqref{Zeqm3} and is (recalling $\lambda=p^2/N$)
\begin{align}\label{Zeqm31}
\begin{split}
Z(\b,\mub) =& \exp\(N\log (2) + \frac12 \mub^2 \sqrt{N}-\frac{1}{12} \mub^4\)\cdot \\[10pt]
& \int_0^\pi \frac{d\theta}{2\pi}\(q,e^{\pm2i\theta};q\)_\infty \exp\[-\frac{2\b J\cos\theta}{\sqrt{1-q}}\exp\(\frac{\lambda}{2}-\frac12\mub^2\sqrt{\lambda}\)\]~,
\end{split}
\end{align}
where $q(\mu)=\exp(-4\lambda \(\cosh\mu\)^2)=\exp\[-4\lambda + O(N^{-1/2} )\]$ so $q=e^{-4\lambda } $.
Note that we reinstated a factor of $2^N$ to the partition function, which was implicit before because of the normalization $\tr~ \mathbb{1}=1$ that was used. We want to evaluate the $\theta$ integral in the above expression for which we use the methods developed in \cite{Berkooz:2018qkz,Berkooz:2018jqr}.
The $q\to 1$ limit of the measure factor can be simplified by change of variables $\theta=\pi-4\lambda y$
\be
\int_0^\pi \frac{d\theta}{2\pi} (q,e^{\pm2i\theta};q)_\infty  \longrightarrow \frac{64\lambda^3(q;q)_\infty^3}{2\pi^2}\int_0^\infty dy~ 2y \sinh(2\pi y)~.
\ee
For notational convenience, let us define the quantity
\begin{align}
\delta \equiv &\exp\(\frac{\lambda}{2}-\frac12\mub^2\sqrt{\lambda}\)~.
\end{align}
Then, the $\theta$-integral in \eqref{Zeqm31} becomes
\be
\frac{64{\lambda}^3(q;q)_\infty^3}{2\pi^2}\int_0^\infty dy~ 2y \sinh(2\pi y) \exp \(\frac{2\b J \cos (4\lambda y)}{\sqrt{1-q}}\delta\)~.
\ee
Assuming ${\lambda}y\ll 1$ we can expand the cosine in the above equation
\begin{align}
\frac{{64 \lambda}^3(q;q)_\infty^3}{2\pi^2} \exp\(\frac{2\b J \delta}{\sqrt{1-q}}\) \int_0^\infty dy~ y \exp\(2\pi y - \frac{\b J \delta}{\sqrt{1-q}}\(4\lambda y\)^2 +\cdots \)~.
\end{align}
The $y$-integral can then be approximated by a saddle point. We find the saddle point is at $y^*=(\pi \sqrt{1-q})/(16\b J \delta \lambda^2)$ and the saddle-point estimate of the $y$-integral is
\be
\frac{64\lambda^3(q;q)_\infty^3}{2\pi^2} \exp\(\frac{2\b J \delta}{\sqrt{1-q}}+\frac{\pi^2 \sqrt{1-q}}{16 \b J \delta \lambda^2}\) \(\frac{\pi \sqrt{1-q}}{16\b J \delta \lambda^2} \)^{3/2}~.
\ee
The above expression is the $q\to 1$ limit of the expression in the second line of \eqref{Zeqm31}. Combining with the terms from the first line and noting that $(q;q)^3_\infty\approx \exp\(-\pi^2/(8\lambda)\)$, we obtain the following low temperature expansion of $\Omega(T,\bar{\mu})$
\begin{align}\label{omegaUS}
\Omega(T,\bar{\mu}) =&-\frac{J}{\sqrt{\lambda}} \exp\(-\frac12\mub^2\sqrt{\lambda}+\frac{\lambda}{2}\) -T \(N\log (2) + \frac12 \mub^2 \sqrt{N}-\frac{1}{12} \mub^4\)\no\\[5pt]
&+\frac{\pi^2 T}{8\lambda}-T\log\(\frac{{64 \lambda}^3}{2\pi^2}\)-\frac{3}{2}  T\log\left( \frac{\pi }{8 J \delta  \lambda ^{3/2} } \right) +O(T\log T) ~.
\end{align}

We now compare this result to the large $p$ result in \cite{Davison:2016ngz}, that was quoted in \eqref{SgPot}, where $v$ there can be expanded
\be
v=1-\frac{2T}{\cJCSYK}+\frac{4T^2}{\cJCSYK^2}+\cdots
\ee
and $\cJCSYK$ is defined in \eqref{curlJDef}. With the scaling $\mu=\frac{\bar{\mu}}{N^{1/4}}$, we find that taking the large $N$ limit first and then doing the low temperature expansion gives
\begin{align}\label{omegaS}
\Omega\(T,\bar{\mu}\)=& -\frac{J}{ \sqrt{\lambda }} \exp\(-\frac{1}{2} \text{$\bar\mu$}^2\sqrt{\lambda } \) -T\(N \log (2)+\frac12 \text{$\bar \mu$}^2 \sqrt{N}-\frac{\text{$\bar \mu$}^4}{12}\)+\frac{\pi ^2 T}{8 \lambda }+O\left(T^2\right) ~.
\end{align}
We see that this indeed matches with \eqref{omegaUS} (up to smaller $O(\lambda^0,\, \log(\lambda))$ terms).

\section{Details of the $U(M)$ model}\label{UMDetails}
Let us introduce the following notation
\be
\Sigma_3= \bigotimes_{i=1}^{N}\sigma_3~.
\ee
A convenient matrix representation of the algebra \eqref{unga} (which is a generalization of Eq.\ \eqref{complex_Clifford_rep}) is
\begin{equation}
\begin{split}
& \psi^{1\a}=\mathbb{1}_{2^{N(\a-1)}}\otimes\big(\sigma_{+}\otimes\sigma_{3}\otimes\cdots\otimes\sigma_{3}\big)\otimes \bigotimes_{i=1}^{M-\a}\Sigma_3\\
& \psi^{2\a}=\mathbb{1}_{2^{N(\a-1)}}\otimes\big(\mathbb{1}_2\otimes\sigma_{+}\otimes\sigma_{3}\otimes\cdots\otimes\sigma_{3}\big)\otimes \bigotimes_{i=1}^{M-\a}\Sigma_3\\
&~~~ \vdots\\
& \psi^{N\a}=\mathbb{1}_{2^{N(\a-1)}}\otimes\big(\mathbb{1}_2\otimes\cdots\otimes\mathbb{1}_2\otimes\sigma_{+}\big)\otimes \bigotimes_{i=1}^{M-\a}\Sigma_3
\end{split}
\end{equation}
where $\a$ is the $U(M)$ fundamental index that runs from 1 to $M$. For $\bar{\psi}$ we replace $\sigma_{+}\to\sigma_{-}$ in the above formulas. In this representation we have
\begin{align}
\begin{split}
\exp\(\sum_{\a=1}^M -\frac{\mu_\a}{2}\sum_{i=1}^N\(\bar{\psi}_{i\a}\psi^{i\a}-\psi^{i\a}\bar{\psi}_{i\a}\)\)= \bigotimes_{\a=1}^M\bigg[\left(
\begin{array}{cc}
 e^{\mu_\a } & 0 \\
 0 & e^{-\mu_\a } \\
\end{array}
\right)^{\otimes N} \bigg]~.
\end{split}
\end{align}
The normalized trace of this quantity is simply
\be
\tr\exp\(\sum_{\a=1}^M -\frac{\mu_\a}{2}\sum_{i=1}^N\(\bar{\psi}_{i\a}\psi^{i\a}-\psi^{i\a}\bar{\psi}_{i\a}\)\)=\prod_{\a=1}^M (\cosh\m_\a)^N ~.
\ee

Next we proceed towards the calculation of the moments \eqref{UMmoments1}. Let $I,J,...$ be index sets of cardinality $p$ with indices arranged in ascending order. It is the set of site indices $i_a$ $(a=1,2,\cdots,p)$. Let $A,B,...$ be set of cardinality $p$. It contains the flavor indices $\a_a$ $(a=1,2,\cdots,p)$. Denote by $\jb_{IA}=\prod_{a=p}^1\jb_{i_a,\a_a}$ and by $\j^{IA}=\prod_{a=1}^p\j^{i_a,\a_a}$. In this notation the Hamiltonian is
\begin{align}
\begin{split}
H& =\sum_{\a_1=1}^M  \cdots \sum_{\a_p=1}^M ~~\sum_{\substack{1\le i_{1}<\cdots<i_{p}\le N\\
1\le j_{1}<\cdots<j_{p}\le N
}
}J_{j_{1}\cdots j_{p}}^{i_{1}\cdots i_{p}}~\bar{\psi}_{i_{p}\a_p}\cdots\bar{\psi}_{i_{1}\a_1}\psi^{j_{1}\a_1}\cdots\psi^{j_{p}\a_p}~,\\[5pt]
&\equiv \sum_{IJA} J^I_{J}~\jb_{IA}\j^{JA}~.
\end{split}
\end{align}
The quantity that we would like to evaluate is the following
\be\label{generic_trace0}
\JUM^k M^{-kp/2}  {N\choose p}^{-k} \sum_{I_1,\cdots I_k}\sum_{\{A,A'\}} \tr\(\jb_{I_1A_1}\j^{I_2 A_1}\cdots \jb_{I_3A_2}\j^{I_4 A_2}\cdots  \jb_{I_2A_1'}\j^{I_1 A_1'}\cdots e^{-\mu_r Q^r}\)
\ee
which contributes to the $k$'th moment of the partition function. The above object can be represented as an oriented chord diagram, with chords of opposite orientation always occurring in pairs. The chords have an orientation that points from $\j^{IA}$ to $\jb_{IA'}$. They represent contraction of the site indices.

Before we calculate these diagrams, let us first determine their sign. The analysis is a straightforward extension of the $U(1)$ analysis.
First note that each chord is associated with the sites index set I, but also with a pair of flavor index sets $A,A'$ in general. However, we will see in the analysis below, that eventually we can restrict to $A=A'$, so that an oriented chord can be associated with a pair of index sets $I,A$.\footnote{In the analysis of the different situations we do, there are in fact cases where this does not happen, such as the case where we need a further restriction on the index sets (overlaps beyond the contraction) in the analysis of diagram \ref{unor21}, as well as the case $a=b$ in diagram \ref{unor11}. However, these exceptions are suppressed in large $N$ and do not contribute anyway.}
For oriented chords that do not intersect, we do not get any minus signs. Consider a pair of oriented chords $IA,JB$ that intersect. For any $(i_a,\a_a)\in IA$ and $(j_b,\b_b)\in JB$ we get a minus sign whenever, $i_a+(\a_a-1)N<j_b+(\b_b-1)N$, $i_a+(\a_a-1)N>j_b+(\b_b-1)N$ and no sign when $i_a+(\a_a-1)N=j_b+(\b_b-1)N$, where $a,b$ range from $1$ to $p$. The last equality is satisfied only when $i_a=j_b$ and $\a_a=\b_b$. Therefore for oriented chords $IA,JB$ that intersect, we get the following sign
\be\label{UMsigns}
(-1)^{p^2}\prod_{a,b=1}^p (-1)^{\delta_{i_a,j_b}\delta_{\a_a,\b_b}}~.
\ee
This is similar to the $U(1)$ case where instead of $p_{ij} $ we have the intersection between the index sets where $(i,\alpha )$ are considered as a single index. As a result, the sign in the partition function will again be positive.

\begin{figure}[h]
\subfloat[]{
\centering
\includegraphics[width=0.3\textwidth]{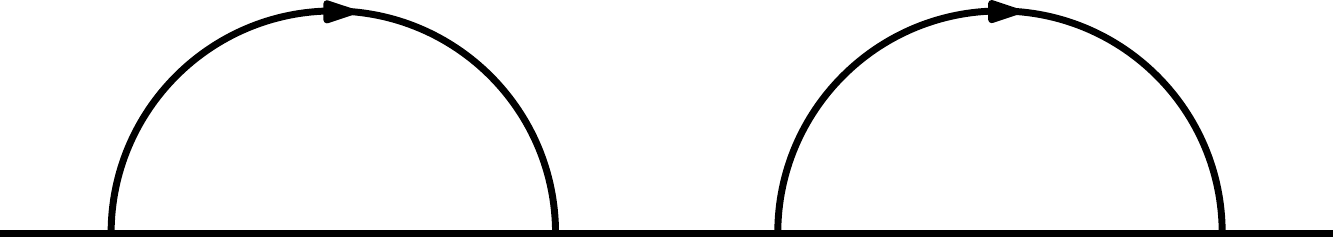}\label{or1}}
\subfloat[]{
\centering
\includegraphics[width=0.3\textwidth]{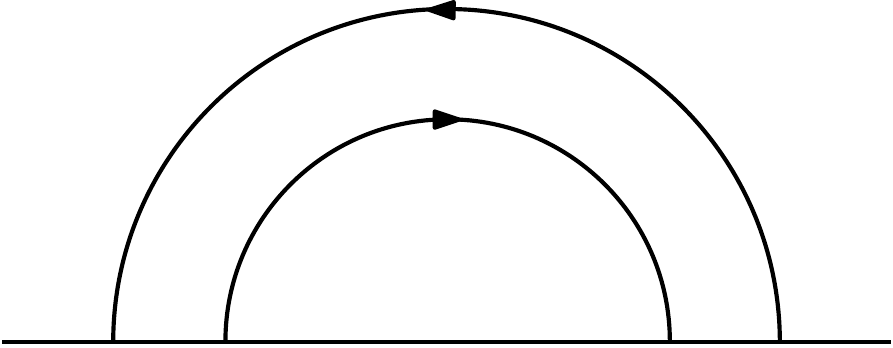}\label{or5}}
\subfloat[]{
\centering
\includegraphics[width=0.3\textwidth]{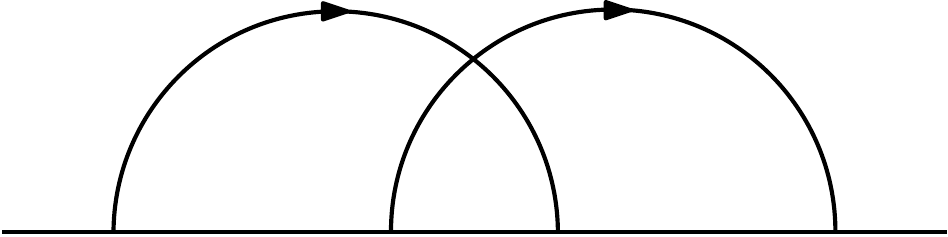}\label{or3}}\\[10pt]
\subfloat[]{
\centering
\includegraphics[width=0.3\textwidth]{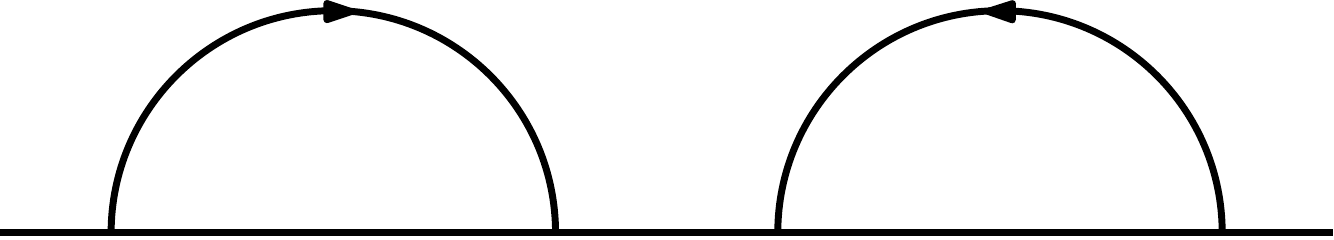}\label{or2}}
\subfloat[]{
\centering
\includegraphics[width=0.3\textwidth]{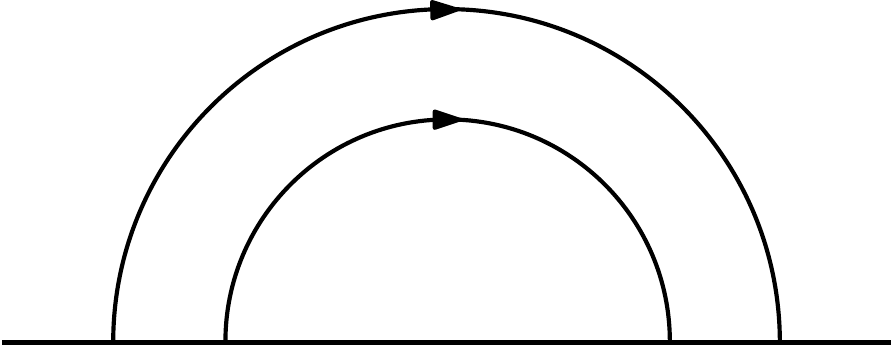}\label{or6}}
\subfloat[]{
\centering
\includegraphics[width=0.3\textwidth]{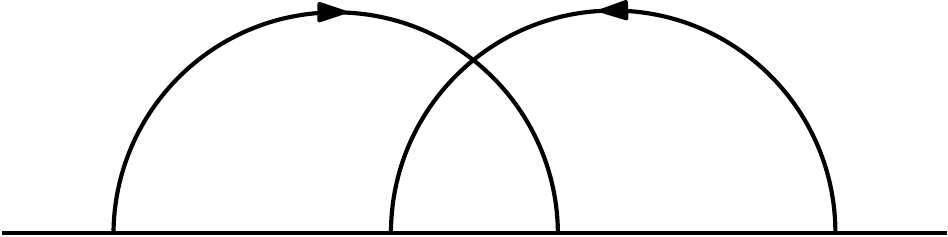}\label{or4}}
\caption{Oriented chord configurations (plus the reversed orientation counterparts) that dominate in the large $N$ limit.}
\label{or_all}
\end{figure}

\begin{figure}[h]
\subfloat[]{
\centering
\includegraphics[width=0.35\textwidth]{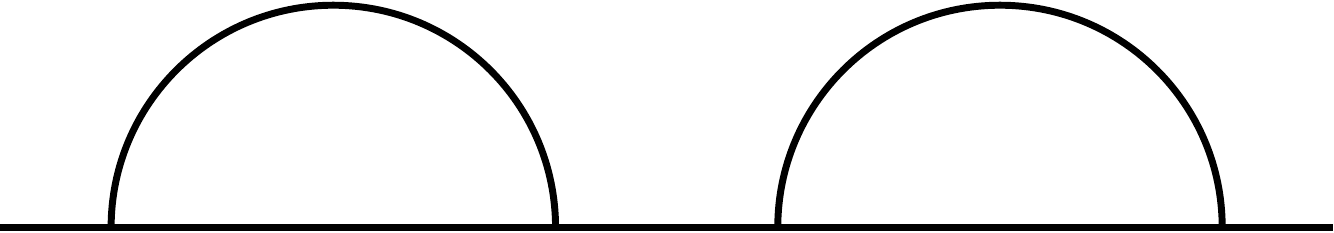}\label{unor21}}
\subfloat[]{
\centering
\includegraphics[width=0.3\textwidth]{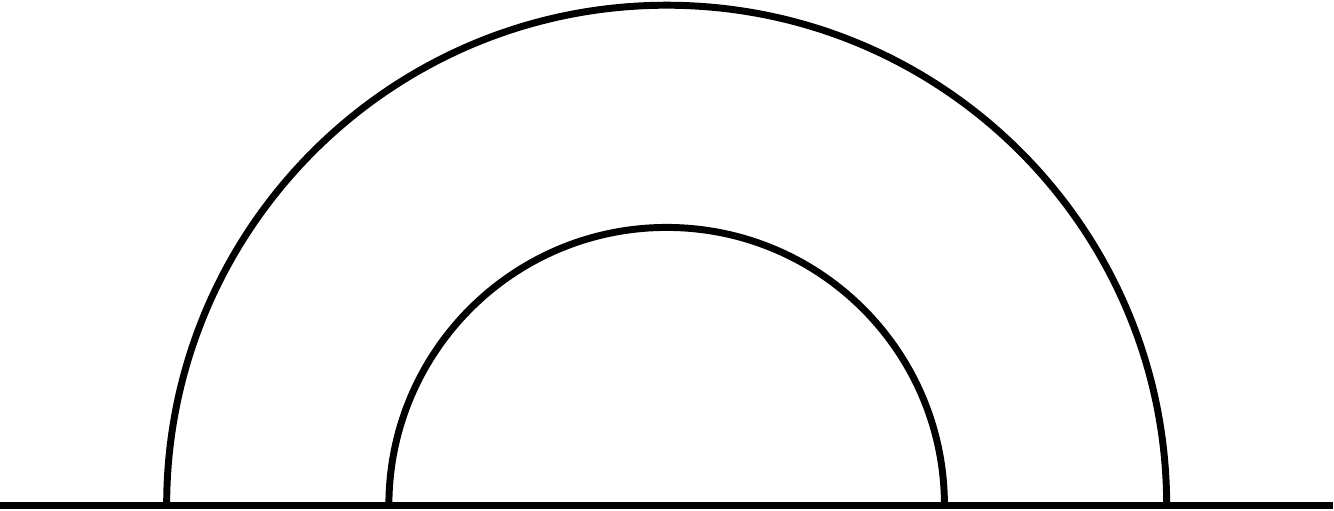}\label{unor31}}
\subfloat[]{
\centering
\includegraphics[width=0.3\textwidth]{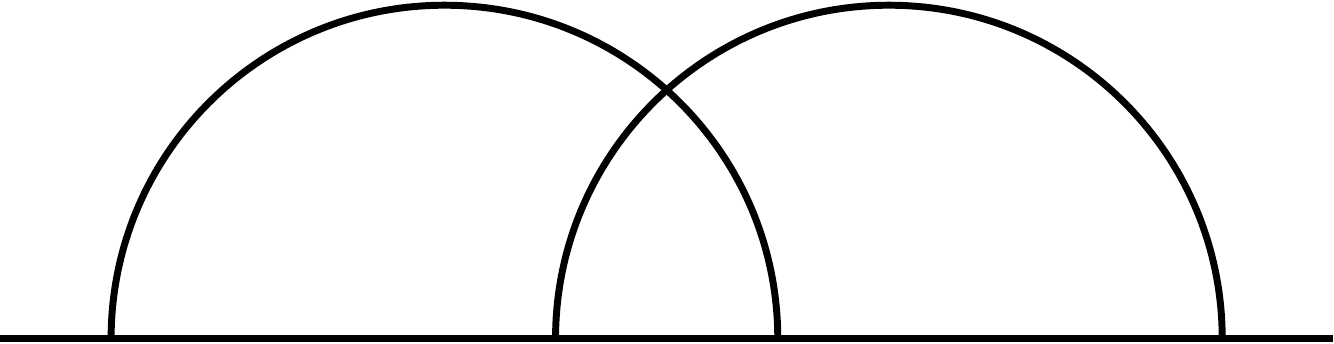}\label{unor41}}\\
\subfloat[]{
\centering
\includegraphics[width=0.35\textwidth]{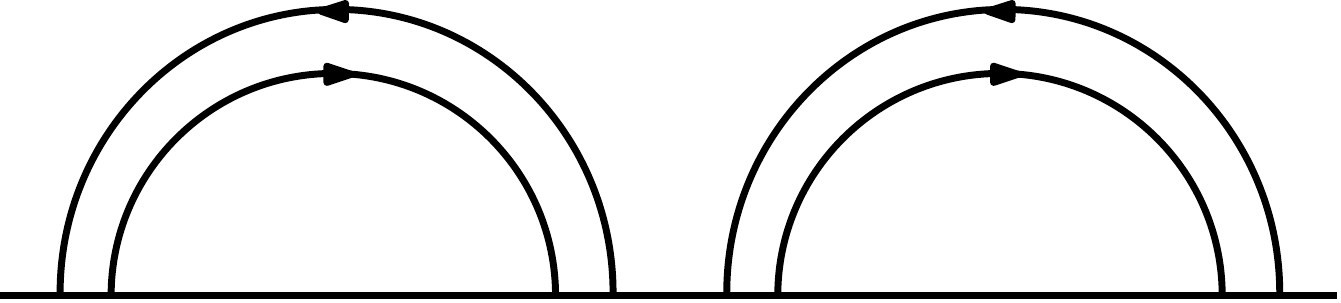}\label{unor20}}
\subfloat[]{
\centering
\includegraphics[width=0.3\textwidth]{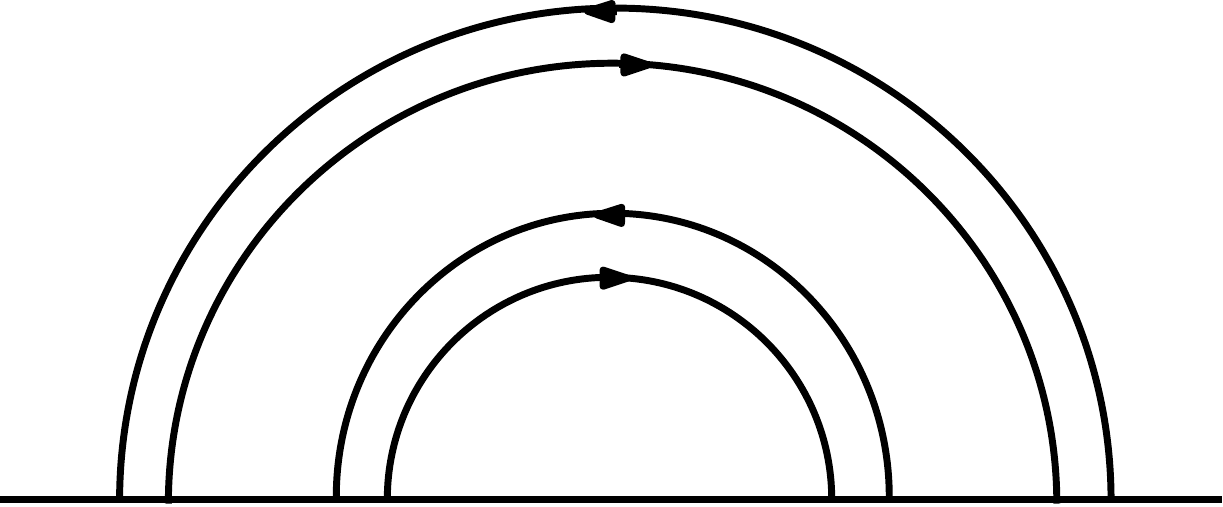}\label{unor30}}
\subfloat[]{
\centering
\includegraphics[width=0.3\textwidth]{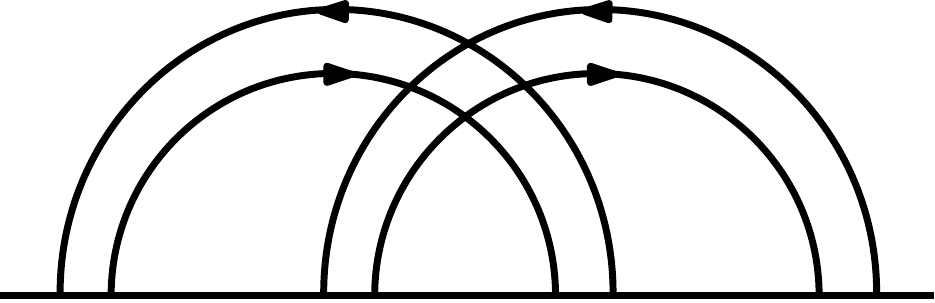}\label{unor40}}\\
\subfloat[]{
\centering
\includegraphics[width=0.25\textwidth]{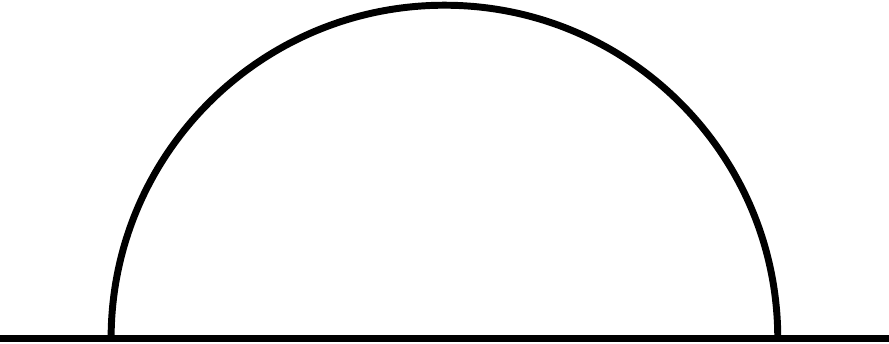}\label{unor11}}
\subfloat[]{
\centering
\includegraphics[width=0.25\textwidth]{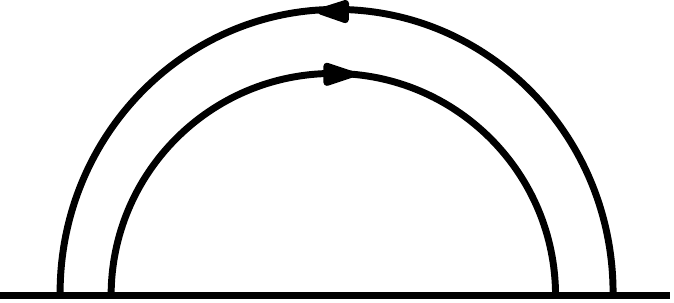}\label{unor10}}
\caption{Unoriented $H$-chord configurations (along with their orientated counterparts) that dominate in the large $N$ limit. Each of these configurations contribute independently to an unoriented chord diagram.}
\label{unor_all}
\end{figure}

As in the $U(1)$ model, the general strategy to evaluate \eqref{generic_trace0} is to first assume that all the site indices are mutually disjoint. In this case, for the trace to be non-zero, all the $A'$ index sets have to coincide with the corresponding $A$ index sets. Every right going chord contributes $\prod\limits_{a=1}^p e^{\mu_{\a_a}}/\cosh\mu_{\a_a}$ and every left going chord contributes $\prod\limits_{a=1}^p e^{-\mu_{\a_a}}/\cosh\mu_{\a_a}$. Since each right going chord is accompanied by a left going chord the net contribution from a single (unoriented) $H$-chord is
\be\label{curlAdef}
\sum_{\a_1=1}^M\cdots \sum_{\a_p=1}^M \prod_{a=1}^p (\cosh\mu_{\a_a})^{-2}\equiv\cA(\mu)^p~,~~~~\cA(\mu) =\sum_{\a=1}^M \(\cosh\mu_{\a}\)^{-2}~.
\ee
Since there are $k/2$ such $H$-chords, we get for the case when the site index sets are mutually disjoint \footnote{If the site index sets have no overlap, then we get a factor of $(-1)^{p^2}$ from \eqref{UMsigns} for every intersection of an oriented chord with another oriented chord. However, as in the case of the $U(1)$ model, since each intersecting $H$ chords involve four oriented chord intersections, there are no factors of minus sign.}
\be
\prod_{\a=1}^M (\cosh\m_\a)^N \cA(\mu)^{kp/2}~.
\ee
The next step is to correct this result for non-zero mutual intersection of the site indices $\{I_i\}$. In the large-$N$ limit the intersections among the index sets  occurs independently in pairs with a Poisson distribution and contribution from triple and higher intersections are subleading which can therefore be ignored. So at the level of chord diagrams, at large-$N$ the dominant contribution to a diagram comes from all possible pairs of oriented chords. These configurations are shown in figure \ref{or_all}.

However, these configurations have a hanging flavor index at the nodes which are not contracted. In a chord diagram, every oriented chord in figures \ref{or_all} is accompanied with a chord of opposite orientation and the flavor indices of these two chords are contracted separately at each node. Therefore we have to simultaneously consider the contribution to the trace from four oriented chords for every pair of unoriented chords and two oriented chords for every unoriented chord. The distinct pairs of unoriented chords and its corresponding oriented version is shown in the figure \ref{unor_all}.

Once we know the contribution of each of the configurations above, we then need to know how many such configurations are there in a given chord diagram. In a chord diagram with $k$ insertions of the Hamiltonian and $\kappa_H$ number of intersections of (unoriented) $H$-chords, there are $k/2$ chords of type \ref{unor11}, a total of ${k/2 \choose 2}-\kappa_H$ chords of type \ref{unor21} and \ref{unor31} combined and $\kappa_H$ number of type \ref{unor41}. In the ensuing subsections, we compute the contribution from each kind of configuration above.

\subsubsection*{Diagram \ref{unor21}}\label{sec:D2.1}
The trace structure for this configuration is
\be\label{eqor11}
\sum_{\{A_1,A_1',A_2,A_2'\}}\tr \(\jb_{I_1 A_1} \j^{I_2A_1} \cdots \jb_{I_2A_1'}\j^{I_1 A_1'} \cdots \jb_{I_3A_2}\j^{I_4A_2} \cdots \jb_{I_4A_2'} \j^{I_3A_2'} \cdots e^{-\mu_r Q^r}\)~.
\ee

The four kind of site intersections that we look at are $I_1\cap I_3,I_1\cap I_4,I_2\cap I_3,I_2\cap I_4$. As explained earlier, we cannot look at these intersections independently. Because of the flavor index contractions, $I_1\cap I_3$ is paired with $I_2\cap I_4$ and $I_1\cap I_4$ is paired with $I_2\cap I_3$. Let the number of intersections between $I_i$ and $I_j$ be $p_{ij}$.

Consider fermions with site index $i_a\in I_1,j_b\in I_2,k_c\in I_3$ and $l_d\in I_4$. If $p_{ij}$ were zero for all $i,j$ that is $I_i\cap I_j=\emptyset$ then the contribution to the trace from the fermions shown above is calculated as follows. Consider a fermion in $\jb_{I_1A_1}$, say $\jb_{i_a\a_a}$ and the fermion with the same site index in $\j^{I_1A_1'}$ which is $\j^{i_a\a'_a}$. The product of the two will give a non-vanishing contribution only when $\a_a=\a'_a$ in which case the result is $e^{-\m_{\a_a}}$. Since this holds for all $a$, the sum over $A_1'$ collapses to $A_1$. Let us now write the previous sum a little more explicitly
\be
\sum_{\a_1,...,\a_p} \tr\(\textstyle\prod\limits_{a=p}^1 \jb_{i_a\a_a}\textstyle\prod\limits_{b=1}^p \j^{j_b\a_b}\textstyle\prod\limits_{c=p}^1 \jb_{j_c\a_c}\textstyle\prod\limits_{d=1}^p \j^{i_d\a_d}\cdot (i\to k,j\to l,\a\to \b)\cdot e^{-\mu_r Q^r}\)~.
\ee
Without the insertion of the fermions from the Hamiltonian, the result of the trace is $\prod_{\a=1}^M (\cosh\m_\a)^N$ as we deduced in the previous section. Now in the above trace, for every $\jb_{i_a\a_a}$ and $\j^{i_a\a_a}$ (in this particular order) a contribution of $\cosh\m_{\a_a}$ is removed and instead a factor of $e^{-\mu_{\a_a}}$ is multiplied to the result. Similarly for every $\j^{j_a\a_a}$ and $\jb_{j_a\a_a}$ (in this particular order) a contribution of $\cosh\m_{\a_a}$ is removed and instead a factor of $e^{\mu_{\a_a}}$ is multiplied to the result. Since each of these cases appears equally, the net result of the trace is
\be \label{eq:UM_disjoint_sets}
\prod_{\a=1}^M (\cosh\m_\a)^N \(\sum_{\a_1=1}^M\cdots\sum_{\a_p=1}^M  \prod_{a=1}^p \(\cosh\mu_{\a_a}\)^{-2}\)^2 \equiv \prod_{\a=1}^M (\cosh\m_\a)^N \cA(\mu)^{2p}
\ee
where we defined the quantity $\cA$ in Eq.\ \eqref{curlAdef}.

If $p_{ij}\neq 0$ then modulo corrections due to intersection of the site indices, the previous result will be modified to
\be
\prod_{\a=1}^M (\cosh\m_\a)^N \cA(\mu)^{2(p-\sum p_{ij})}~.
\ee
Next we calculate the correction to this result that comes from the intersection of the site indices. First, we look at indices in $I_1\cap I_3$; say $i_a=k_b$. Without loss of generality we set $a=b=1$ (if $a\neq b$, we can always permute the fermions around --- the net sign is always positive in any permutation because they are done in pairs --  and relabel the flavor and the site summation indices). Since these fermions carry a flavor index $\a$ we also write below the accompanying fermion in the $I_2$ and $I_4$ index sets with the same flavor index but generically different site index $j_1\in I_2$ and $l_1\in I_4$
\be
\sum_{\{\a_1,\a_1',\b_1,\b_1'\}}\tr \(\cdots \jb_{i_1 \a_1} \j^{j_1\a_1} \cdots \jb_{j_1\a_1'}\j^{i_1 \a'_1} \cdots \jb_{i_1\b_1}\j^{l_1\b_1} \cdots \jb_{l_1\b'_1} \j^{i_1\b'_1} \cdots e^{-\mu_r Q^r}\)~.
\ee
The summation over the flavor indices gives rise to three and only three distinct cases $(\a_1=\a_1')\neq(\b_1=\b_1')$, $(\a_1=\b_1')\neq(\b_1=\a_1')$ and $(\a_1=\a_1'=\b_1=\b_1')$ for which the trace is non-zero. For each case, the trace is easy to calculate. The basic rules are: (1) for every pair  $\j^{i_a\a_a}$ and $\jb_{i_a\a_a}$ (in this particular order) there is a contribution of $e^{\mu_{\a_a}}/\cosh\m_{\a_a}$ (2) for every pair $\jb_{i_a\a_a}$ and $\j^{i_a\a_a}$ (in this particular order) there is a contribution of $e^{-\mu_{\a_a}}/\cosh\m_{\a_a}$ (3) for every quadruple $\j^{i_a\a_a}...\jb_{i_a\a_a}...\j^{i_a\a_a}...\jb_{i_a\a_a}$ there is a contribution of $2e^{\mu_{\a_a}}/\cosh\m_{\a_a}$ and finally (4) for every quadruple $\jb_{i_a\a_a}\cdots \j^{i_a\a_a}\cdots \jb_{i_a\a_a}\cdots \j^{i_a\a_a}$ there is a contribution of $2e^{-\mu_{\a_a}}/\cosh\m_{\a_a}$. With these rules, the result of the trace is
\begin{align}
&\sum_{(\a_1=\a_1')\neq(\b_1=\b_1')} \frac{1}{\(\cosh\m_{\a_1}\cosh\m_{\b_1}\)^2}+ \sum_{(\a_1=\b_1')\neq(\b_1=\a_1')} \frac{\delta_{j_1}}{\(\cosh\m_{\a_1}\cosh\m_{\b_1}\)^2}+\no\\
&\hspace{.3in}\sum_{\a} \frac{2e^{-\m_\a}}{\cosh\m_\a}\(\frac{e^{2\m_\a}}{(\cosh \m_\a)^2} +\delta_{j_1}(\cdots) \)
\end{align}
where by $\delta _{j_1} $ we mean that $j_1$ necessarily should intersect at least one of the other sets (such as $I_1,I_3,I_4$). However, the probability of $\delta _{j_1} $ happening in fact goes to zero at large $N$; there is a finite probability that a set of size $p$ intersects another set of size $p$, but imposing that a single site intersects a set of size of the order of $p$ is already suppressed (going as $1/\sqrt{N}$). Therefore we can ignore the $\delta _{j_1} $ terms in the above expression.

Since the site index set $I$ has distinct indices which are ordered, this procedure can be repeated for every $i_a$ and $k_b$ in $I_1\cap I_3$ and the net correction factor to \eqref{eq:UM_disjoint_sets} due to non-vanishing $p_{13}$ is $\cC(\mu)^{p_{13}}$ (all the corrections here and below are with respect to \eqref{eq:UM_disjoint_sets})
\be\label{defC}
\cC(\mu)\equiv \frac{1}{\cA(\m)^2} \(\sum_{(\a_1=\a_1')\neq(\b_1=\b_1')} \frac{1}{\(\cosh\m_{\a_1}\cosh\m_{\b_1}\)^2}+\sum_{\a} \frac{2e^{\m_\a}}{\cosh\m_\a^3}\)~.
\ee

Next, we look at indices in $I_2\cap I_4$. The calculation is identical to the previous case and the net result is simply $\cC(-\mu)^{p_{24}}$.

Finally, we look at the indices in $I_1\cap I_4$ and $I_2\cap I_3$. Again the calculation is the same as in the previous cases except that there is vanishing contribution from the case $\a_1=\a_1'=\b_1=\b_1'$. So let us define
\be
\cD(\mu)\equiv \frac{1}{\cA(\m)^2} \(\sum_{(\a_1=\a_1')\neq(\b_1=\b_1')} \frac{1}{\(\cosh\m_{\a_1}\cosh\m_{\b_1}\)^2}\)~.
\ee
Then, the correction factors that we get from non-vanishing $p_{23}$ and $p_{14}$ is $\cD^{p_{14}+p_{23}}$. Doing the sum over the $p_{ij}$
\be
\sum_{p_{ij}=0}^\infty \frac{\lambda^{p_{ij}}}{p_{ij}!} e^{-\lambda} (\#)^{p_{ij}}= e^{\lambda(\#-1)}
\ee
we get that the contribution coming from the diagram \ref{unor21} is
\be
\exp\[\lambda\(\cC(\m)+\cC(-\m)+2\cD(\mu)-4\)\]~.
\ee
We can further simplify the expression
\begin{align}
&\(\cC(\m)+\cC(-\m)+2\cD(\mu)-4\)\no\\
=&\frac{4}{\cA(\mu)^2}\bigg( \sum_{(\a_1=\a_1')\neq(\b_1=\b_1')} \frac{1}{\(\cosh\m_{\a_1}\cosh\m_{\b_1}\)^2} + \sum_{\a} \frac{1}{(\cosh\m_\a)^2} \bigg)-4\no\\
=&\frac{4}{\cA(\mu)^2} \(\sum_{\a} \frac{1}{(\cosh\m_\a)^2} -\sum_{\a}\frac{1}{(\cosh\m_\a)^4}\) \no\\
=&\frac{4}{\cA(\mu)^2} \sum_{\a} (\tanh \m_\a \text{sech} \m_\a)^2~.
\end{align}

\subsubsection*{Diagram \ref{unor31}}
The trace structure for this configuration is
\be\label{eqor22}
\sum_{\{A_1,A_1',A_2,A_2'\}}\tr \(\jb_{I_1 A_1} \j^{I_2A_1} \cdots \jb_{I_3A_2}\j^{I_4A_2} \cdots \jb_{I_4A_2'} \j^{I_3A_2'} \cdots \jb_{I_2A_1'}\j^{I_1 A_1'}  \cdots e^{-\mu_r Q^r}\)~.
\ee

The calculation of this trace is identical to that in the previous case and gives the same result. Since the result from these two diagrams is the same we can conclude that for every pair of the unoriented $H$ chords that do not intersect we have a factor of
\be\label{UMrule22}
\exp\(\frac{4\lambda}{\cA(\mu)^2} \sum_{\a} (\tanh \m_\a \text{sech} \m_\a)^2\)~.
\ee

\subsubsection*{Diagram \ref{unor41}}
The trace structure for this configuration is
\be\label{eqor33}
\sum_{\{A_1,A_1',A_2,A_2'\}}\tr \(\jb_{I_1 A_1} \j^{I_2A_1} \cdots \jb_{I_3A_2}\j^{I_4A_2} \cdots \jb_{I_2A_1'}\j^{I_1 A_1'}  \cdots \jb_{I_4A_2'} \j^{I_3A_2'} \cdots e^{-\mu_r Q^r}\)~.
\ee
First of all we note that the net sign of this chord configuration is positive on account of the fact that there are four oriented chord intersections. Again, if there are no intersections (all $p_{ij}=0$), we get a factor of $\cA^{2p}$. For non-empty intersections we get $\cA^{2(p-\sum p_{ij})}$ times corrections which is evaluated in the same manner as in the subsection \ref{sec:D2.1}. The only difference now is that for every index in an intersection (say $i_a=k_b$ for example when looking at indices in $I_1\cap I_3$), whenever all the associated flavor indices are the same (that is $\a_a=\a_a'=\b_b=\b_b'$) we always get a zero no matter which pair of chord intersection we are looking at. However, when flavor indices are not all equal we get a factor of $\cD^{p_{ij}}$. Since there are four such $p_{ij}$'s after summing over the $p_{ij}$ variables we get a contribution of
\be
\exp(4\lambda(\cD-1))~.
\ee
The factor of $\cD-1$ can be simplified a bit
\begin{align}
\cD-1&=\frac{1}{\cA(\m)^2} \(\sum_{(\a_1=\a_1')\neq(\b_1=\b_1')} \frac{1}{\(\cosh\m_{\a_1}\cosh\m_{\b_1}\)^2}\)-1\no\\[10pt]
&= -\frac{1}{\cA(\mu)^2}\sum_{\a=1}^M\frac{1}{\(\cosh\mu_\a\)^4}~.
\end{align}
Hence, we arrive at the rule that for every pair of unoriented $H$ chords that intersect, we have a factor of
\be\label{UMrule33}
\exp\(-\frac{4\lambda}{\cA(\mu)^2}\sum_{\a=1}^M\frac{1}{\(\cosh\mu_\a\)^4}\)~.
\ee
\subsubsection*{Individual $H$-chord contribution: diagram \ref{unor11}}
For individual chords, we have the following trace structure:
\be
\sum_{I_1I_2AA'} \tr \( \jb_{I_1A}\j^{I_2A} \cdots \jb_{I_2A'}\j^{I_1A'} \cdots e^{-\mu_r Q^r}\)~.
\ee
If there are no intersections $I_1\cap I_2=\emptyset$ (that is $p_{12}=0$), we get a factor of $\cA^{p}$. For $I_1\cap I_2\neq \emptyset$, this factor receives corrections that we evaluate next. For simplicity, we first look at $p_{12}=1$ in which we have $i_a=j_b$ for some $i_a\in I_1$ and $j_b\in I_2$. Unlike the previous diagrams, there are two distinct cases $a=b$ and $a\neq b$. For the diagrams we considered previously, the case $a\neq b$ is the same as $a= b$ because we could relabel the various summation indices freely for the reason that we were looking at those oriented chords intersections that do not end in the same Hamiltonian. When we consider intersections between two oriented chord intersections that end in the same Hamiltonian, we are constrained by the flavor summation indices and the case $a=b$ distinct from the case $a\neq b$. Below, we look at these two cases in turn.

First, consider the case $a=b$. Without loss of generality say $a=b=1$. The fermions outside the intersection contribute $\cA(\m)^{p-1}$. For the remaining fermions we have
\be
\sum_{\a_1\b_1} \tr (\cdots\jb_{i_1\a_1} \j^{i_1\a_1} \cdots \jb_{i_1\b_1}  \j^{i_1\b_1} \cdots ~e^{-\mu_r Q^r})\no~.
\ee
For $\a_1\neq\b_1$ we get a contribution of $e^{-\m_{\a_1}-\m_{\b_1}}/(\cosh\mu_{\a_1}\cosh\mu_{\b_1})$ and for $\a_1=\b_1$ we get $2e^{-\m_{\a_1}}/\cosh\mu_{\a_1}$. Therefore for the case when $a=b$ we have the result
\be
\cA(\mu)^{p-1} \( \sum_{\a_1\neq\b_1}\frac{e^{-\m_{\a_1}-\m_{\b_1}}}{\cosh\mu_{\a_1}\cosh\mu_{\b_1}} +\sum_{\a_1} \frac{2e^{-\m_{\a_1}}}{\cosh\mu_{\a_1}} \)~.
\ee
Now consider the case $a\neq b$. Without loss of generality assume $a=1$ and $b=2$ that is $j_2=i_1$. Start by, considering the following fermions inside the trace
\be
\sum_{\a_1,\a_2,\a'_1,\a'_2} \tr(\cdots \jb_{i_2\a_2} \jb_{i_1\a_1}  \j^{j_1\a_1} \j^{i_1,\a_2} \cdots  \jb_{i_1\a'_2} \jb_{j_1\a'_1}  \j^{i_1\a'_1} \j^{i_2\a'_2} \cdots ~e^{-\mu_r Q^r})~.\no
\ee
From the above expression, to get a non-zero trace we see that the contracted $i_2$ index restricts $\a_2=\a'_2$ and the contracted $j_1$ index restricts $\a_1=\a'_1$. When $\a_1\neq\a_2$ the contribution to the trace is $(\cosh\m_{\a_1})^{-2}(\cosh\m_{\a_2})^{-2}$ and when $\a_1=\a_2$ the contribution to the trace is $2e^{-\m_{\a_1}}/\(\cosh \m_{\a_1}\)^3$. Therefore for the case $a\neq b$ we have the result
\be
\cA(\mu)^{p-2} \( \sum_{\a_1\neq\a_2}(\cosh\m_{\a_1})^{-2}(\cosh\m_{\a_2})^{-2} +\sum_{\a_1} \frac{2e^{-\m_{\a_1}}}{\(\cosh \m_{\a_1}\)^3}\)~.
\ee
As a check, we see that for $M=1$ where there is no difference between the cases $a=b$ and $a\ne b$ and therefore the two results give the same answer.

Next let us estimate how many cases are there like $a=b$ and $a\neq b$. Since $a$ and $b$ take values $1,2,\cdots,p$, there are a total of $p^2$ pairs of which $p$ are of $a=b$ type and $p(p-1)$ are of $a\neq b$ type. This implies that the probability that $a=b$ is $1/p$ and $a\neq b$ is $(p-1)/p$. Therefore, for large $p$, cases like $a\neq b$ are much more likely to occur than cases like $a = b$.

For generic $p_{12}$, since each site index sets have distinct indices, this procedure can be repeated independently for every pair $\(\{i_a,i_b\},\{j_a,j_b \}\)$ with $i_a$ and $j_b$ in $I_1\cap I_2$ and the net correction factor due to non-vanishing $p_{12}$ is
\begin{align}
\cA(\mu)^{-2p_{12}} \( \sum_{\a_1\neq\a_2}(\cosh\m_{\a_1})^{-2}(\cosh\m_{\a_2})^{-2} +\sum_{\a_1} \frac{2e^{-\m_{\a_1}}}{\(\cosh \m_{\a_1}\)^3}\)^{p_{12}}\equiv \cC(-\mu)^{p_{12}}
\end{align}
where, we have used the definition in \eqref{defC}. Summing over $p_{12}$, we get that for every $H$-chord, there is an associated factor of
\be
\exp\(\lambda(\cC(-\mu)-1)\)~.
\ee
The factor of $\cC(-\m)-1$ can be simplified further
\begin{align}
\cC(-\mu)-1&=\frac{1}{\cA(\mu)^2}\( \sum_{\a_1\neq\b_1}\frac{1}{(\cosh\mu_{\a_1}\cosh\mu_{\b_1})^{2}} +\sum_{\a_1} \frac{2e^{-\m_{\a_1}}}{\(\cosh\mu_{\a_1}\)^3}- \cA(\mu)^2 \)\no\\
&=\frac{1}{\cA(\mu)^2} \sum_{\a=1}^M \frac{e^{-2\m_\a}}{\(\cosh\mu_\a\)^4}~.
\end{align}
Hence, we arrive at the rule that for every $H$-chord, there is an associated factor of
\be\label{UMrule11}
\exp\(\frac{\lambda}{\cA(\mu)^2} \sum_{\a=1}^M \frac{e^{-2\m_\a}}{\(\cosh\mu_\a\)^4} \)~.
\ee
The results in \eqref{UMrule11}, \eqref{UMrule22} and \eqref{UMrule33} give rise to the chord diagram rules summarized in \eqref{UMrule1}, \eqref{UMrule2} and \eqref{UMrule3}.

The sum over chord diagrams can then be evaluated using the same transfer matrix techniques. The appearance of the usual structure of chords originates in the form of the random couplings in the Hamiltonian, as the chord structure comes from the contraction of the latter. The group theory structure only changes the weights that chords receive. Summing the moments for chemical potentials scaling as $O(N^0)$ to obtain the partition function suffers from the same problem as in the $U(1)$ model due to the appearance of a $k^2$ exponent in  \eqref{genericUMmoment}, but other scalings can similarly be analyzed.

\section{Details of the 2-point and 4-point functions} \label{app:details_2pf}

In the calculation of the 2-point function and the 4-point function in the presence of a chemical potential (sections \ref{charged operator} and \ref{sec:4pf_chemical_potential}), we need to map pairs of unoriented chords to pairs of oriented chords, and use fig.\ \ref{fig:allowed_overlaps} in order to evaluate the latter. This mapping is as follows for the 2-point function:
\begin{enumerate}
\item There are $k/2$ $H$-chords, each giving a pair of oriented chords of fig.\ \ref{fig:allowed_overlaps_1}. For each of those we should assign
\begin{equation}
\exp\left[\frac{p^2}{N} \left(2e^{-\mu } \cosh(\mu )-1\right)\right].
\end{equation}
\item There are $\binom{k/2}{2} -\kappa _H=\frac{k(k-2)}{8} -\kappa _H$ pairs of non-intersecting $H$-chords, each giving one of fig.\ \ref{fig:allowed_overlaps_3} and \ref{fig:allowed_overlaps_4}, or fig.\ \ref{fig:allowed_overlaps_1} and \ref{fig:allowed_overlaps_2}. So anyway for each of those we have
\begin{equation}
\exp\left[\frac{p^2}{N} \left(4\cosh(\mu )^2-2\right)\right] .
\end{equation}
\item There are $\frac{k(k-2)}{4} +2\kappa _H$ remaining pairs of oriented solid chords, each giving $e^{-p^2/N} $.
\item Now we move on to include the dashed chord. There are $\frac{k_1-\kappa _{HM} }{2} $ pairs of fig. \ref{fig:allowed_overlaps_2} with the upper chord dashed for $p_M$, and the same number of pairs of fig.\ \ref{fig:allowed_overlaps_1} with the upper chord dashed for $\bar p_M$. So we get the following expression to the power $\frac{k_1-\kappa _{HM} }{2} $
\begin{equation}
\begin{split}
 \exp\left[\frac{pp_M}{N} (2e^{\mu } \cosh(\mu )-1 )\] \exp\left[\frac{p\bar p_M}{N} (2e^{-\mu } \cosh(\mu )-1 )\] .
\end{split}
\end{equation}
Note that in the current convention, $k_1$ is the number of Hamiltonian nodes enclosed by the dashed chord.
\item There are $\frac{k_2-\kappa _{HM} }{2} $ pairs of: fig.\ \ref{fig:allowed_overlaps_3} where the left chord is dashed and of size $p_M$, and fig.\ \ref{fig:allowed_overlaps_4} where the left chord is dashed and of size $\bar p_M$, the product of which is
\begin{equation}
\begin{split}
\exp\left[\frac{pp_M}{N} (2e^{-\mu } \cosh(\mu ) -1)\] \exp\left[\frac{p\bar p_M}{N} (2e^{\mu } \cosh(\mu ) -1)\] .
\end{split}
\end{equation}
\item There are $\frac{k}{2} +\kappa _{HM} $ remaining pairs of the $p_M$ dashed chord with an oriented solid chord, and the same for $\bar p_M$, not allowed to have common indices, contributing
\begin{equation}
\exp\left[- \frac{p(p_M+\bar p_M)}{N}  \right]
\end{equation}
to the power $\frac{k}{2} +\kappa _{HM} $.
\item The pair of oriented dashed chords is of the form fig.\ \ref{fig:allowed_overlaps_2} (with dashed chords) and so gives
\begin{equation}
\begin{split}
\exp\left[  \frac{p_M \bar p_M}{N} \left( 2e^{\mu } \cosh(\mu )-1\right) \right] .
\end{split}
\end{equation}
\end{enumerate}
For the 4-point function the counting is the following:
\begin{enumerate}
\item Overall factor of $(\cosh \mu )^{N-kp-2p_{M1} -2p_{M2} } $ for the entire chord diagram.
\item
$k/2$ unoriented solid chords, each giving $\exp\left[ \frac{p^2}{N} (2e^{-\mu } \cosh \mu -1)\right] $.
\item
$\binom{k/2}{2} -\kappa _H$ pairs of non-crossing unoriented solid chords, each giving $\exp\left[ \frac{p^2}{N} \left(  4\cosh(\mu )^2-2\right) \right] $.
\item
The remaining number of pairs of oriented solid chords is $\frac{k(k-2)}{4} +2\kappa _H$, each giving $\exp\left[ -p^2/N\right] $.
\item
The first unoriented dashed chord does not intersect $k/2-\kappa _{HM1} $ unoriented solid chords, each giving $\exp\left[ \frac{pp_{M1}}{N} \left( 4\cosh(\mu )^2-2\right) \right] $. Same for the second dashed chord replacing $1 \to 2$.
\item There are remaining $k+2\kappa _{HM1} $ pairs of oriented first dashed chord - oriented solid chord, each giving $\exp\left[ -\frac{pp_{M1} }{N} \right] $. Similarly for the second one.
\item
The dashed chords among themselves give
\begin{equation}
\exp\left[ \frac{p_{M1} ^2}{N} \left( 2e^{\mu } \cosh \mu  -1\right) \right]  \exp \left[ \frac{p_{M2} ^2}{N} \left( 2e^{\mu } \cosh \mu  - 1\right) \right] \exp \left[ -4 \frac{p_{M1} p_{M2} }{N} \right] .
\end{equation}
\end{enumerate}

\newpage
\bibliography{CSYK}
\bibliographystyle{JHEP}

\end{document}